\documentclass[fleqn,usenatbib]{mnras}

\usepackage{newtxtext,newtxmath}

\usepackage[T1]{fontenc}

\DeclareRobustCommand{\VAN}[3]{#2}

\let\VANthebibliography\thebibliography
\def\thebibliography{\DeclareRobustCommand{\VAN}[3]{##3}\VANthebibliography}

\usepackage{graphicx}	
\usepackage{amsmath}	
\usepackage{ulem}
\usepackage{CJKutf8}
\usepackage{ragged2e}  

\newcommand{\hdcc}{{\tt hd-cc}}
\newcommand{\mhdcc}{{\tt mhd-cc}}

\newcommand{\mhd}{{\tt mhd-fid}}
\newcommand{\mhdb}{{\tt mhd-tf5}}
\newcommand{\mhda}{{\tt mhd-tf6}}
\newcommand{\mhdc}{{\tt mhd-fid}}
\newcommand{\mhdd}{{\tt mhd-tc7}}
\newcommand{\mhde}{{\tt mhd-tc6}}
\newcommand{\mhdf}{{\tt mhd-05pl}}
\newcommand{\mhdg}{{\tt mhd-ti7}}
\newcommand{\mhdcd}{{\tt mhd-cd-fid}}

\newcommand{\mhdcda}{{\tt mhd-cd-v1}}
\newcommand{\mhdcdb}{{\tt mhd-cd-fid}}
\newcommand{\mhdcdc}{{\tt mhd-cd-v20}}
\newcommand{\mhdcdd}{{\tt mhd-cd-dr1}}
\newcommand{\mhdcde}{{\tt mhd-cd-dr100}}
\newcommand{\mhdcdf}{{\tt mhd-cd-tceil6}}
\newcommand{\mhdcdg}{{\tt mhd-cd-v2}}
\newcommand{\mhdcdh}{{\tt mhd-cd-v10}}

\newcommand{\kai}{\,{\rm K}}
\def\gsim{\;\rlap{\lower 2.5pt
 \hbox{$\sim$}}\raise 1.5pt\hbox{$>$}\;}
\def\lsim{\;\rlap{\lower 2.5pt
   \hbox{$\sim$}}\raise 1.5pt\hbox{$<$}\;}

\title[]{{\it Eppur Si Muove}: Self-Sustained Streaming Motions in Multi-Phase MHD}
\author[Wang et al.]{
Chaoran Wang,$^{1}$ 
S. Peng Oh,$^{1}$
Yan-Fei Jiang(姜燕飞)$^{2}$,
Ish Kaul$^{1}$
\\
${}^{1}$ Department of Physics, University of California, Santa Barbara, CA 93106, USA. \\
$^{2}$Center for Computational Astrophysics, Flatiron Institute, New York, NY 10010, USA\\
}

\date{Accepted XXX. Received YYY; in original form ZZZ}

\pubyear{2022}

\begin{document}
\begin{CJK*}{UTF8}{gbsn}
\label{firstpage}
\pagerange{\pageref{firstpage}--\pageref{lastpage}}
\maketitle

\end{CJK*}

\def\lsim{\;\rlap{\lower 2.5pt
   \hbox{$\sim$}}\raise 1.5pt\hbox{$<$}\;}


\begin{abstract}
Radiative cooling can drive dynamics in multi-phase gas. A dramatic example is hydrodynamic `shattering', the violent, pressure-driven fragmentation of a cooling cloud which falls drastically out of pressure balance with its surroundings. We run MHD simulations to understand how shattering is influenced by magnetic fields. In MHD, clouds do not `shatter' chaotically. Instead, after initial fragmentation, both hot and cold phases coherently `stream' in long-lived, field-aligned, self-sustaining gas flows, at high speed ($\sim 100 \, {\rm km \, s^{-1}}$). MHD thermal instability also produces such flows. They are due to the anisotropic nature of MHD pressure support, which only operates perpendicular to B-fields. Thus, even when $P_{\rm B} + P_{\rm gas} \approx$const, pressure balance only holds perpendicular to B-fields. Field-aligned gas pressure variations are unopposed, and results in gas velocities $v \sim (2 \Delta P/\rho)^{1/2}$ from Bernoulli's principle. Strikingly, gas in adjacent flux tubes {\it counter-stream} in opposite directions. We show this arises from a cooling-induced, MHD version of the thin shell instability. Magnetic tension is important both in enabling corrugational instability and modifying its non-linear evolution. Even in high $\beta$ hot gas, streaming can arise, since magnetic pressure support grows as gas cools and compresses. 
Thermal conduction increases the sizes and velocities of streaming cloudlets, but does not qualitatively modify dynamics. 
These results are relevant to the counter-streaming gas flows observed in solar coronal rain, as well as multi-phase gas cooling and condensation in the ISM, CGM and ICM.  

\end{abstract}




\section{Introduction}

How does a multi-phase medium develop? The classic mechanism is thermal instability: slightly overdense gas cools, loses pressure, and undergoes compression and runaway cooling, until it reaches a new equilibrium \citep{field65}. Multi-phase gas is ubiquitous in astrophysics, and accordingly a plethora of studies have investigated thermal instability in environments ranging from the interstellar medium (ISM), circumgalactic medium (CGM), intracluster medium (ICM), and solar corona (indeed, the last was the motivation for the original study of thermal instability by \citealt{parker53}). Often, radiative cooling does not occur in isolation, but alongside other processes which act to suppress the development of a multi-phase medium. For instance, buoyant oscillations (operating on a buoyancy timescale $t_{\rm buoy} \sim t_{\rm ff}$) suppresses thermal instability in a stratified medium, while turbulence (operating on the eddy turnover time $t_{\rm eddy}$) suppresses multi-phase gas via mixing. In this case, the ratio of these timescales $t_{\rm cool}/t_{\rm ff}$ \citep{mccourt12,sharma12} in stratified gas and $t_{\rm cool}/t_{\rm turb}$ \citep{gaspari13,tan21} in turbulent gas governs the development of a multi-phase medium. 

A key issue is physical processes which affect the morphology and dynamics of the cooler phase, in the non-linear saturated state of thermal instability. For instance, classical diffusive thermal conduction suppresses thermal instability on scales smaller than the Field length $\lambda_{\rm F}$, which potentially sets a characteristic scale for cool gas. Note, however, that the Field length along the magnetic field by Spitzer conduction at $T\sim 10^4$K is {\it much} smaller ($\sim 10^{-4}$pc) than observed cold filament lengths in galaxy clusters ($\sim 10$ kpc), so additional physical processes must be at play \citep{sharma10}. Self-sustained turbulence generated by thermal instability can also arise in hydrodynamic simulations \citep{vazquez-semadeni00,kritsuk02,iwasaki14}, though at a level likely sub-dominant to extrinsically driven turbulence. When thermal conduction is present, evaporation and condensation flows dominate, with cloud disruption occurring due to the Darrieus-Landau instability \citep{jennings21}. If extrinsic turbulence is present, the competition between turbulent fragmentation and cooling-induced coagulation \citep{gronke22-coag} can create a scale-free power law distribution of cloud masses $dn/dm \propto m^{-2}$ \citep{gronke22-turb,tan23-cloud-atlas,das23}, with equal mass per logarithmic interval. Magnetic fields can significantly change morphology and dynamics \citep{sharma10,choi12,hennebelle19,jennings21,das23}, due to both MHD forces as well as the field-aligned nature of energy and momentum transport from conduction and viscosity, given that gyro-radii are much smaller than collisional mean free paths. 

One key fragmentation process that has been identified in recent years is the development of strong thermal pressure gradients due to radiative cooling. If a cooling fragment maintains sonic contact with its surroundings, it will cool isobarically. However, if the cooling time falls far below the sound crossing time $t_{\rm cool} \ll t_{\rm sc}$, the cloud falls drastically out of pressure balance, and the subsequent cloud-crushing shock can {\it shatter} the cloud into tiny fragments \citep{mccourt18}. These authors argued that a crucial lengthscale in pressure-confined clouds was the cooling length $c_{\rm s} t_{\rm cool}$ (evaluated at its minimum at $T \sim 10^4$K), when $t_{\rm cool} \sim t_{\rm sc}$, similar to the Jeans length $\lambda_{\rm G} \sim c_s t_{\rm ff}$ in gravitationally confined clouds, when $t_{\rm ff} \sim t_{\rm sc}$. The cloud is strongly compressed by its surroundings, overshoots, then explodes into many small pieces in a rarefaction wave \citep{gronke20-mist,yao25}. Shattering, or the closely related splattering from acoustic oscillations \citep{waters19-linear} can occur in linear thermal instability \citep{gronke20-mist,das21}, or at radiative shocks, where pre-existing cold gas is compressed \citep{mellema02,mandelker19-shatter}. Shattering, alongside turbulent fragmentation \citep{gronke22-turb}, and hydrodynamic instabilities \citep{cooper09,sparre19,liang20}, can contribute to the myriad observed small-scale cold gas structure \cite{mccourt18}. Shattering has even been observed in simulations of cosmic sheets \citep{mandelker19-shatter, mandelker21}, and suggested to be important in the formation of Lyman limit systems. 

Surprisingly, the impact of magnetic fields on shattering is unexplored. Shattering is driven by extreme thermal pressure gradients, which develop as gas cools. However, magnetic fields provide non-thermal pressure support which is {\it amplified} by flux freezing during cooling and compression; plasma $\beta$ can fall by orders of magnitude in the cool phase. In principle, cosmic rays can also contribute non-thermal pressure support, but the fact that they can diffuse or stream out of cooling gas strongly dilutes their effect \citep{huang22}. By contrast, B-fields are tied to the plasma by flux freezing. Importantly, unlike thermal or cosmic ray pressure, MHD forces are {\it anisotropic}: magnetic tension prevents field lines from bending, while magnetic pressure operates perpendicular to field lines. Since fluid elements can slide freely along field lines, cold gas has a characteristic field-aligned filamentary morphology in MHD thermal instability, in initially high $\beta$ plasma \citep{sharma10,xu19}, although cold filaments can form both along or perpendicular to B-fields in strong magnetic fields \citep{jennings21}. Simulations which have explored shattering have thus far been purely hydrodynamic. In addition, simulations of MHD thermal instability have only been run in regimes where strong thermal pressure gradients do not develop.   

In this paper we demonstrate a novel phenomenon where the cold filaments that condense out of the thermally unstable hot medium stream along magnetically dominated flux tubes. 
The streaming motion are fast, $\sim 100 {\rm km \, s^{-1}}$, which is highly supersonic relative to the sound speed of $T \sim 10^{4}$K gas, and comparable to turbulent velocities expected in the multiphase CGM and ICM. They could thus contribute significantly to non-thermal line broadening observed in the CGM. In addition, the predicted velocities are in good agreement with observed velocities ($\sim 70-80 \, {\rm km \, s^{-1}}$) of $T\sim 10^{4}$K cold gas streaming in the hot ($T \sim 10^{6}$K) solar corona \citep{alexander13}. Such `siphon flows' are also thought to arise as a result of cooling-driven pressure gradients \citep{fang15,claes20}.  

The outline of this paper is as follows. In \S\ref{sec:methods}, we outline our simulation methodology. In \S\ref{sec:results}, we summarize our main simulation results. In \S\ref{sec:streaming}, which is the heart of the paper, we explain the mechanisms behind streaming, with particular focus on the physical origins of counter-streaming flows. In \S\ref{sec:parameters}, we discuss parameter dependence of streaming flows on factors such as conduction, cooling and magnetic field strength. In \S\ref{sec:discussion}, we connect with previous work, and discuss physical and observational implications of our work. Finally, we conclude in \S\ref{sec:conclusions}.   


\section{Methodology}
\label{sec:methods}
\begin{table*}
	\centering
	\label{tab:listOfSims}
	\begin{tabular}{lccccccr} 
		\hline
            \hline
            \multicolumn{8}{c}{\bf Non-conduction runs}\\
        Name       & $(T_{\rm floor}, T_{\rm ceil}, T_{\rm init})^a$ & n$^b$  & $\beta^c$ &
        $\alpha^d$ & $\Delta x ({\rm pc})^e$ & $ n_x\times n_y(\times n_z)^f$ & Setup$^g$ \\
		\hline 
    \hdcc  & $(10^4, 10^8, 10^7)$& $n_{\rm hot}=0.05~{\rm cm^{-3}}, \chi=10$& $\infty$ & -- & 9.77 & $1024\times1024$ & CC\\
    \mhdcc & $(10^4, 10^8, 10^7)$& $n_{\rm hot}=0.05~{\rm cm^{-3}}, \chi=10$& 1        & -- & 9.77 & $4096\times1024$ & CC\\
    \mhda  & $(10^6, 10^8,2\times10^6)$& $\langle n_i \rangle=0.25~{\rm cm^{-3}}$ & 1  & 0 & 9.77 & $1024\times1024$         & TI\\
    \mhdb  & $(10^5, 10^8,2\times10^6)$& $\langle n_i \rangle=0.25~{\rm cm^{-3}}$ & 1  & 0 & 9.77 & $1024\times1024$         & TI\\
    \mhdc  & $(10^4, 10^8,2\times10^6)$& $\langle n_i \rangle=0.25~{\rm cm^{-3}}$ & 1  & 0 & 9.77 & $1024\times1024$         & TI\\
    \mhdd  & $(10^4, 10^7,2\times10^6)$& $\langle n_i \rangle=0.25~{\rm cm^{-3}}$ & 1  & 0 & 9.77 & $1024\times1024$         & TI\\
    \mhde  & $(10^4, 2.5\times10^6,2\times10^5)$& $\langle n_i \rangle=0.25~{\rm cm^{-3}}$& 1 & 0 & 9.77 & $1024\times1024$  & TI\\
\mhdf  & $(10^4, 10^8,2\times10^6)\ (\Lambda\propto T^{0.5}$)& $\langle n_i \rangle=0.25~{\rm cm^{-3}}$&1&0&9.77&$1024\times1024$ &TI\\
    \mhdg  & $(10^4, 10^8,10^7)$       & $\langle n_i \rangle=0.25~{\rm cm^{-3}}$ & 1  & 0 & 9.77 & $1024\times1024$         & TI \\
\texttt{mhd-256} & $(10^4, 10^8,2\times10^6)$&$\langle n_i \rangle=0.25~{\rm cm^{-3}}$& 1     & 0   &39.1& $256\times256$    & TI \\
\texttt{mhd-2048}& $(10^4, 10^8,2\times10^6)$&$\langle n_i \rangle=0.25~{\rm cm^{-3}}$& 1     & 0   &4.88& $2048\times2048$  & TI  \\
\texttt{mhd-b10} & $(10^4, 10^8,2\times10^6)$&$\langle n_i \rangle=0.25~{\rm cm^{-3}}$& 10    & 0   &9.77& $1024\times1024$  & TI \\
\texttt{mhd-b100}& $(10^4, 10^8,2\times10^6)$&$\langle n_i \rangle=0.25~{\rm cm^{-3}}$&100    & 0   &9.77& $1024\times1024$  & TI \\
\texttt{mhd-bxy} & $(10^4, 10^8,2\times10^6)$&$\langle n_i \rangle=0.25~{\rm cm^{-3}}$&1(diag)& 0   &9.77& $1024\times1024$  & TI \\
\texttt{mhd-a167}& $(10^4, 10^8,2\times10^6)$&$\langle n_i \rangle=0.25~{\rm cm^{-3}}$&1      & -5/3&9.77& $1024\times1024$  & TI \\
\texttt{mhd-3d}  & $(10^4, 10^8,2\times10^6)$&$\langle n_i \rangle=0.25~{\rm cm^{-3}}$&1      & 0   &39.1& $256\times256\times256$&TI\\
    \hline
    \multicolumn{8}{c}{\bf Conduction runs}\\
    Name       & $(T_{\rm floor}, T_{\rm ceil}, T_{\rm init}, \alpha_{\parallel}/\alpha_{\rm FID}, \alpha_{\parallel}/\alpha_{\perp})^a$
    & density$^b$ & $\beta^c$ &
        $\alpha^d$ & $\Delta x ({\rm pc})^e$ & $n_x\times n_y^f$ & Setup$^g$ \\
    \hline
    \texttt{hd-vs-cd} & $(10^5,10^8,10^7,0.033, 1)$ & $n_{\rm hot}=0.05~{\rm cm^{-3}}, \chi=10$ & $\infty$& -- & 9.77 & $1024\times1024$ &VS\\
    \texttt{mhd-vs-cd}& $(10^5,10^8,10^7,1    ,30)$ & $n_{\rm hot}=0.05~{\rm cm^{-3}}, \chi=10$ & 1       & -- & 9.77 & $1024\times1024$ &VS\\
    \mhdcda & $(10^4, 10^8,2\times10^6, 0.2, 30)$ &$\langle n_i \rangle=0.25~{\rm cm^{-3}}$&1& 0 & 9.77 & $1024\times1024$         & TI\\
    \mhdcdg & $(10^4, 10^8,2\times10^6, 0.4, 30)$ &$\langle n_i \rangle=0.25~{\rm cm^{-3}}$&1& 0 & 9.77 & $1024\times1024$         & TI\\
    \mhdcdb & $(10^4, 10^8,2\times10^6, 1, 30)$ &$\langle n_i \rangle=0.25~{\rm cm^{-3}}$&1& 0 & 9.77 & $1024\times1024$         & TI\\
    \mhdcdh & $(10^4, 10^8,2\times10^6, 2, 30)$&$\langle n_i \rangle=0.25~{\rm cm^{-3}}$&1& 0 & 9.77 & $1024\times1024$         & TI\\
    \mhdcdc & $(10^4, 10^8,2\times10^6, 4, 30)$&$\langle n_i \rangle=0.25~{\rm cm^{-3}}$&1& 0 & 9.77 & $1024\times1024$         & TI\\
    \mhdcdd & $(10^4, 10^8,2\times10^6, 0.033, 1)$  &$\langle n_i \rangle=0.25~{\rm cm^{-3}}$&1& 0 & 9.77 & $1024\times1024$         & TI\\
    \mhdcde & $(10^4, 10^8,2\times10^6, 3.3, 100)$&$\langle n_i \rangle=0.25~{\rm cm^{-3}}$&1& 0 & 9.77 & $1024\times1024$         & TI\\
    \mhdcdf & $(10^4,2.5\times10^6,2\times10^6, 1, 30)$&$\langle n_i \rangle=0.25~{\rm cm^{-3}}$&1&0&9.77&$1024\times1024$       & TI\\
\texttt{mhd-cd-256} & $(10^4,10^8,2\times10^6,1,30)$&$\langle n_i \rangle=0.25~{\rm cm^{-3}}$&1&0&39.1&$256\times256$ & TI\\
\texttt{mhd-cd-2048}& $(10^4,10^8,2\times10^6,1,30)$&$\langle n_i \rangle=0.25~{\rm cm^{-3}}$&1&0&4.99&$2048\times2048$ &TI\\
        \hline
        \hline
	\end{tabular}
	\\ \ \\
\caption{List of simulations. Following the superscripts in the top row: \\
a) $T_{\rm init}$ is the initial gas temperature, $T_{\rm ceil}$ is the ceiling temperature, and $T_{\rm floor}$ is the floor temperature. For the conduction runs listed in the lower panel, the heat diffusivity parallel to the magnetic fields, $\alpha_{\parallel}$ and the ratios of the anisotropic diffusivity along the magnetic fields to the isotropic component, $\alpha_\parallel/\alpha_{\perp}$ are listed. $\alpha_{\parallel}$ is scaled by the fiducial value, $\alpha_{\rm FID}=1.5\times10^{28}~{\rm cm^2\, s^{-1}}$. The \citet{SutherlandDopita93} cooling function is adopted by default; and \mhdf\ uses the $\Lambda(T) \propto T^{0.5}$ power law cooling function as noted. \\
b) Initial gas number densities. For the cooling cloud and vertical slab set up, initial hot gas density ($n_{\rm hot}$) and cloud overdensity $\chi$ is given; for the thermal instability setup, the average initial gas density ($\langle n_{\rm i} \rangle$) for the almost uniform medium is given. \\
c) Initial plasma beta ($\beta=\infty$ stands for purely hydrodynamic runs). \\
d) Power law index of the spectrum of the seed perturbation. \\
e) Spatial resolution of the simulation.\\
f) Number of grids along each dimension. For 2D runs $n_x\times n_y$ are shown; and for 3D runs, $n_x\times n_y\times n_z$ is given.\\
g) Type of setup: ``CC'' stands for a cooling clouds setup where cold clouds with large overdensity are manually placed in the hot medium at the beginning of the simulation. ``VS'' stands for a vertical slab setup which is the same as CC but initially the shape of the cold cloud is a vertical slab. ``TI'' stands for a thermal instability setup where cold clumps grow from small initial perturbations.}
\end{table*}

We perform numerical simulations using the magnetohydrodynamics (MHD) code Athena++ \citep{2020ApJS..249....4S}. We adopt the HLLC and HLLD Riemann solvers for the hydrodynamic and MHD runs, respectively. 
We solve the MHD equations: 
\begin{align}
&\frac{\partial \rho}{\partial t} + \nabla \cdot (\rho \mathbf{v}) = 0;\\
&\frac{\partial (\rho \mathbf{v})}{\partial t} + \nabla \cdot \left(\rho \mathbf{v} \mathbf{v} + P_{\rm tot} \boldsymbol{I} - \frac{\mathbf{B} \mathbf{B}}{4\pi}\right) = 0; \\
& \frac{\partial e_{\rm tot}}{\partial t} + \nabla \cdot \left[\left(e_{\rm tot} + P_{\rm tot}\right) \mathbf{v} + \mathbf{F}_{\parallel} + \mathbf{F}_{\rm iso} - \frac{(\mathbf{v} \cdot \mathbf{B})\mathbf{B}}{4\pi}
\right] = - n^2 \mathcal{L} ;\\ 
&\frac{\partial \mathbf{B}}{\partial t} - \nabla \times (\mathbf{v}\times \mathbf{B}) = 0;
\end{align}
 where the total pressure
 \begin{equation}
     P_{\rm tot} = P_{\rm th} + P_{\rm B} = nk_B T + \frac{B^2}{8\pi}
 \end{equation} 
 is the sum of gas thermal and magnetic pressure; $n=\rho/\mu m_p$ is the gas number density; and $m_p$ is the proton mass. The total energy density
\begin{equation} 
  e_{\rm tot}=e_{\rm k}+e_{\rm th}+e_{\rm B} = \frac{1}{2} \rho v^2 + \frac{P_{\rm th}}{\gamma-1} + \frac{B^2}{8\pi}
\end{equation}
  is the sum of gas kinetic, thermal and magnetic energy density, where $\gamma=5/3$ is the gas adiabatic index. The terms $\mathbf{F}_{\parallel}, \mathbf{F}_{\rm iso}$ denote the thermal conductive flux parallel to B-field lines and a smaller isotropic component respectively. They will be described below. We do not implement viscosity; thermal instability simulations which do so find that it regulates flow speed and thus the rate at which structures form, though not the final non-linear outcome \citep{jennings21}.  
\begin{figure}
\includegraphics[width=\linewidth]{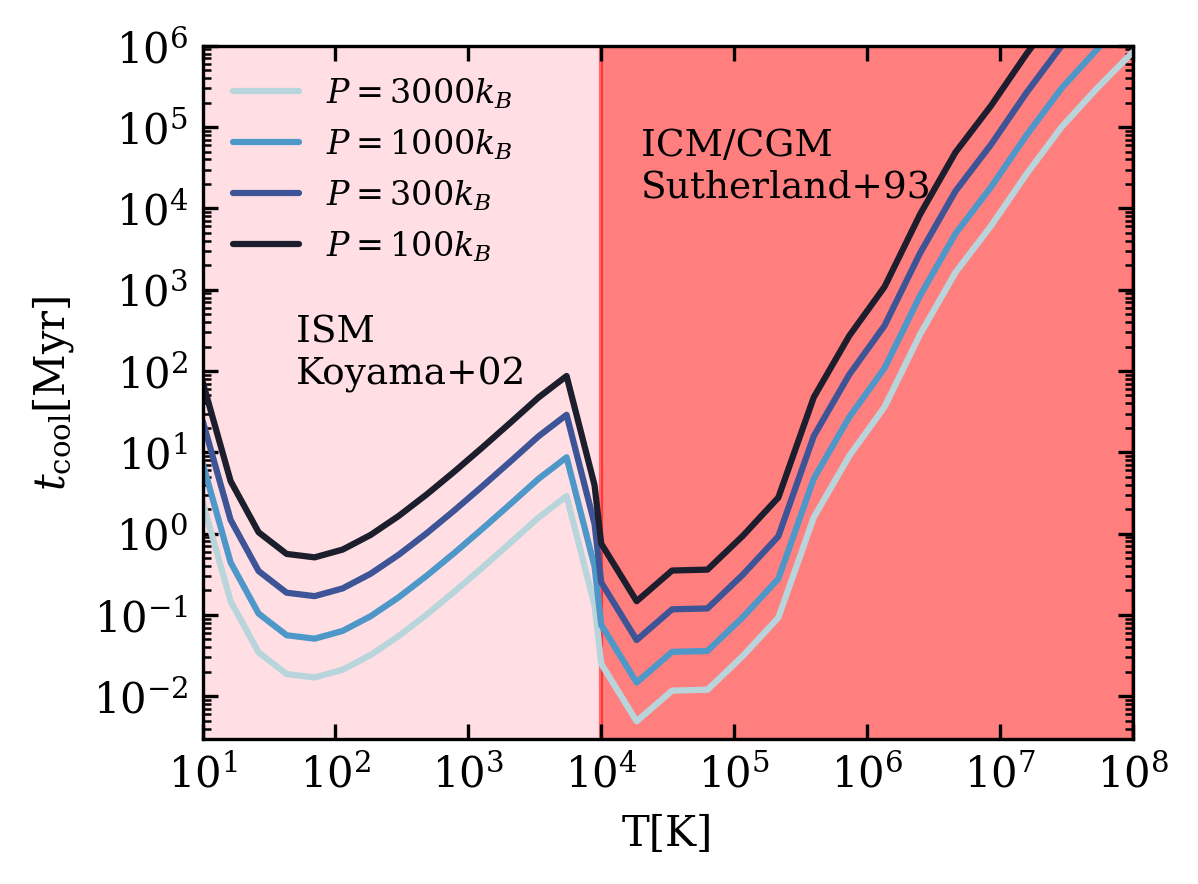}
    \caption{Analytic cooling time as a function of temperature as derived from \citealt{SutherlandDopita93} for the ICM and \citealt{KoyamaInutsuka02} for the ISM, for a range of ambient pressures.}
    \label{fig:ism_tcool}
\end{figure}

  The source term 
  \begin{equation} 
  n^2 \mathcal{L} = n_e n_i \Lambda(T) - n\Gamma
  \end{equation} 
  represents the net cooling rate, where $n^2 \Lambda(T)$ describes radiative cooling and $n\Gamma$ describes heating. 
The majority of our similations mimic conditions in the CGM and ICM. In this case, the gas is always fully ionized; we implement a temperature floor of $T= 10^{4}$K to mimic the effects of photoionization. For these simulations, we adopt the collisional ionisation equilibrium cooling functions $\Lambda(T)$ calculated by \citet{SutherlandDopita93} with solar metallicity, 
with $n_i = \rho /\mu_i m_p$, $n_e=\rho /\mu_e m_p$, $n=\rho/\mu m_p$ are the ion, electron and total particle number density, where we adopt the particle weights $\mu=0.61$, $\mu_e=1.18$, and $\mu_i=1.26$. Fig. \ref{fig:ism_tcool} shows the cooling time across the entire range of temperatures we consider in our simulations for different ambient pressures, as derived from the analytic cooling curves for the ICM and ISM.
  In these idealized simulations, we do not simulate realistic heating processes for the CGM and ICM conditions (e.g., supernovae and AGN feedback). Instead, we employ idealized heating which enforces global thermal equilibrium by fiat. 

At every time-step, the total cooling in the box is calculated, and the average cooling rate is equated to the average heating rate, so that $\Gamma = \langle n_e n_i \Lambda(T) \rangle/\langle n \rangle$. Similar prescriptions have been widely used in idealized simulations of thermal instability in stratified environments \citep{mccourt12,sharma12}, with results similar to simulations which incorporate more realistic heating prescriptions such as AGN feedback \citep{gaspari13,li14}. Note that this prescription assumes density weighted heating. We have also used volume-weighted heating (i.e., $n^{2} \mathcal{L} = n_{e} n_{i} \Lambda(T) - \Gamma$, so that the heating rate is identical for every grid cell) and found our results to be relatively insensitive to this change, as has previously been found \citep{sharma10}.

Thermal conduction results in a net heat flux from hot to cold gas, $F = - \kappa \nabla T$. It is well known that in multi-phase environments, convergence in cold gas morphology requires that the Field length $\lambda_{\rm F} \sim \sqrt{\kappa(T) T/(n^2 \Lambda(T))}$ 
of {\it cold} gas (which is typically very small) must be resolved \citep{koyama04,sharma10}. Note that convergence in other bulk quantities, such as the temperature PDF, may not require that the Field length be resolved. For instance, in turbulent mixing layers, \citet{tan21} show that conduction has little impact on mass flux $\dot{m}$ from hot to cold phases, if turbulent heat diffusion dominates over thermal conduction (which is generally true for transonic shear). Lower numerical resolution, which does not resolve the Field length, does not create any systematic biases in $\dot{m}$, although it increases its variance \citep{tan21}. However, because in this study we {\it are} interested in small scale gas dynamics, we implement thermal conduction to have numerically converged gas morphology. We implement field-aligned thermal conduction: 
\begin{equation}
\mathbf{F}_{\parallel} = - \kappa_{\parallel} \hat{\mathbf{b}} (\hat{\mathbf{b}} \cdot \nabla) T. 
\end{equation}
as well as a (smaller) isotropic component of thermal conduction, intended to model cross-field heat diffusion: 
\begin{equation}
\mathbf{F}_{\rm iso} = - \kappa_{\rm iso} \nabla T. 
\end{equation}
We set $\kappa_{\parallel}/\kappa_{\rm iso} =30$ as our fiducial value and show how our results change as we vary this ratio.

In ionized gas, the Spitzer conduction coefficient is appropriate: 
\begin{equation}
\kappa_{\parallel} = {5 \times 10^{-7}} T^{5/2} {\rm erg \, s^{-1} \, K^{-7/2} \, cm^{-1}}. 
\label{eq:spitzer}
\end{equation}
However, this gives a strongly temperature dependent Field length. In terms of column density, which is a density independent number, the Field length is: 
\begin{equation}
N_{\rm F} \sim n \lambda_{\rm F} \sim f_{\rm s} 10^{18}  \, T_{6}^2 \, {\rm cm^{-2}} 
\end{equation}
where we have assumed Spitzer conduction (equation \ref{eq:spitzer}), $\Lambda(T) \propto T^{-0.5}$, as crudely appropriate for $10^{4} \, {\rm K}< T < 10^{6}$K, and $f_{\rm S} \lsim 1$ is a numerical factor which encapsulates supression below Spitzer conductivity (e.g., due to tangled magnetic fields, or scattering other than Coulomb scattering). For isobaric cooling, this implies $\lambda_{\rm F} \propto T^{3}$, i.e. the Field length in the $T \sim 10^4\,$K cold gas will be $\sim 6$ orders of magnitude smaller than in the $T \sim 10^6\,$K hot gas. Resolving this is numerically infeasible \citep{sharma10}. Instead, we alter the temperature dependence of $\kappa(T)$ and adopt constant heat diffusivities $\alpha_{\parallel}, \alpha_{\rm iso}$, where the conductive heat flux ${\mathbf F} = - \alpha \nabla e_{\rm th}$, so that $\alpha$ has the dimensions of a spatial diffusion coefficient, $[ \alpha ] \sim L^2 T^{-1}$. 
The relation between $\kappa$ and $\alpha$ is\footnote{Since the conductive heat flux is ${\mathbf F} = {\mathbf F}_{\parallel} + {\mathbf F}_{\rm iso}$, the definitions below ensure that ${\mathbf F}_{\parallel} = - \alpha_{\parallel} \nabla e_{\rm th}$.} 

\begin{align}
&\kappa_\parallel = \frac{3}{2}(\alpha_{\parallel} -\alpha_{\rm iso}) \frac{ k_{\rm B} \rho}{\mu m_{\rm p}}; \\
&\kappa_{\rm iso} = \frac{3}{2}\alpha_{\rm iso} \frac{ k_{\rm B} \rho}{\mu m_{\rm p}}.
\end{align}

\begin{figure*}
     \centering
\includegraphics[width=\textwidth]{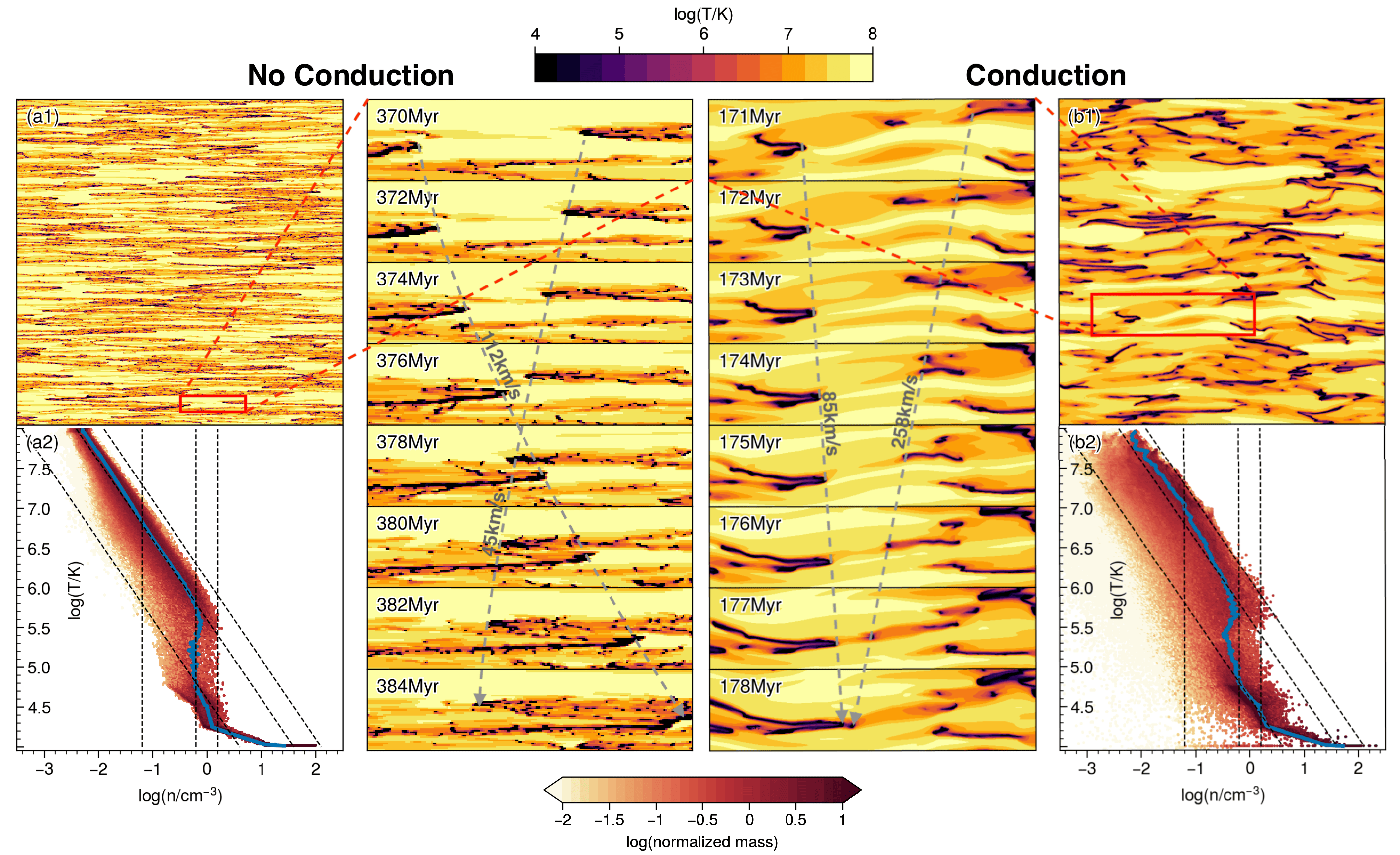}
       \caption[]{The streaming motions and thermal properties of multiphase gas arising from linear thermal instability. Panels (a1) and (b1) show gas temperature maps of \mhd\ at $t=370~{\rm Myr}$ and \mhdcd\ at $t=171~{\rm Myr}$, respectively. The corresponding $n-T$ phase plots are shown in the panels (a2), (b2), in which the dashed black lines annotate the isobaric (diagonal lines) and isochoric (vertical lines) tracks. The blue lines show the mass-weighted median gas density in each temperature bin. The two middle panel columns show how the gas within the red rectangular regions in panel (a1) and (b1) evolves with time. The gray dashed arrows sketch the streaming motion of the cold clouds by tracking the streaming heads. Each arrow is labelled by the average horizontal velocities of the head.}
\label{fig:rtp_all}
\end{figure*}

Under isobaric conditions, a constant $\alpha$ is equivalent to $\kappa(T) \propto \rho \propto T^{-1}$. It also mimics the effects of numerical diffusion, albeit in a controlled way \footnote{See \citet{tan21-lines} for simulations which explore how altering $\kappa(T)$ alters the temperature PDF in turbulent mixing layers.}. The Field length has a weaker temperature dependence $\lambda_{\rm F} \propto (n^2 \Lambda(T))^{-1/2} \propto T^{5/4}$, or $N_{\rm F} \propto T^{1/4}$ under isobaric conditions\footnote{We shall find that gas does not cool isobarically, since it is supported by magnetic pressure at low temperatures; thus the temperature dependence is even weaker.}. 
Thus, our choice of a constant diffusivity is a numerical convenience to ensure that the balance between thermal conduction and cooling is spatially resolved at most temperatures. For our fiducial parameters,  the Field length is:
\begin{equation} 
\lambda_{\rm F} \sim (\alpha t_{\rm cool})^{1/2} \sim 4 \, {\rm pc} \, \left( \frac{\alpha}{\alpha_{\rm FID}} \right)^{1/2} T_{4.3}^{1/2} \Lambda_{-21.7}^{-1/2} \left( \frac{n}{1 \,{\rm cm^{-3}}} \right)^{-1/2} 
\label{eq:field}
\end{equation}
where $\Lambda_{-21.7} = \Lambda(T)/10^{-21.7} \, {\rm erg \, s^{-1} \, cm^3}$. We have evaluated quantities at the minimum of $t_{\rm cool}$, i.e. when cooling peaks. The smallest relevant Field length thus is comparable to the grid scale in our simulations, which have a resolution of $\Delta x \sim 10$pc. The Field length in the cross-field direction (which is less important for our purposes) is a factor $(\alpha_{\rm iso}/\alpha_{\rm FID})^{1/2} \sim (30)^{-1/2} \sim 0.2$ smaller, and not resolved. In fact, we shall find that in conduction runs, cool blobs are considerably larger than the Field length at $T \sim 10^4$K, and thus are numerically resolved (\S\ref{sec:conduction}). Note that in our MHD simulations, the gas does not cool isobarically, since it is supported by magnetic pressure; it is lower density than might be expected. The density of $n \sim 1 \, {\rm cm^{-3}}$ chosen above is the typical density of gas as it cools isochorically (see phase plots in Fig \ref{fig:rtp_all}). We vary conduction coefficients and simulation resolution to see the effect of resolving the Field length, and also run simulations with no conduction.

As thermal conduction is a diffusive process, it is usually computationally expensive to implement. We employ a two moment approximation method for conduction similar to the approach used to treat cosmic rays in \cite{jiang18}. This is done by introducing a second equation: 
\begin{equation}
  \frac{1}{V_{\rm m}^2} \frac{\partial F}{\partial t} + \rho\nabla\left(\frac{e_{\rm th}}{\rho}\right) = -\frac{k_{\rm B} \rho F}{\mu m_{\rm p} (\gamma -1)\kappa} ,
\end{equation}
with an effective propagation speed $V_{\rm m}$. The latter represents the ballistic velocity of free electrons, which is $\sim \sqrt{m_{p}/m_{e}} \sim 45$ times larger than the gas sound speed\footnote{The analogous quantity in \citet{jiang18} is the reduced speed of light for free-streaming cosmic rays.}. In the limit that $V_{\rm m}$ goes to infinity, the equation reduces to the usual equation for heat conduction. As long as $V_{\rm m}$ is large compared to other characteristic velocities in the simulation, the solution is a good approximation to the true solution. We adopt $V_{\rm m} =1000~{\rm km/s}$ and also check that our results are converged with respect to $V_{\rm m}$, 
The timestep of this approach scales as $O(\Delta x)$, compared to traditional explicit schemes which scale as $O(\Delta x^2)$. Implicit schemes, which also have a linear scaling with resolution, are constrained by the fact that they require matrix inversion over the whole simulation domain, which can be slow and hinders parallelization. The module employs operator splitting to compute the transport and source terms, using a two step van-Leer time integrator; the  source term is  added implicitly. The algorithm was also used in \cite{tan21}, which presents some code tests.

In our fiducial setup, the initial B-fields are straight and horizontally oriented. Our fiducial setup initially had diagonal B-fields, which has also been adopted in some other works, but we found that this introduced numerical artifacts; our results were not rotationally invariant (see Appendix \ref{app:bfields}). For the sake of speed and to enable rapid exploration of parameter space, our simulations are 2D. We run some 3D simulations to ensure that our conclusions are not sensitive to dimensionality. Our simulations adopt periodic boundary conditions. 

Our fiducial simulations are initialized with seed isobaric fluctuations on top of a uniform gas background with $T_{\rm int}$ and $n_{\rm init}$. The gas temperature in the simulation has the upper and lower limit $T_{\rm ceil}$ and $T_{\rm floor}$. 
Table~1 lists all simulations performed in which we explore parameter space\footnote{In \S\ref{sec:counter-streaming}, we run specialized `slab' setups to understand counter-streaming motions, which we describe separately there.} . In addition to the setup that thermal instabilities grow from the initial seeds as mentioned above (denoted as ``TI'' in the ``Setup'' column of Table~1), we include simulations with the cooling cloud setup (denoted as ``CC'' in the ``Setup'' column of Table~1) where cold clouds are manually positioned in the domain, which once again has purely horizontal magnetic fields. These cooling cloud runs are included to better illustrate the streaming motions that we find. In addition, they are direct MHD analogs of hydrodynamic `shattering' simulations \citep{mccourt18,gronke20-mist}. We begin by discussing them in \S\ref{sec:mhd-shattering}.

\begin{figure*}
     \centering
\includegraphics[width=\textwidth]{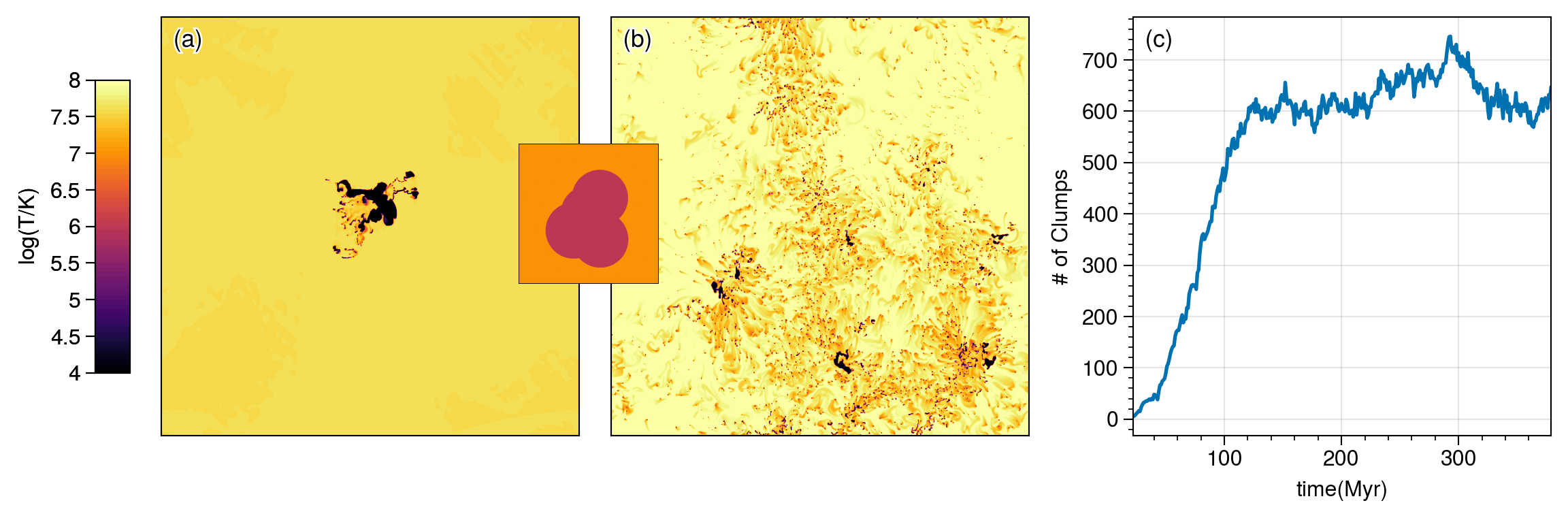}
       \caption[]{ 
       Cloud shattering in the run \hdcc. Panel (a) and (b) show the temperature maps at the end of cloud contraction and the subsequent explosion, respectively.
       The inset between the two panels shows the initial temperature distribution. 
       Panel (c) show the time evolution of number of  cold clumps.
       }
\label{fig:hd_cc}.
\end{figure*}

\begin{figure*}
     \centering
\includegraphics[width=\textwidth]{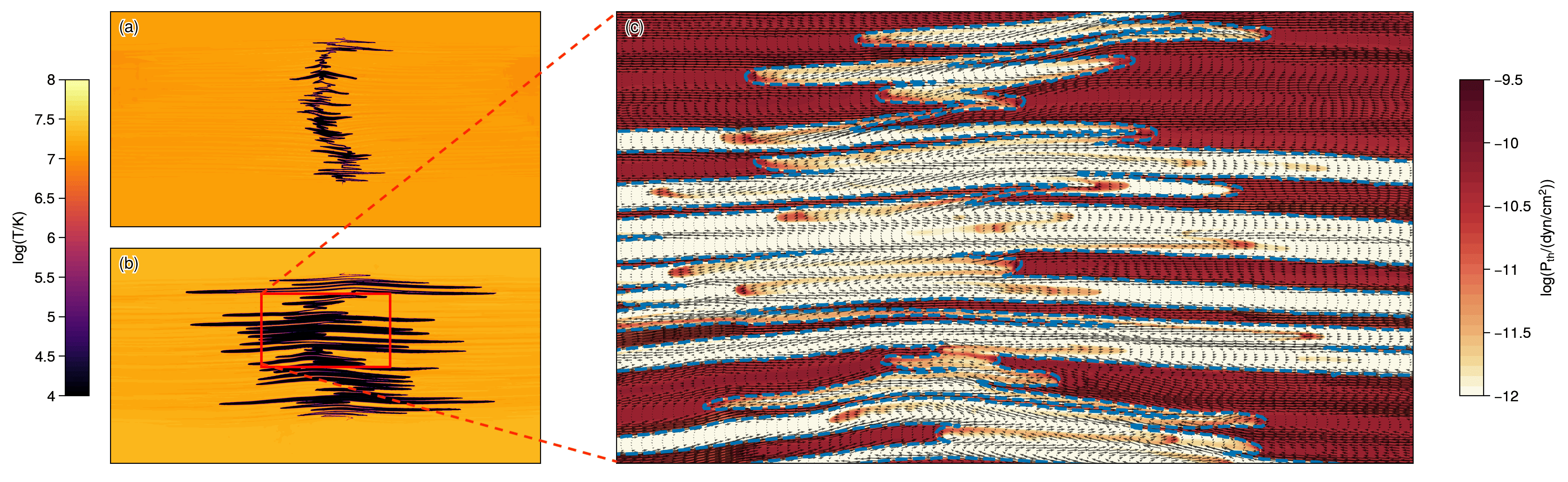}
       \caption[]{ Panel (a) and (b) show the temperature maps of the central $20\times10~{\rm kpc^2 }$ region in \mhdcc. The cloud extends horizontally with growing corrugation. Panel (c) shows gas thermal pressure map in a zoomed-in region as denoted by the red box in panel (b). The cold cloud lays in the low-pressure region enclosed by the blue-dashed lines; and the arrows annotate gas velocity. 
       }
\label{fig:mhd_cc}.
\end{figure*}

\section{Main Results}
\label{sec:results}

\subsection{Magnetized Shattering}
\label{sec:mhd-shattering}
We begin by discussing simulations with the cooling cloud setup. We do so because these simulations are particularly illustrative of the key physics, and relatively simple; there is initially only one cooling blob in the simulation domain.  
It is analogous to hydrodynamic simulations of pressure-driven cloud fragmentation \citet{mccourt18,waters19-linear,gronke20-mist,das21}, but now in MHD. The hydrodynamic simulations show that a cooling cloud whose cooling time $t_{\rm cool}$ is much shorter than its sound-crossing time $t_{\rm sc} \sim R/c_s$ (where $c_s$ is the sound speed) will fall out of sonic contact with its surroundings, leading to rapid cloud contraction and rebound which causes the cloud to explode into small pieces -- a process dubbed ``shattering''. In hydrodynamics, the criterion for cooling driven fragmentation $t_{\rm cool} \ll t_{\rm sc}$ is equivalent to a size requirement, that the cooling blob significantly exceed the cooling length $R \gg l_{\rm cool} \sim c_s t_{\rm cool}$. This is analogous to the requirement for gravitational driven fragmentation, $t_{\rm ff} \ll t_{\rm sc}$, which dictates that condensing blobs must be larger than the Jeans length, $R \gg L_{\rm J} \sim c_s t_{\rm ff}$. The cooling length $l_{\rm cool} \sim c_s t_{\rm cool}$ is evaluated at its minimum, which in CGM/ICM contexts is $T \sim 10^4$K. In addition, \citet{gronke20-mist} found that `shattering' only took place if the final overdensity of the cloud once it regains pressure balance $\chi_{\rm final} = \rho_{\rm c}/\rho_{\rm h} >300$, which for two-phase medium where the cold gas has $T_{\rm c} \sim 10^4$K, requires $T_{\rm h} \sim 3 \times 10^6$K. Physically, `shattering' arises when cloudlets flung out by rebound have enough inertia to resist cooling induced coagulation \citep{gronke22-coag}, which requires that they be above a critical overdensity.  

Our 2D hydrodynamic run, \hdcc, reproduces the cloud shattering seen in previous simulations. The simulation is initialized with a single cold cloud with initial temperature $T_{\rm cl} = 0.1 T_{\rm h}$ and overdensity $\chi_{\rm i}= \rho_{\rm i}/\rho_{\rm h} = 10$ placed in the hot medium, which has $T_{\rm h} = 10^7$K, $n_{\rm h} = 0.05 {\rm cm^{-3}}$.  The shape of the cloud is bounded by four randomly overlapping circles with radius of $r_{\rm cl, init}=2~{\rm kpc}$, (which corresponds to a cloud size at the floor temperature $T\sim 10^{4}$ K of $r_{*} \sim (\rho_{\rm i}/\rho_{\rm c})^{1/3} r_i \sim (T_{\rm c}/T_{\rm h})^{1/3} r_{\rm i} \sim 200 {\rm pc} \sim 3 \times 10^5 l_{\rm cool}$, for a cloud which remains in pressure balance, 
as shown by the the inset panel of Fig.~\ref{fig:hd_cc}. This geometry avoids an excessively symmetric setup, allowing hydrodynamic instabilities to develop efficiently, and also avoiding numerical (carbuncle) instabilities. These initial conditions provide a large non-linear overdensity initially in pressure balance with its surroundings, but which cools rapidly and quickly falls out of thermal pressure equilibrium.
The panels (a) and (b) demonstrate the process of cloud shattering: the cold cloud first contracts rapidly as it falls out of pressure balance with surroundings. Then contraction overshoots, the cloud is compressed and over-pressured; it subsequently rebounds and breaks into grid-scale clumps. As pressure balance is restored, gas velocities decay. The number of clumps saturates and does not decline (panel c), indicating that shattering dominates over coagulation. The cold cloud in \hdcc\ satisfies the two criteria for shattering: besides $r_{\rm cl} \gg l_{\rm shatter}$, as previously noted, 
the final overdensity of the cold clouds, $\chi_{\rm final}=T_{\rm cl}/T_{\rm floor}\chi_{\rm init}=10^3$ for the initial hot gas temperature at $t=0$; and increases to $T_{\rm ceil}/T_{\rm floor}=10^4$ when the hot phase gets heated to $T_{\rm ceil} \sim 10^8$K.  

However, in the MHD run \mhdcc, where the initial plasma beta $\beta =P_g/P_{\rm B} \sim 1$, the anisotropic pressure support of magnetic fields prevents cloud shattering. As shown by panel (a) of Fig.~\ref{fig:mhd_cc}, the initial contraction occurs along the field lines, resulting in a narrow, quasi-1D structure perpendicular to B-field lines at the point of maximum contraction. 
However, instead of overshooting and exploding, the cloud continues to grow in mass, forming a zig-zag structure with growing corrugation. The peaks of the corrugation stream away from both sides of the cloud (Fig.~\ref{fig:mhd_cc} panel b). 

In these numerical experiments, one must take care with boundary conditions/and or box size. In an MHD run with a square domain with periodic boundary conditions (same as \hdcc), the cold cloud does retain thermal pressure balance with the hot gas that is in the same lanes set by the magnetic fields. 
However, this is because the pressure of the entire box drops during the cooling process: the limited amount of hot gas condenses onto the cloud and cools adiabatically as its density falls. If one adopts inflow rather than periodic boundary conditions, then this adiabatic cooling of the hot gas does not occur; the cold gas loses pressure balance and streaming motions result. However, one can only follow these motions for a limited amount of time, before cold gas streams out of the box. 
In the run \mhdcc\ we adopt a wider simulations box so there is more hot gas and the pressure drop of the hot phase is smaller. In this case the cold cloud remains underpressured (see Fig.~\ref{fig:mhd_cc} panel c).

The anisotropic pressure support from magnetic fields is the key factor behind the thermal pressure gradients which drive these streaming motions. While total pressure ($P_{\rm tot}=P_{\rm th}+P_{\rm mag}$) equilibrium is maintained, thermal pressure gradients along the field lines, $\nabla P_{\rm th, \parallel}$, 
are not balanced since magnetic fields only provide pressure support perpendicular to the field lines. The unbalanced $\nabla P_{\rm th, \parallel}$  drives gas motion. Panel (c) of Fig.~\ref{fig:mhd_cc} gives a detailed view of the pressure structures of the streaming cloud and illustrates how the pressure gradients drive cloud motions. Most part of the cold cloud (enclosed by the blue dashed line) is highly under-pressured -- it has thermal pressure two orders of magnitudes lower than that of the surrounding hot medium. Driven by such large pressure differences, hot gas flows towards the cloud, as shown by the annotated velocity arrows. Although the cloud cold gas is mostly at the floor temperature of $T\sim 10^4$K, it occupies a broad range of densities and pressures. The high density cloud `head' in a flux tube has gas pressure comparable to the the surrounding hot medium, while gas in the `tail' is at lower pressure. Hot gas flows towards the low pressure cold gas tail, causing a `wind' which leads to the cold gas moving in the direction of the head, which is compressed to high densities.  

Apart from the fact that loss of thermal pressure balance causes `streaming' rather than `shattering', perhaps the most striking feature of the MHD case is the fact that flows are {\it counter-streaming.} Both hot and cold gas flows are staggered, with flows in adjacent flux tubes streaming in opposite directions. Since the initial setup is static, the fact that there is no net direction to the flow must be true from momentum conservation. Nonetheless, the physics of this effect is sufficiently rich that we devote an entire section (\S\ref{sec:counter-streaming}) to understanding it. In particular, since gas flows towards the under-pressured cold gas from both the left and right, one might expect that the end state should be static compressed cold gas in the middle of the simulation box. The breaking of this symmetry, so that both hot and cold gas in an individual flux tube only streams in one direction, is a pivotal effect which lies at the heart of the development of streaming motions.

\subsection{MHD Thermal Instability}

The `magnetized shattering' setup in the previous section might appear somewhat specialized. In particular, the initial conditions invoke large non-linear perturbations, with an overdensity $\chi_{\rm i} \sim \rho_i/\rho_{\rm h} \sim 10$. How generic is MHD streaming motions, and does it arise in other settings? Perhaps the most general setting involves the formation of multi-phase gas via linear thermal instability \citep{field65}, where the initial density perturbations are small $\chi_{\rm i} \ll 1$. Although MHD thermal instability was already discussed in Field's seminal paper, and there have been decades of both analytic and simulation work since then, to our knowledge there is no discussion of long-lived, self-sustained streaming motions in this setting (although see discussion of `siphon flows' in \S\ref{sec:corona}). 

Our fidicial MHD thermal instability setup, \mhdc, is a 2D $1024^2$ simulation which starts from an initial background of $T_i \sim 2 \times 10^6$K, $n_i \sim 0.25 \, {\rm cm^{-3}}$ gas with isobaric density perturbations drawn from a Gaussian distribution assigned to each pixel, and an initial plasma beta $\beta_i =1$, with purely horizontal magnetic fields. Thus, the perturbations have a white noise power spectrum. Given that MHD and our perturbation spectrum are scale free, the only scales imposed come from non-ideal processes. In our fiducial setup, the only such process is radiative cooling. Since the cooling curve is not a scale-free power law, the range over which it operates is important. Here, $T_{\rm floor} \sim 10^4$K, and $T_{\rm ceil} \sim, 10^8$K. All lengthscales can be scaled to the cooling length $l_{\rm cool}(T) \sim c_{\rm s} t_{\rm cool}$, which is a function of temperature in general, though a notable reference value is the cooling length at $T \sim 10^4$K, $l_{\rm cool, min} \sim 10^{17} (n_{\rm c}/1 \, {\rm cm^{-3}})^{-1} \, {\rm cm}$, which is the minimum value of $l_{\rm cool}(T)$ when $T_{\rm floor} \sim 10^4$K \citep{mccourt18}, as is typical for photoionized gas. In addition, we contrast \mhd\, with a simulation with thermal conduction \mhdcd, where $\alpha_{\parallel} \sim \alpha_{\rm FID} = 1.5 \times 10^{28} {\rm cm^2 \, s^{-1}}$, and a perpendicular heat diffusion coefficient $\alpha_{\perp} \sim \alpha_{\rm iso} \sim \alpha_{\rm FID}/30$. In simulations with thermal conduction, the Field length (equation \ref{eq:field}) is another important scale. Note that the ratio of these two lengthscales, $\eta(T) \equiv \lambda_{\rm F}(T)/l_{\rm cool}(T)$ is an important dimensionless parameter, which has a different temperature dependence with Spitzer conduction (see discussion in \S\ref{sec:conduction}).

Indeed, we find long-lived, self-sustained streaming motions to be a robust outcome of MHD thermal instability in both \mhd \,and \mhdcd. Density fluctuations cool and condense out as cold droplets which consistently stream along field lines, both merging and breaking up during this process. After some time ($t > 100 \, {\rm Myr}$ in these setups), the velocity field and the cold gas mass stabilize as the simulations approach the saturated non-linear state of thermal instability. 
Panels (a1) and (b1) of Fig.~\ref{fig:rtp_all} show the gas temperature map during the stable stages of \mhd\ and \mhdcd.
As shown by the time series plots in the left and right columns of Fig.~\ref{fig:rtp_all}, the cold clouds stream horizontally at a typical speed of $\sim100{\rm km/s}\gg c_{\rm s, floor}$ -- far in excess of their internal sound speed, and roughly the same order of magnitude as the sound speed of the hot medium. The two middle panel columns show how the gas within the red rectangular regions in panel (a1) and (b1) evolve with time. The gray dashed lines (labeled by the horizontal velocity) allow the reader to track the motion of individual cloudlets.  Including thermal conduction increases the sizes of clouds, enabling them to be resolved; clouds are close to grid scale in non-conduction runs. Conduction does not alter the nature of the streaming motions, although characteristic velocities do appear to increase with conduction. Panels (a2) and (b2) of  Fig.~\ref{fig:rtp_all} show 
the $n-T$ gas phase plot. Initially, the gas cools isobarically, following the isobars given by the black dashed lines. However, at $T \sim 10^{5.7}$K, the gas falls sharply out of pressure balance, and cools isochorically at $n \sim 1 \, {\rm cm^{-3}}$ from $T \sim 10^{5.7}$K down to $T \sim 10^4$K, during which time it is strongly out of thermal pressure balance with its surroundings. As we have alluded to (and analyze further in \S\ref{sec:streaming}), it is this deviation from thermal pressure balance which drives streaming. The transition to isochoric cooling arises from the sharp drop in cooling time once $T < 10^{5.7}$K in these setups. In other setups, cooling does not necessarily transition from isobaric to isochoric, but clear deviation from isobaric cooling is always seen in setups where streaming flows occur. Note also the remarkable similarity in phase plots between the conduction and non conduction runs. The close similarity in dynamics and phase structure between the conduction and non-conduction runs, even though the cold gas is unresolved in non-conduction runs, suggests that multi-phase MHD gas streaming is fairly robust to the (highly uncertain) strength of thermal conduction. We now delve further into the physical origins of streaming.

\section{Understanding Gas Streaming}
\label{sec:streaming}

\begin{figure*}
     \centering
\includegraphics[width=\textwidth]{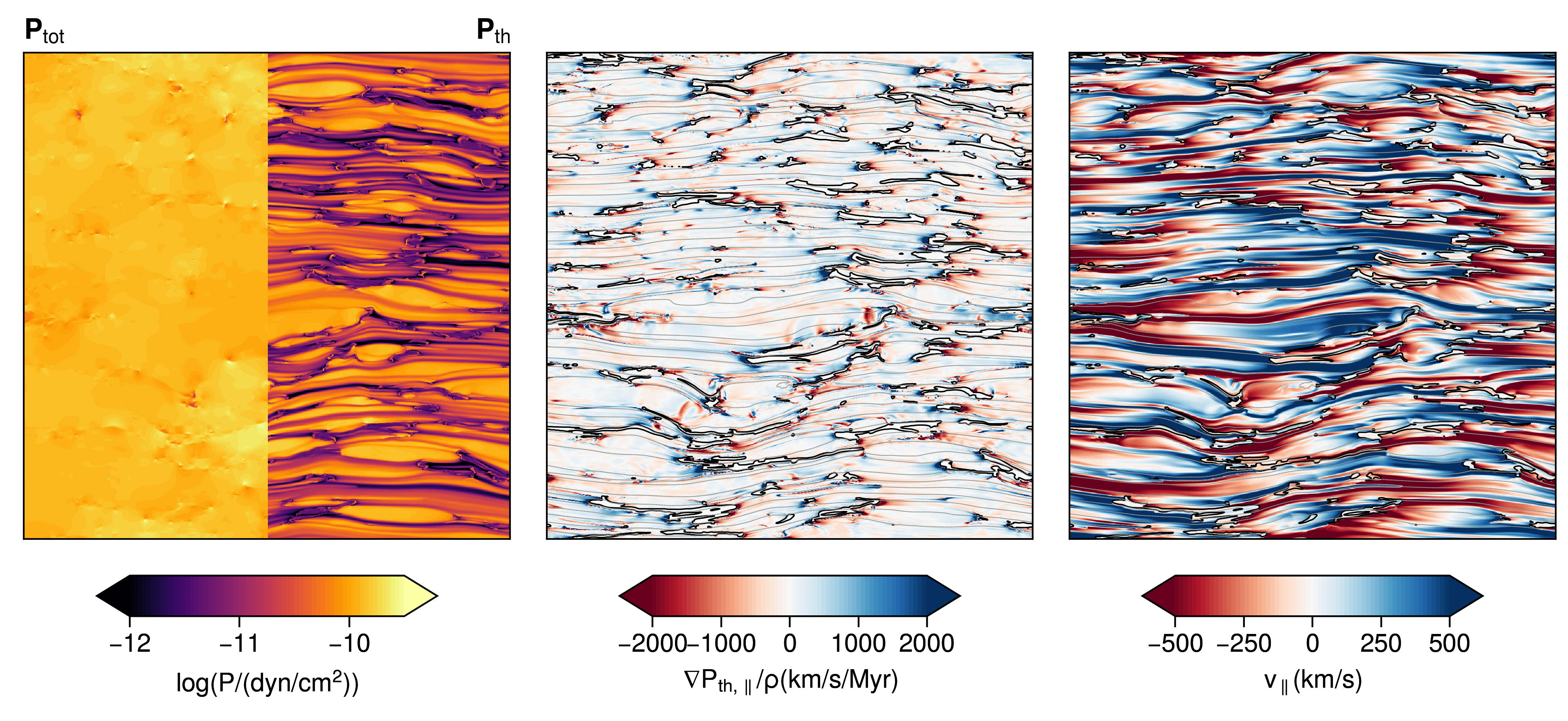}
       \caption[]{Anisotropic MHD pressure support results in field-aligned thermal pressure gradients which drive gas streaming. {\it Left panel}: the left and right halves shows the total pressure ($P_{\rm tot} = P_{\rm th}+P_{\rm mag}$) and thermal pressure $P_{\rm th}$ respectively in the \mhdcd\ run. While the total pressure is roughly constant, there are strong fluctuations in thermal pressure, which creates pressure gradient forces along field lines where magnetic pressure gradients vanish. Middle: the acceleration due to thermal pressure gradients along the magnetic field lines. The black contours enclose cold gas filaments. Right: gas velocity along the magnetic field lines. Note the large dynamic range in velocity, which can reach $v_{\parallel} \sim 500 {\rm km \, s^{-1}}$. In the middle and right panels, blue (red) respresents positive (negative) values. 
       }
\label{fig:drive}.
\end{figure*}

\begin{figure}
     \centering
\includegraphics[width=0.8\columnwidth]{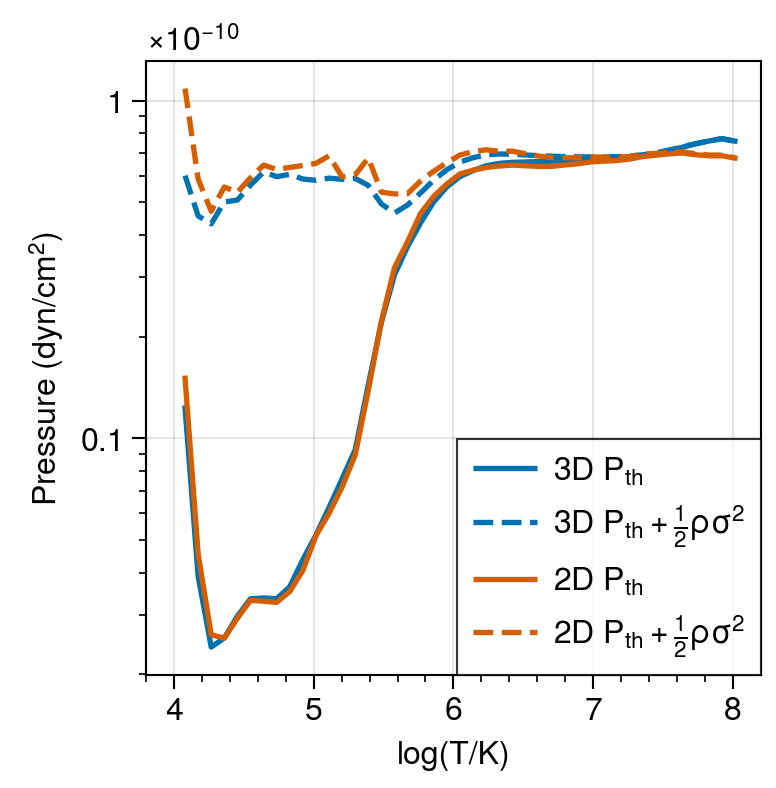}
       \caption[]{Median pressure as a function of temperature in both 2D (\texttt{mhd-256}) and 3D (\texttt{mhd-3d}) runs. The solid lines show thermal pressure; and the dashed lines show the Bernoulli constant $P_{\rm th} + \rho \sigma^2/2$. The latter is indeed roughly constant (except for a spike near $T \sim 10^4$K; see text for details), implying that the drop in thermal gas pressure is compensated by a rise in kinetic energy. All profiles are obtained by averaging snapshots from 50 to 70~Myr. 
       }
\label{fig:3d_pressure}.
\end{figure}

\begin{figure*}
     \centering
\includegraphics[width=\textwidth]{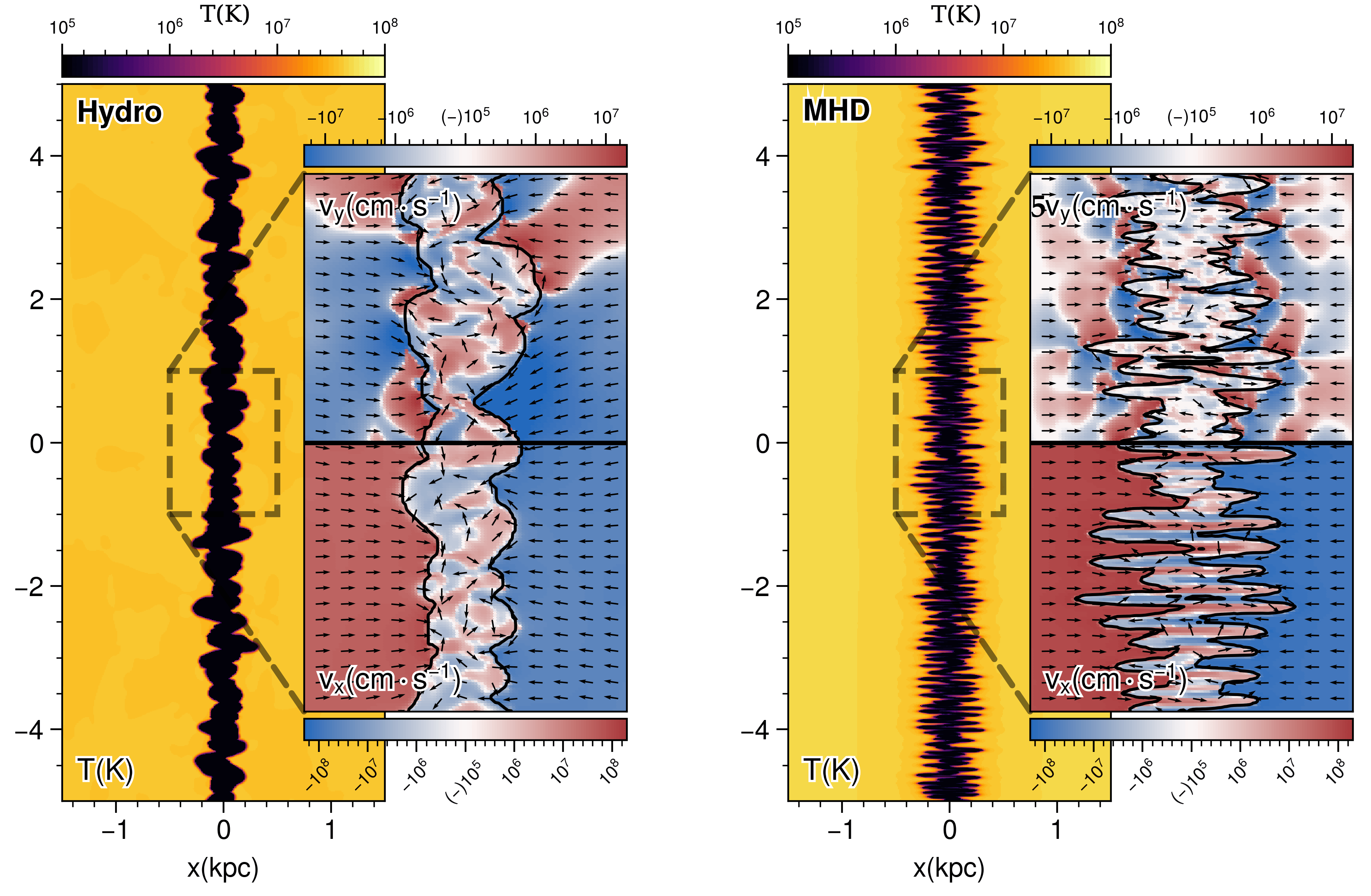}
       \caption[]{Non-linear development of the thin shell instability in hydrodynamics (left) and MHD (right), starting with a cooling slab of cold gas. The large panels present temperature maps at $t=23~{\rm Myr}$, when the thin shell instability is well developed. Velocity fields in the central $1\times2~{\rm kpc^2}$ region are shown in the inset panels, with color maps indicating $v_x,v_y$ respectively. Red (blue) colors represents positive (negative) velocities on a logarithmic scales, with a cut-off at $\pm 10^5~{\rm cm\cdot s^{-1}}$, with right (up) as denoted positive for $v_x,v_y$ respectively. 
       Directions of the velocity vectors are overplotted. In the MHD case, the vectors show $(v_x, 5v_y)$ instead to better illustrate the vertical variations. In the hydrodynamic case, the thin shell instability saturates as the velocity shear generates turbulence inside the cold cloud. In the MHD case, the horizontal B-fields suppress turbulence and maintains ordered motions so that corrugations continue to grow. 
       }
\label{fig:deflection}.
\end{figure*}

\begin{figure*}
     \centering
\includegraphics[width=\textwidth]{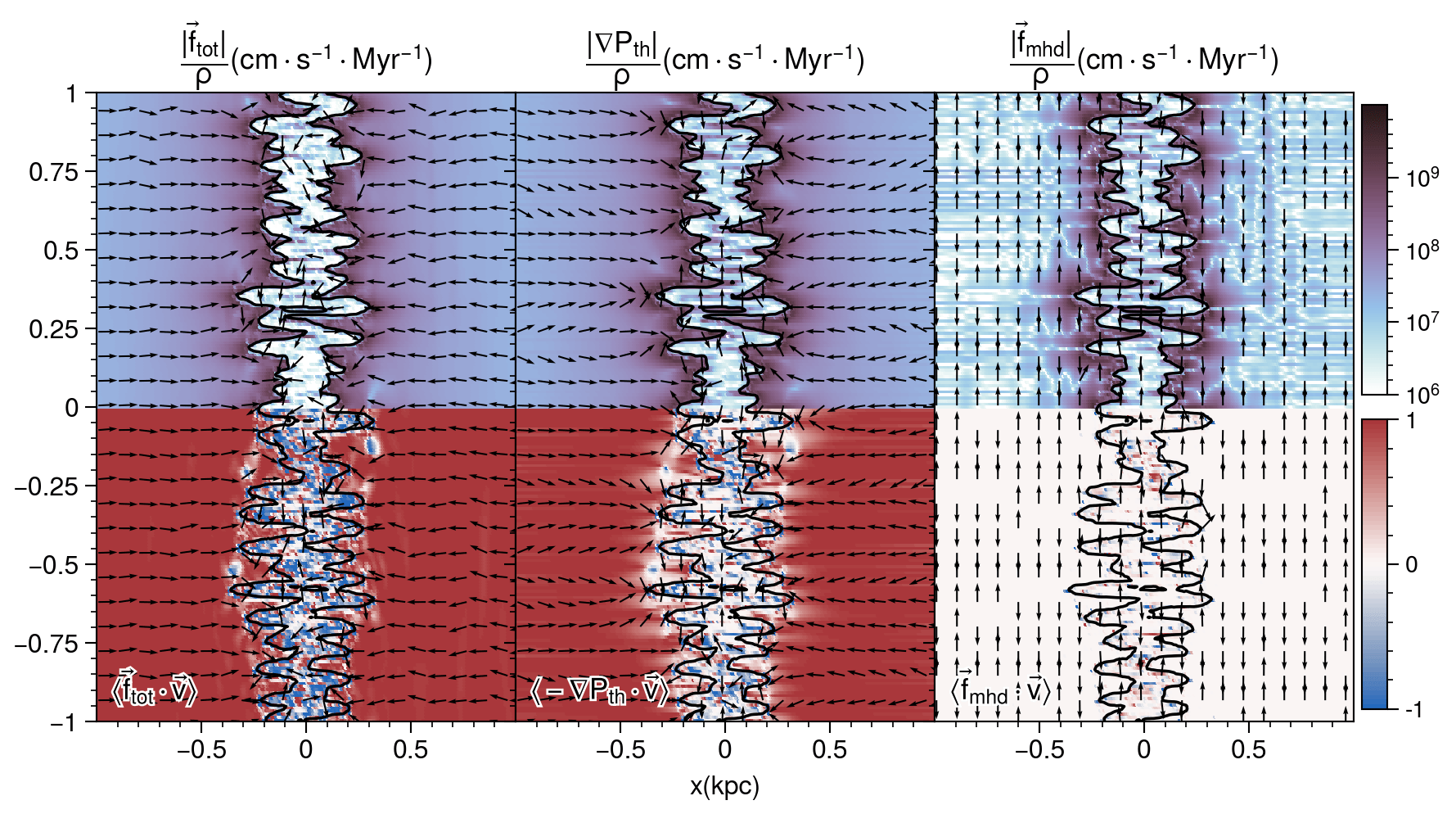}
       \caption[]{Source terms of the Euler momentum equation and   their cross-correlation with gas velocity. The momentum source terms are shown as ${\bf f}_{\rm tot}$, (the total force, left panel), $- \nabla P_{\rm th}$ (thermal pressure, middle panel), and ${\bf f}_{\rm mhd}$ (the MHD force, right panel), where ${\bf f}_{\rm tot}={\bf f}_{\rm mhd} - \nabla P_{\rm th}$, and ${\bf f}_{\rm mhd}=\nabla\cdot({\rm BB})/4\pi - \nabla P_{\rm mag}$. The top half of each panel shows the magnitude of the acceleration caused by each component; and the bottom half shows their cross-correlation with gas velocity, $\langle{\bf f\cdot v}\rangle={\bf f\cdot v}/|{\bf f}||\bf v|$. Directions of the force vectors are overplotted. The force analysis indicates that (1) gas acceleration reaches a maximum at the transition between hot and cold phase, where thermal pressure gradients are largest and the field lines have the largest distortion; (2) the inflow of hot gas onto both sides of the cold slab is driven by the thermal pressure gradient, which dominants the horizontal component of the total force. 
     }
\label{fig:euler}.
\end{figure*}

\begin{figure}
     \centering
\includegraphics[width=\columnwidth]{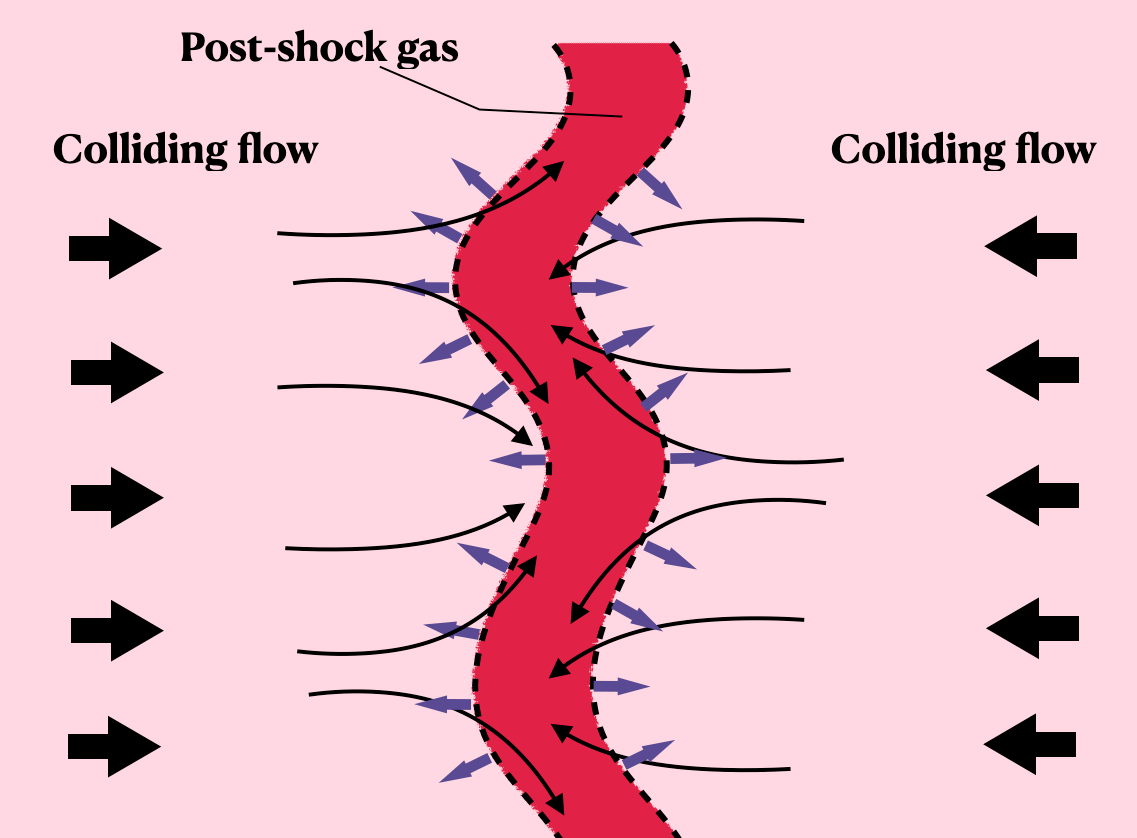}
       \caption[]{Sketch of the classic non-linear thin shell instability (NTSI). Two head-on flows represented by the thick black arrows collide in the middle. The collision creates high pressure shocked gas (red; shock fronts are depicted by dashed lines). Thermal pressure gradients ($-\nabla P_{\rm th}$) point to the pre-shock gas and is normal to the shock fronts, as shown by the purple arrows. The thin black arrows depict how the colliding flows are deflected due to the misalignment between the pressure gradient and ram pressure. These deflections cause corrugations to grow. At convex surfaces (where the force vectors diverge) flows are deflected away, decreasing ram pressure and allowing the head to stream faster, while at the concave surfaces (where the flow vectors converge) the ram pressure is higher, increasing the concavity and pushing the opposing convex head from behind.
       }
\label{fig:cartoon-ntsl}.
\end{figure}

\begin{figure*}
     \centering
\includegraphics[width=\textwidth]{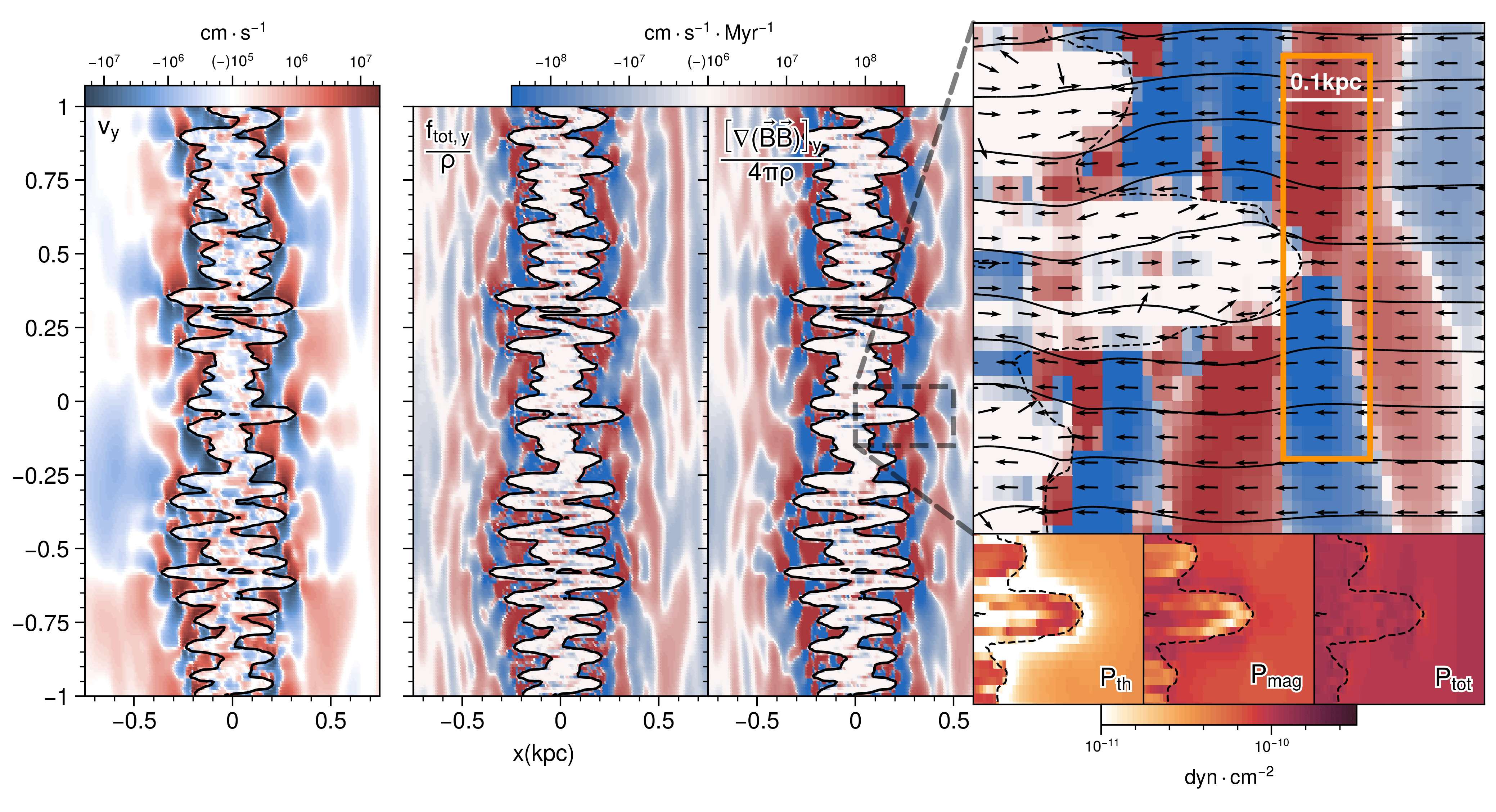}
       \caption[]{Deflection of the streaming flow. From the left to right panel: $v_y$, the vertical component of gas velocity, $f_{\rm tot,y}/\rho$, the total vertical acceleration, and $\left[ \nabla {\mathbf B} {\mathbf B} )\right]_{y}/4 \pi \rho$, the vertical acceleration due to magnetic tension. The black contour encloses the cold gas ($T<10^{5.5}~{\rm K}$). The major inset zooms in the magnetic tension force panel of an individual cold gas head streaming to the right. 
       Directions of velocity vectors are shown as arrows; and the streamlines depict the magnetic fields. To enable better visualization,  vertical deflections (which are small and hard to see) are exaggerated: hot gas velocity vectors show directions of $(v_x, 4v_y)$, and $B_y$ is enlarged by 20\% when drawing streamlines. 
       The three minor inset panels below show maps of thermal pressure, magnetic pressure and the total pressure of the same region, respectively. Most of the vertical acceleration arises from magnetic tension, which deflects gas flows away from convex heads and into concave tails, allowing cold clumps to be pushed from behind. 
       }
\label{fig:vertical}.
\end{figure*}

\begin{figure}
     \centering
\includegraphics[width=\columnwidth]{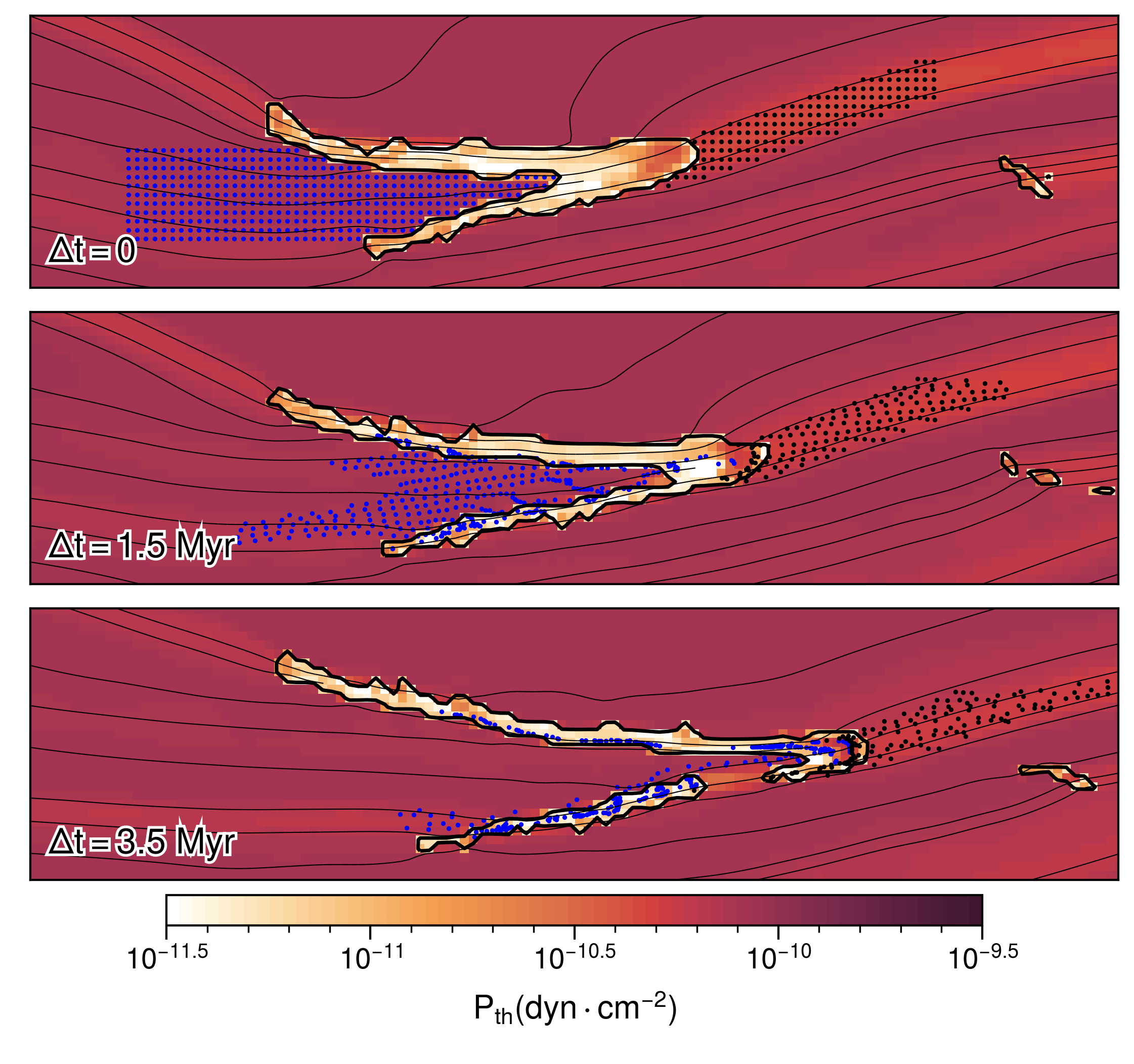}
       \caption[]{The asymmetric hot gas inflow around a streaming cold cloud revealed by tracer particles. Color maps show thermal pressure distribution in a $1.2\times0.3~{\rm kpc^2}$ region centered on a streaming cold cloud, which is enclosed by the black contours and is streaming to the right. The magnetic fields are shown as the stream lines. 
       Particles passively tracing the hot gas in the upstream (downstream) of the cloud are shown in blue (black).
       The three panels are in chronological order, labelled with time since particle injection ($\Delta t$). Particles in the rear of the cloud are incorporated into the cloud, which is pushed from behind. By contrast, only a small fraction of the upstream particles make it into the cloud.
       }
\label{fig:tracer}
\end{figure}

\begin{figure}
     \centering
\includegraphics[width=\columnwidth]{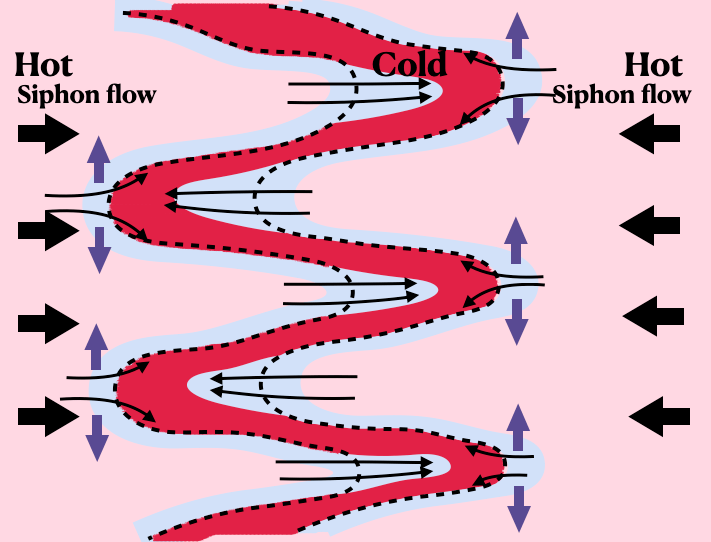}
       \caption[]{Sketch of how corrugations grow in the cooling-induced MHD thin shell instability. 
       The dashed curves shows the interface between hot and cold gas. The values of thermal pressure is represented by colors, where P(blue)$\ll$P(red) $<$ P(pink): the cold gas (red) is underpressured relative to the hot gas (pink), while intermediate temperature gas at the interfaces (blue) has the lowest pressures.  
       Thermal pressure gradients drive the quasi-uniform horizontal streaming flows from hot to cold gas, as illustrated by the thick black arrows. The deflection of streaming flows near the interfaces are depicted by the thin black arrows. Similar to the convex shock fronts in the classic NTSI (Fig.~\ref{fig:cartoon-ntsl}), streaming flows are deflected away from the convex cold gas heads and direct toward concave interfaces in adjacent flux tubes. This amplifies the corrugation and leads to counter-streaming flows. The divergent magnetic tension force at convex heads (represented by the purple arrows) is responsible for the deflection.
       }
\label{fig:cartoon-mhd}.
\end{figure}

The main result of this paper is the self-sustained, long-term counter-streaming motions seen in both cold and hot gas in multi-phase MHD simulations with radiative cooling. These are not present in hydrodynamic simulations with otherwise identical initial conditions. They arise in both `thermal instability' and `magnetized shattering' setups, where initial density perturbations are linear and non-linear respectively. In this section, we elucidate the physical origin of such gas flows. We pay particular attention to the puzzling origin of counter-streaming motions, which turns out to be crucial for understanding the streaming mechanism. In \S\ref{sec:parameters}, we then turn to the dependence of MHD gas streaming on cooling properties, thermal conduction, plasma $\beta$, initial perturbations, and numerical resolution. 

\subsection{Streaming is Due to Thermal Pressure Gradients} 
\label{sec:thermal-pressure}

Previously, we asserted that anisotropic MHD pressure support (which only operates perpendicular to field lines) results in field-aligned thermal pressure gradients which drive gas streaming. We first show this in large-scale thermal instability simulations before specializing to a custom setup where we can clearly resolve the forces at play. Fig.~\ref{fig:drive} shows (from left to right) total and thermal pressure maps, field-aligned thermal pressure gradients, and velocity maps of \mhdcd. Although $P_{\rm tot}$ is generally constant (left panel), both the magnetic $P_{\rm B}$ and thermal $P_{\rm th}$ pressure fluctuate. The hot gas has high thermal pressure and low magnetic pressure, while cold gas has high magnetic pressure (due to flux-freezing, B-fields are amplified when hot gas cools and compresses) and low thermal pressure. Since $P_{\rm tot} \approx$const, there is force balance in the cross-field (vertical) direction, where magnetic pressure operates, and $\nabla (P_{\rm th} + P_{\rm B}) \approx 0$. However, {\it along} the magnetic field, there are no magnetic pressure forces, but nonetheless thermal pressure gradients still exist. In general, for fluctuating $P_{\rm th}, P_{\rm B}$, it is impossible to achieve a force free configuration both parallel and perpendicular to field lines. Since the parallel component of gas thermal pressure is not balanced by magnetic pressure, it drives cloud streaming and deformation. As shown by the middle panel of Fig.~\ref{fig:drive}, $\nabla P_{\rm th, \parallel}$ is largest at the cloud interface between hot and cold gas (the left and right edges of the clouds). The pressure gradient drives hot gas flows that accelerate and entrain the cold gas in their flow, akin to the entrainment of cold gas in galactic winds. The right panel of Fig.~\ref{fig:drive} shows the ensuing gas flows, where gas streams along field lines. Note the alternating blue and red colors, which show counter-streaming gas flows in the $x$ and $-x$ directions in adjacent flux tubes. Also note that large dynamic range in velocity, which can reach $v_{\parallel} \sim 500 \, {\rm km \, s^{-1}}$. We address these features shortly.

One worry might be that these streaming effects are a 2D phenomenon, and might vanish in a 3D setting, where there are more degrees of freedom. We perform a 3D run, \texttt{mhd-3d}, to check this, and contrast it with a 2D run, \texttt{mhd-256}, which has an otherwise identical setup. Clump streaming is still present in \texttt{mhd-3d}, and the velocity and pressure structure of the resulting multi-phase medium is similar with its 2D counterpart. In Fig. \ref{fig:3d_pressure}, we show the median thermal pressure and Bernoulli constant $P_{\rm th} + 1/2 \rho \sigma^2$ as a function of temperature in both the 2D and 3D runs. These are remarkably similar between the 2D and 3D runs, despite morphological differences: compared with the 2D run, there are more small-scale cold clumps in the 3D run, which remain as under-pressured single grid cells due to poor resolution. Nonetheless, the pressure decrements and streaming velocities (which are $\sim 50~{\rm km/s}$ for the cold gas) are comparable in both cases. This suggests that the MHD streaming effect is robust, just as hydrodynamic `shattering' appears both in 2D and 3D. Both are primarily driven by the sharp drop in $t_{\rm cool}$ with temperature. We defer more extensive 3D simulations to future work. 

If gas flows are driven by the thermal pressure gradient, this suggests that: 
\begin{equation}
P_{\rm th} + \frac{1}{2}\rho v^2 \sim P_{\rm isobar}\sim{\rm const},     
\label{eq:bernoulli}
\end{equation}
i.e. that there is a conserved Bernoulli constant along magnetic field lines. As shown by the dashed lines in Fig. \ref{fig:3d_pressure}, this indeed holds in {\it both} 2D and 3D simulations, except for an upturn near the floor temperature. In \S\ref{sec:parameters}, we show that equation \ref{eq:bernoulli} still holds despite changes in temperature range, cooling function, conduction, and resolution.

Equation \ref{eq:bernoulli} allows us to understand characteristic streaming velocities. It implies the velocity for the $P_{\rm th}\ll P_{\rm isobar}$ gas should be
\begin{equation}
v(T) \sim \left(\frac{2 \Delta P_{\rm th}}{\rho(T)} \right)^{1/2}
\label{eq:v_T}
\end{equation} 
Later (in Figs. \ref{fig:tc0_anly} and \ref{fig:tc_anly}) we show this predicted gas velocity for both the non-conduction and conduction cases respectively, which agrees well with the actual gas velocities. It also allows us to understand why the streaming velocity is roughly constant across the temperature range $10^{4.3} < T < 10^{5.5}$K. The gas cools roughly isochorically, $\rho(T) \sim$const, over this temperature range (panels (a2) and (b2) of Fig. \ref{fig:rtp_all})). Moreover, the thermal pressure decrement is large, $P_{\rm th} \ll P_{\rm isobar}$, so that $\Delta P \sim P_{\rm isobar}$. Thus, the characteristic streaming velocity across this temperature range is $v \sim (P_{\rm isobar}/\rho(T\sim 10^{5.5})^{1/2} \sim c_{\rm s}(T\sim 10^{5.5} \, {\rm K}) \sim 100 \, {\rm km \, s^{-1}}$. Since the cooling time is short, most of the gas ends up close to the temperature floor, with the gas streaming highly supersonically relative to its internal sound speed. There are two potential ways of thinking about this. One is that that the cool cloud has become `entrained' in a hot wind (along a flux tube) of high velocity. Another is that the cool cloud is highly underdense relative to expectations from isobaric thermal cooling $n(T) \ll P_{\rm isobar}/ k_{\rm B} T$, and thus the characteristic velocity which can arise from pressure gradients, $v(T) \sim (P_{\rm isobar}/\rho(T))^{1/2}$, is much higher. Close to the floor temperature, the gas is no longer isochoric, but is isothermally compressed, spanning a broad range of densities at fixed temperature $T \sim T_{\rm floor}$ (panels (a2) and (b2) of Fig. \ref{fig:rtp_all}). Some work has to be done to perform this compression, and hence the gas slows down, i.e. $v(T)$ undergoes a downturn. However, the broad range in densities means that $\rho(T\sim T_{\rm floor})$ is no longer well-characterized by a single number, and equation \ref{eq:bernoulli} is no longer accurate. The upturn in $P + 1/2 \rho v^2$ close to the floor temperature can be thought of as the consequence of a transition to a momentum conserving flow.




\subsection{Counter-Streaming Motions are Due to the Thin Shell Instability}
\label{sec:counter-streaming}

In order to fully resolve the interface region and the forces at play, we conduct `slab' simulations, where we place a thin slab in the center of a long simulation box. Similar to the cloud shattering setup (\S\ref{sec:mhd-shattering}), the slab is initially at a temperature $T_i \sim 0.1 T_{\rm h}$, and in thermal pressure balance with its surroundings. As the slab quickly cools to the temperature floor $T \sim 10^4$K, it loses pressure balance with surrounding gas, and the interface develops strong pressure gradients. 
Fig. \ref{fig:deflection} shows slabs in both hydrodynamic and MHD runs, soon after initialization. In both cases, corrugations in the slab develop. However, in the MHD run, streaming motions soon ensue, whereas in the hydro run, the cold cloud develops turbulence but still stays intact; it does not develop strong streaming motions. 

Now that the interface region is much better resolved, we can analyze in detail the forces responsible for MHD streaming. The bottom panels of Fig. \ref{fig:euler} show the cross-correlation between hydrodynamic/MHD forces and gas velocities, ${\rm cos}\theta = \langle \mathbf{f} \cdot \mathbf{v} \rangle/(\langle \mathbf{f} \rangle \langle \mathbf{v} \rangle)$. The left panel shows a high correlation between the total force (i.e., the sum of thermal and magnetic pressure gradients, and magnetic tension forces) and gas velocities in the hot gas ${\rm cos}\theta_{\rm tot} = \langle \mathbf{f_{\rm tot}} \cdot \mathbf{v} \rangle/(\langle \mathbf{f_{\rm tot}} \rangle \langle \mathbf{v} \rangle) \approx 1$, whereas forces and velocities are decorrelated in the cold gas ${\rm cos}\theta_{\rm tot} \approx 0$. 
The middle and right panels show that this strong correlation is almost entirely due to thermal pressure gradients in the hot gas, which act in the horizontal direction, along the magnetic field. 
By contrast, MHD forces act mostly in the vertical direction, perpendicular to the B-field, and do not correlate with gas velocity. 
Unlike the hot gas, forces and gas velocities are {\it uncorrelated} in cold gas. Thus, direct acceleration via hydrodynamic/MHD forces is not responsible for the streaming of cold gas cloudlets. Instead, as we shall show, cold gas is `entrained' in the hot wind of each flux tube.   

Even more puzzling than the presence of gas motions is the fact that gas in adjacent flux tubes stream in opposite directions. Of course, since the initial setup was static, conservation of momentum requires that there is no net bulk motion. However, the systematic, staggered anti-symmetric flow in adjacent flux tubes demands an explanation. Since cold gas creates a minimum in thermal pressure, hot gas streams towards it from {\it both} left and right in a flux tube. What breaks the symmetry, so that gas in a flux tube streams in one direction? Naively, one might expect the cold gas to simply be compressed by the ram pressure of inflows until it reaches thermal pressure balance. The right zoom-in panel of Fig.~\ref{fig:deflection} offers some clues. Although hot gas flows are primarily horizontal, at the cold gas head, there are small vertical components ($v_y$) that deflect the hot flows away from the head. These small deflections turn out to extremely important. They produce small vertical shifts which break the left-right symmetry of hot inflowing gas to produce counter-streaming flows. By contrast, velocity flows in the cold gas are chaotic, in line with the force analysis above. This indicates that cloud motion arises from momentum transfer from the hot gas wind, rather than forces internal to the cloud. 

It turns out that counter-streaming flows develop due to a corrugational instability closely related to the non-linear thin shell instability (NTSI; \citealt{vishniac83}). The NTSI is a corrugational instability seen in colliding flows, typically shock fronts. It arises from a misalignment between the ram pressure of the flow and thermal pressure of the dense postshock gas, which is overpressured compared to its surroundings. Thermal pressure gradients are always normal to the dense gas surface, whereas the ram pressure of inflowing gas is fixed by the flow direction. Any corrugation of the shock surface leads to misalignment between ram pressure and opposing thermal pressure,
which deflects the incoming flow and amplifies the perturbation, causing it to grow. 
Fig \ref{fig:cartoon-ntsl} depicts how the vertical deflection of colliding flows enhances corrugation of the post-shock gas in the case of classic NTSI. 
The purple arrows illustrate the force due to thermal pressure gradient ($-\nabla P_{\rm th}$). In the corrugated shock fronts, force vectors diverge at the convex surfaces and converge at the concave surfaces. Consequently the colliding flows are deflected away from the convex heads, which continue to extrude further, while at the concave surfaces the concentrated inflows push the the opposing convex head from behind and enhance the corrugation further. 
The NSTI is a non-linear instability, requiring finite amplitude perturbations of order the shell-thickness or larger; it eventually saturates due to the increasing thickness of the shell. The classic NTSI creates vertical shear and clumping in the dense shell. While there are MHD simulations of colliding flows that show the thin shell instability operating (e.g., \citealt{heitsch07,fogerty17}), to our knowledge none report the streaming motions that we see. Why not? 

The cooling-induced MHD thin shell instability we simulate has three important differences. Firstly, in the classic NTSI, the post-shock gas is over-pressured. Instead, as we have shown, the cooling dense gas in our setup  is under-pressured. Thus, thermal pressure gradients do not oppose ram pressure, but act in the same direction. Indeed, thermal pressure gradients cause the inflows in the first place, and the instability grows spontaneously without initial colliding flows. 
Secondly, it is the magnetic tension force that accounts for instability growth instead of the thermal pressure gradient. Since the interfaces are under-pressured, the sign of the thermal pressure gradient is opposite to that of the classic NTSI. 
As shown in the left panel of Fig.~\ref{fig:deflection}, in our hydro run the corrugations quickly saturate, and the cold cloud is dominated by turbulent motions. In the MHD run, however, magnetic tension forces deflect the streaming flows in a similar manner as thermal pressure gradients in classic NTSI, so the corrugation is able to grow.  
The deflection pattern in the $v_y$ map (left panel of Fig.~\ref{fig:vertical}) agree with the vertical components of the total force ($f_{\rm tot, y}$, middle panel) and the magnetic tension force ($\nabla_y ({\bf BB})/4\pi$, right panel), indicating the dominant role of magnetic tension in deflecting the streaming flows. Note that red (blue) indicates upwards (downwards). The details of this deflection are shown in the inset panel of the tension force map, which zooms in to an individual cold gas ``head'' streaming to the right. Magnetic field lines drape around the streaming head. 
The diverging field lines in front of the streaming head (i.e., the region enclosed by the orange rectangle in Fig.~\ref{fig:vertical}) produces magnetic tension forces which deflect gas flows away from the head. This deflected gas is directed to the rear of another cold cloud, pushing it in the same direction. 
We can verify that the tension force is large enough to cause this deflection. The {\it vertical} velocity increment after passing the deflection zone is: 
\begin{align}
    \Delta v_y \sim& a_{\rm tot,y} t_{\rm cross} = \frac{f_{\rm tot, y}}{\rho}\frac{L_{\rm zone}}{v_x} \nonumber \\ 
    \approx& 100~{\rm km\cdot s^{-1}}\left(\frac{a_{\rm tot,y}}{10^8{\rm cm/s/Myr}}\right)\left(\frac{L_{\rm zone}}{\rm 0.1kpc}\right)\left(\frac{v_x}{\rm 10^8cm/s}\right),
\end{align} 
where $t_{\rm cross}$ is the time the streaming flows take to cross the deflection; and $L_{\rm zone}$ is the zone width. To order of magnitude, the estimate above corresponds to the simulated values. 

Why do the field lines bend in this way? The issue boils down fundamentally to an asymmetry in plasma $\beta$ (and hence magnetic field curvature) between the head and tail of a cold cloud. The streaming `head' is 
compressed by ram pressure forces and has high thermal pressure, 
leading to weaker magnetic pressure in the `head', for total pressure to remain constant. The pressure distributions are shown in three panels on the lower right of Fig.~\ref{fig:vertical}. 
Effectively, the field lines are `expelled' from the streaming `head', causing them to diverge. Note that once it is magnetically supported at low temperature, gas is mostly compressed {\it along} the field; such compression does not increase the field strength due to flux freezing\footnote{Of course, gas does compress perpendicular to the B-field at higher temperatures, before it is magnetically supported.}. 


Although these field line deflections play a crucial role in streaming, they are small. Indeed, a key requirement in streaming is that B-fields remain relatively straight. MHD streaming flows are generally sub-Alfvenic $\mathcal{M_{\rm A}} \sim v/v_A \lesssim 1$ at the cold/hot gas interface. Gas is magnetically dominated at low temperatures, with $\beta = (c_s/v_A)^2 \lesssim 1$ (as we see in \S\ref{sec:beta}, this is true even if the magnetic field is relatively weak in the hot gas, $\beta_{\rm hot}
\gtrsim 10$; it is simply a consequence of flux freezing). From equation \ref{eq:v_T}, hot gas flows are at most transonic ($\mathcal{M_{\rm s}} \sim v/c_s \sim 1$), which implies $\mathcal{M_{\rm A}} \sim \mathcal{M_{\rm s}} \beta^{1/2} \lesssim 1$. Thus, magnetic fields remain relatively straight and serve as `wave guides' for gas flows, with only small deflections. 
The effect of magnetic fields can be seen in the significant differences between the hydrodynamic and MHD cases in Fig \ref{fig:deflection}. In both cases, cooling induced thermal pressure gradients cause hot gas to stream into the cold gas, and corrugations develop due to the thin shell instability. However, in the hydrodynamic case, this creates relatively disordered vortical motion in the cold thin shell. By contrast, in the MHD case, due to magnetic tension the vertical deflection arising from the thin shell instability is relatively minor, and the motion is largely still horizontal. 
In other words, the non-linear development of the thin-shell instability is arrested by magnetic tension. The small amount of vertical deflection allows opposing streams to avoid colliding. Each cold gas stream is propelled by hot gas which enters from the rear. From the bottom right panel of Fig \ref{fig:deflection}, it can be seen from the color scheme that the cold gas stream has the same sense of direction as the hot gas in its rear -- consistent with being `pushed' from behind. 

We can verify this picture directly in tracer particle simulations, which we run in FLASH \citep{fryxell00} to make use of the tracer particle module. The thermal instability setup is identical to our ATHENA++ simulation \texttt{mhd-fid}. Once the simulation has reached its non-linear steady-state, with steady counter-streaming gas flows, we stop the simulation and inject two distinct sets of tracer particles, one upstream and another downstream of a cold cloud. We then restart the simulation. The results are shown in Fig~\ref{fig:tracer}. We can see that hot gas tracer particles entering the cold cloud from the rear are much more likely to mix and cool, imparting their mass and momentum. The cloud acceleration by mixing and cooling of hot wind material is similar to that invoked for cloud acceleration in galactic winds \citep{gronke18}. By the last snapshot, all the (blue) tracer particles in the rear of the cloud have become incorporated into the cloud. By contrast, only a relatively small fraction of the (black) tracer particles entering from the front become incorporated into the stream; those which are not deflected vertically in time are swept up with the streaming head. This asymmetry in momentum deposition means that the cloud is pushed from behind. Thus, the clump in Fig. \ref{fig:tracer} streams toward the right.

It is worth mentioning an alternative hypothesis for driving gas motions and determining its direction, which we closely examined before ruling it out: thermal conduction \footnote{In our non-conduction runs, this would simply correspond to numerical diffusion.}. In hydrodynamic simulations, evaporation and condensation due to thermal conduction in a multi-phase medium can drive cloud disruption, gas motions and turbulence, for similar reasons as the Darrieus–Landau instability in premixed combustion \citep{nagashima05,inoue06,kim13,iwasaki14,jennings21}. There, the curvature of the hot-cold gas interface determines the sign of energy transfer and gas motions. Cold gas is conductively heated and evaporates into the hot phase at cold gas convex interfaces, while hot gas undergoes conductive cooling and condenses onto the cold phase at cold gas concave interfaces. These flows drive turbulent motions. Why does the curvature of the surface matter? In hydrodynamic simulations with isotropic conduction, the direction of heat flux is determined only by temperature gradients. Thus, convex (concave) cold gas surfaces have converging (diverging) heat flux, and drive evaporation (condensation) respectively. By contrast, in MHD simulations with field-aligned thermal conduction, where the magnetic fields are relatively ordered, we have not found conductive cooling or heating to depend on the curvature of the interface. Instead, from examining the divergence of the heat flux, we generally see conductive cooling of hot gas at both the convex head and concave tail of filaments \footnote{This does not mean that the cloud is growing monotonically in mass. In streaming filaments, other processes such as shear induced disruption,  mixing and radiative cooling are more important in mass flows.}. In field aligned conduction, the heat flux has to follow magnetic field lines, which in our simulations are mostly straight; the divergence of the heat flux does not change sign depending on the curvature of the cold gas surface. The fact that gas motions in the hydrodynamic case are driven by the Darrieus-Landau instability might lead one to think that the Darrieus-Landau instability is responsible for MHD streaming, but this is in fact not the case. The fact that streaming is not significantly different in simulations with and without thermal conduction (e.g., see Fig \ref{fig:rtp_all}) is also consistent with this view.  

We can summarize our main conclusions for the development of counter-streaming flows in Fig. \ref{fig:cartoon-mhd}, which should be contrasted with Fig. \ref{fig:cartoon-ntsl} for the standard NTSI. Far away from an underpressured cooling cloud, hot gas flows are symmetric. However, near the cloud, small vertical flows due to magnetic tension forces cause the mostly horizontal flows to diverge away from convex heads and converge toward the concave tails. In particular, magnetic fields draped around the convex head deflect gas to the concave rear of a neighbor, which is thus pushed from behind. This amplifies the corrugation, and sets up counter-streaming flows, where neighboring flux tubes have winds in opposite directions. 




\section{Physical Parameter Dependence }
\label{sec:parameters}

In this section, we discuss the influence of various parameters and pieces of physics on streaming: the temperature range and cooling function (\S\ref{sec:non-conduction}), thermal conduction (\S\ref{sec:conduction}), and plasma $\beta$ (\S\ref{sec:beta}). 
Finally, in \S\ref{sec:converge}, we discuss numerical convergence. This is a first pass at exploring parameter dependencies, which we intended to revisit in more detail in a follow-up paper. 

\subsection{When Does Streaming Occur?: Dependence on Temperature Range and Cooling Function}
\label{sec:non-conduction}

\begin{figure*}
    \centering  \includegraphics[width=\textwidth]{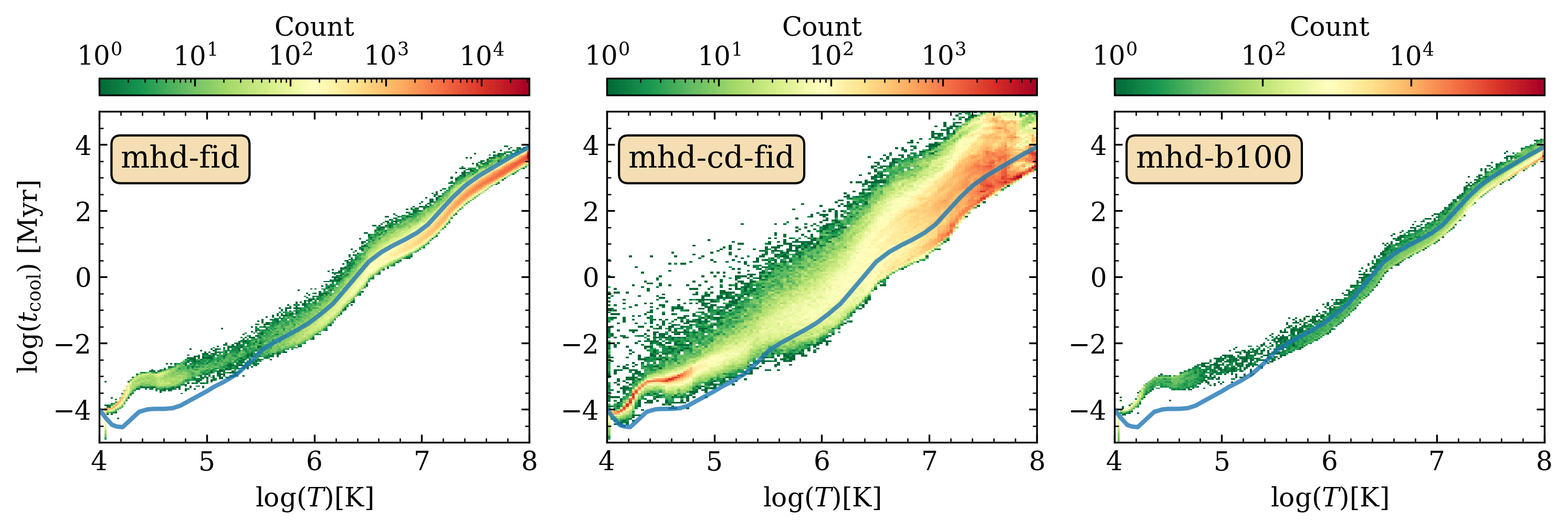}
       \caption{Scatterplot of $t_{\rm cool}$ as a function of temperature for all points within a simulation snapshot, for \texttt{mhd-fid}, \texttt{mhd-cd-fid} and \texttt{mhd-b100}. The solid blue lines in the plots indicate the isobaric cooling expectation. Most of the gas in all three simulations largely follows the expected $t_{\rm cool}$ curves for $T>10^{5.5}$ K, below which isochoric cooling takes place. The addition of conduction makes the scatterplot thicker as more cells occupy intermediate temperature and densities.}
       \label{fig:t_cool_scatter}
\end{figure*}

\begin{figure*}
     \centering  \includegraphics[width=\textwidth]{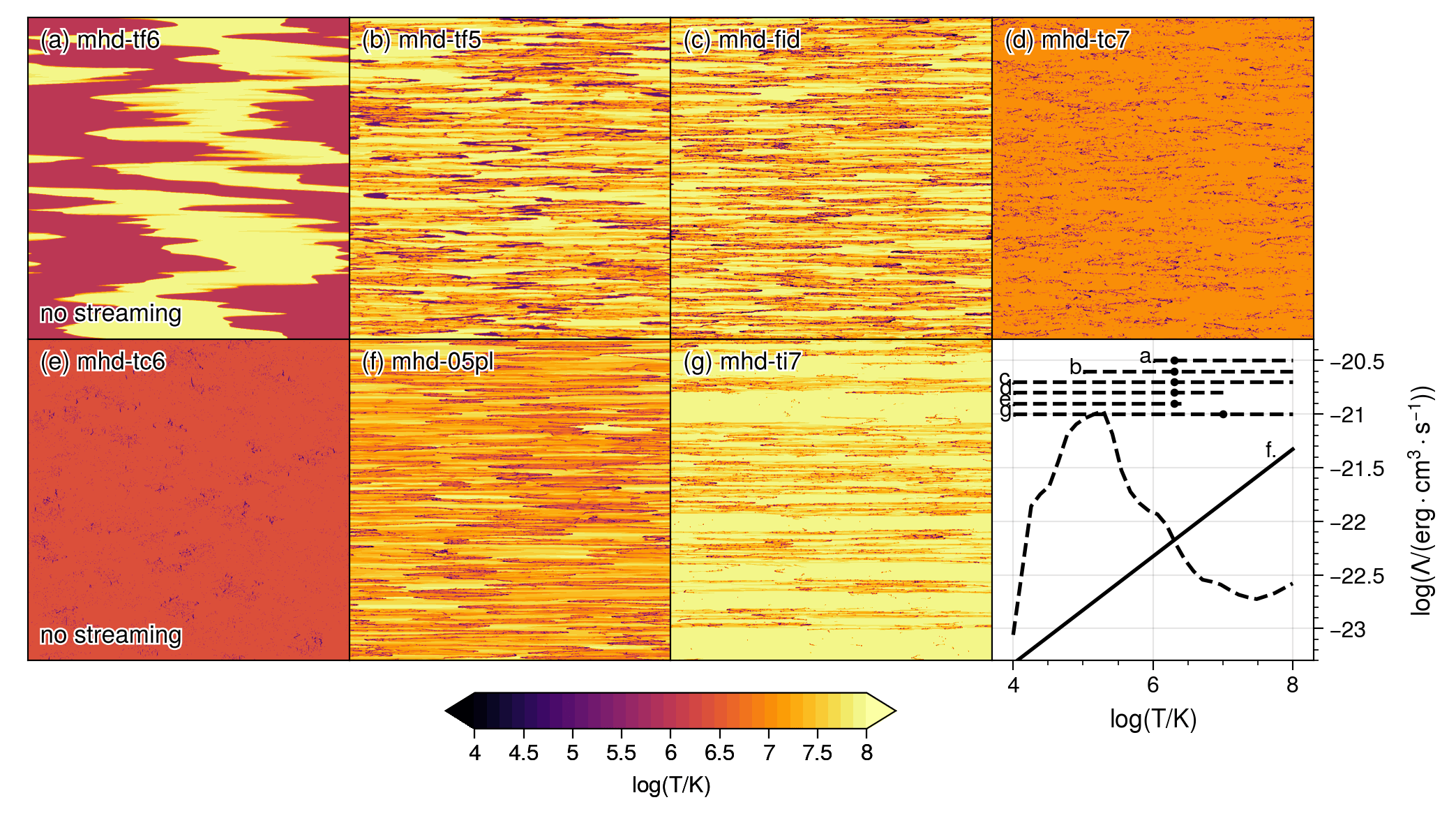}
       \caption[]{Runs without thermal conduction with varying properties of the cooling curve. Panels (a)~--~(g) show snapshots of gas temperature at times when the simulation evolves to the stable stages, i.e., when the cold gas mass and velocity do not significantly change with time. The panel (h) shows the configurations of these runs. Runs (a)~--~(e) and (g) adopt the \cite{SutherlandDopita93} cooling function as shown by the dashed line and have varying floor, ceiling and initial temperatures $T_{\rm floor}$, $T_{\rm ceil}$ and $T_{\rm init}$, which are indicated by the span of the horizontal dashed lines and the black points in between. The run (f) uses a single power law cooling function with power index of 0.5, as shown by the solid line.
       } 
     \label{fig:tsl_tc0_cc}
\end{figure*}

\begin{figure*}
     \centering  \includegraphics[width=\textwidth]{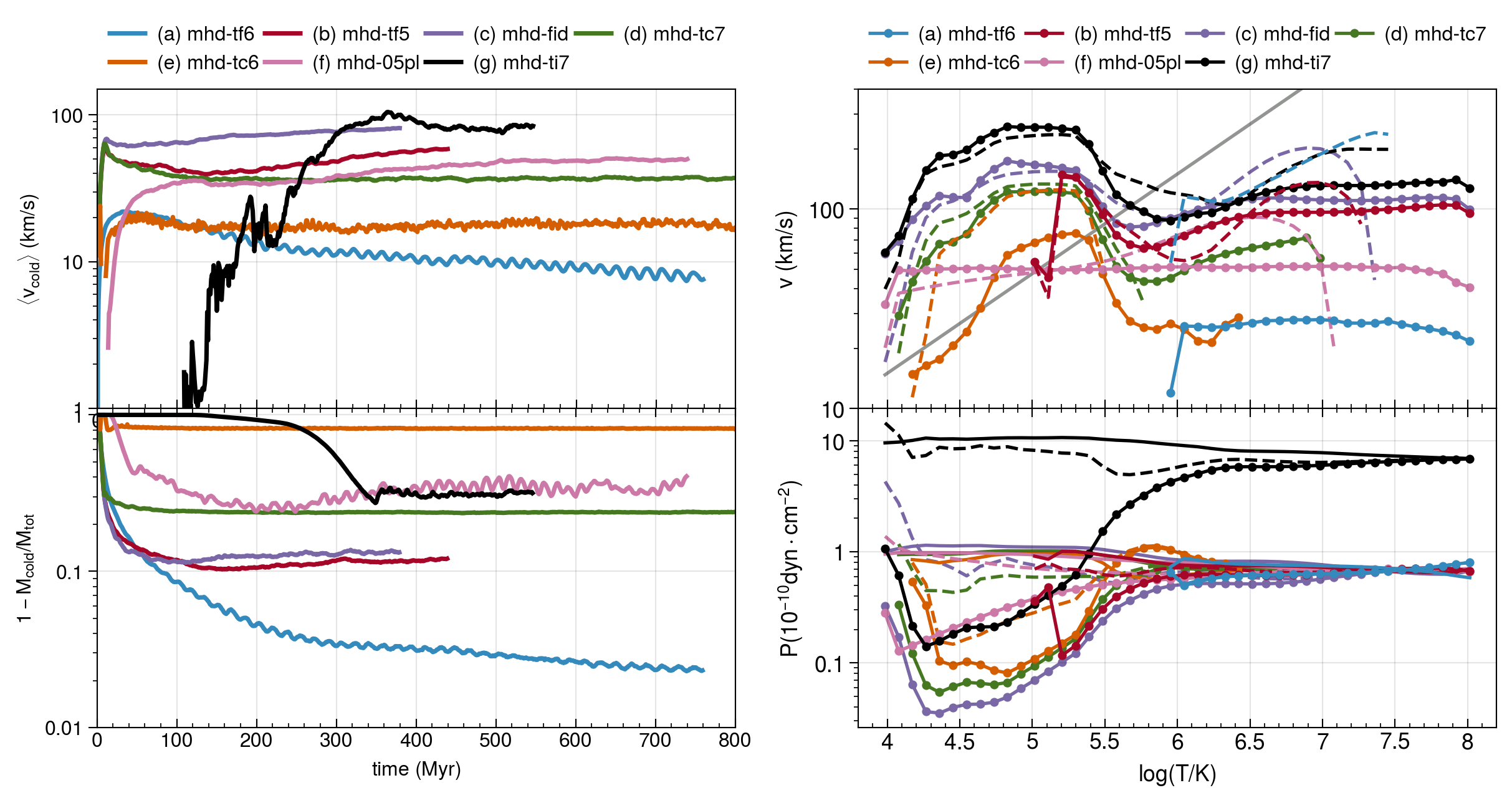}
       \caption[]{ Left: time evolution of cold gas rms velocities (top) and cold gas mass fraction (bottom) of runs (a)~--(g) as labeled in Fig.~\ref{fig:tsl_tc0_cc}. Note that the definitions of ``cold'' gas are different in runs with different $T_{\rm floor}$: $T<10^{4.5}~\kai$ if $T_{\rm floor}=10^{4}~{\kai}$;  $T<10^{5.2}~\kai$ if $T_{\rm floor}=10^{5}~{\kai}$; and  $T<10^{6.2}~\kai$ if $T_{\rm floor}=10^{6}~{\kai}$. Right: temperature dependence of gas rms velocities (top) and pressure components (bottom). The profiles are calculated by averaging the stacked simulation snapshots every $1~{\rm Myr}$ from 0.2 to 0.3~{\rm Gyr}. The dashed lines in the top right panel show the velocity profiles predicted by the thermal pressure decrements, $v_{\rm predict} = \sqrt{ \frac{2\Delta P_{\rm th}}{\rho} }$; and the gray solid line denotes the adiabatic sound speed.
       The pressure components shown in the bottom right panel are: thermal pressure, $P_{\rm th}$ (the dotted lines), the sum of thermal pressure and gas kinetic energy, $P_{\rm th}+\frac{1}{2}\rho v^2$ (the dashed lines), and the magnetic pressure $P_{\rm mag}$ (the solid lines).
       } 
     \label{fig:tc0_anly}
\end{figure*}


Previously, in \S\ref{sec:thermal-pressure}, we showed that streaming is associated with loss of thermal pressure balance in directions parallel to the local magnetic field. This is generally associated with a sharp drop in the cooling time as a function of temperature, $t_{\rm cool}(T)$, and eventual deviation from isobaric cooling expectations. This is shown for three simulations (\texttt{mhd-fid}, \texttt{mhd-cd-fid} and \texttt{mhd-b100}) in Fig. \ref{fig:t_cool_scatter}.
We now investigate this dependence, by studying how streaming changes if we change the temperature range (for realistic cooling curves), or if we artificially change the cooling curve. Here, we do this for simulations without conduction, and in \S\ref{sec:conduction}, we include thermal conduction. 

Here, we run simulations with different values of the floor, ceiling and initial temperature respectively: $T_{\rm floor}, T_{\rm ceil}$, and $T_{\rm init}$. In one case, we also change the shape of the cooling curve. Changing ${\rm T}_{\rm ceil}$ and $T_{\rm init}$ are obviously well-motivated. Different astrophysical environments, such as clusters, groups, and galaxy halos, have different hot gas temperatures ${\rm T}_{\rm ceil}$, which are formally thermally unstable but have long cooling times. Changing $T_{\rm init}$ simply explores sensitivity to initial conditions. 
Changing $T_{\rm floor}$ might seem more artificial, since $T_{\rm floor}$ should always correspond to a stable phase, which is dictated by the cooling curve (typically $T\sim 10^4$K, or $T\sim 10-100$K). However, we do so to explore the underlying physics of streaming, and also to make contact with other simulations which vary the temperature floor (e.g., \citealt{sharma10}), typically so that the highly temperature sensitive Field length can be resolved. The cooling curves adopted are shown in panel (h) of Fig.~\ref{fig:tsl_tc0_cc}. The fiducial run,
\mhdc\ has $T_{\rm floor}=10^{4}~\kai$, $T_{\rm init}=2\times10^{6}~\kai$, and $T_{\rm ceil}=10^{8}~\kai$ with the \citet{SutherlandDopita93} cooling curve. The rest of the runs are variations on \mhdc. The runs \mhda, \mhdb, and \mhdc\ have progressively lower $T_{\rm floor} = 10^{6}~\kai, 10^{5}~\kai, 10^{4}~\kai$ respectively; the runs \mhdc, \mhdd, and \mhde\ have progressively lower $T_{\rm ceil} = 10^{8}~\kai, 10^{7}~\kai, 2.5 \times 10^{6}~\kai$ respectively; the run \mhdg\ has $T_{\rm init}=10^{7}~\kai$; and the run \mhdf\ has a power-law cooling curve, $\Lambda\propto T^{0.5}$, which is normalized such that $\Lambda(T_{\rm init})$ is unchanged. This last cooling curve is chosen so that the isobaric cooling time $t_{\rm cool}(T)$ is a scale-free power law of temperature. This allows us to see if the point at which gas falls out of pressure balance is associated with a characteristic feature in the cooling curve. Note that gas with this cooling curve should be linearly stable to isochoric cooling, as isochoric thermal instability requires $d \,({\rm ln \Lambda(T)}/d \,({\rm ln \, T}) < 0$ \citep{field65}. This allows us to test if the transition to isochoric cooling (e.g., as in Fig \ref{fig:rtp_all}), which we have claimed is due to magnetic pressure support, is instead due to isochoric thermal instability. Finally, we also run a simulation with an ISM/molecular cloud cooling curve, between $T \sim 10~\kai-10^{4} ~\kai$, which we discuss separately at the end of this subsection.   

In brief, we find that streaming motions are present in runs \mhdb, \mhdc, \mhdd, \mhdf, and \mhdg, but absent in runs \mhda, \mhde\, (i.e., runs where the floor and ceiling temperature respectively are $T \sim 10^6$K). Panels (a)--(g) of Fig.~\ref{fig:tsl_tc0_cc} show temperature maps of these runs after they have reached their stable non-linear state 
when cold gas velocities and mass do not change significantly with time. The left column of Fig. ~\ref{fig:tc0_anly} shows the time evolution of the cold gas mass fraction and the velocity of cold gas, while the right column of Fig.~\ref{fig:tc0_anly} shows the gas velocity and pressure components as functions of temperature. 
The latter are obtained by averaging the stacked simulation snapshots every $1~{\rm Myr}$ during the stable stages, from 0.2 to 0.3~{\rm Gyr}. 

In runs where the cold clouds stream, they shatter down to small scales (in these non-conduction runs, they are under-resolved, and their size is resolution dependent) and stream consistently along the field lines (panel b, c, d, f, and g of  Fig.~\ref{fig:tsl_tc0_cc}). Consistent with results from our fiducial run as reported in \S\ref{sec:results}, gas velocity and pressure deficits are tightly related. In these runs, the dashed lines in the lower right panel of Fig.~\ref{fig:tc0_anly} show (consistent with equation \ref{eq:bernoulli}) that the sum of gas thermal pressure and kinetic energy, $P_{\rm th}+\frac{1}{2}\rho v^2 \sim$constant, even over the temperature range $10^{4.5}~\kai<T<10^{5.5}~\kai$ where gas thermal pressure drops significantly due to cooling. Thus, the velocity predicted from equation \ref{eq:v_T} matches simulation results well, as seen in the upper right panel of Fig.~\ref{fig:tc0_anly} (compare the dashed and solid lines for all runs except (a) and (e)). 

Notably, the run with the power-law cooling curve, $\Lambda \propto T^{0.5}$, \mhdf, still shows vigorous streaming, with $v \sim 50 \, {\rm km \, s^{-1}}$. The scale free nature of the cooling curve is reflected in the fact that the velocity and pressure as a function of temperature (Fig.~\ref{fig:tc0_anly} right panels, pink points) are relatively featureless, and do not show the minimum in pressure (and peak in gas velocity) associated with the minimum in cooling time at $T \sim 10^{4.5}~\kai$ for the  standard cooling curve. 
 This confirms that the loss of thermal pressure balance is not associated with linear isochoric instability, since gas with this cooling curve is isochorically stable. We also note that streaming motions in this temperature range are not sensitive to the initial temperature; the pressure dips and streaming velocities of \mhdc ($T_{\rm init} \sim 2 \times 10^6\,\kai$; purple lines) and \mhdg ($T_{\rm init} \sim 10^7\,\kai$; black lines) in Fig.~\ref{fig:tc0_anly} are very similar\footnote{This is not a fully apples-to-apples comparison, since the initial density of the two runs was held fixed, and thus the pressure of \mhdg is 10 times larger. Otherwise, the close similarity of the curves in Fig.~\ref{fig:tc0_anly} support that our broad conclusion that $T_i$ does not matter, as long as it is in the thermally unstable range.}. Raising $T_i$ only stretches out the initial evolution, since the initial cooling time is longer.

  What about the non-streaming runs? In \mhda, gas at the temperature floor merges into a single cloud whose scale is comparable to the size of the simulation box (Fig.~\ref{fig:tsl_tc0_cc}, panel a). The cloud does not have any consistent velocity, it only seems to oscillate slightly, with low gas velocities $v\sim 10 \, {\rm km \, s^{-1}}$, of order the cold gas sound speed. As we can see from the right panel of Fig.~\ref{fig:tc0_anly} (see blue curves), gas remains isobaric, and thus there are no sustained pressure gradients to drive coherent motion. By contrast, in \mhde, cold gas is much sparser; each cold gas parcel occupies of order one grid cell (Fig.~\ref{fig:tsl_tc0_cc}, panel e), with no streaming motion observed. Although-- from the orange curves in Fig.~\ref{fig:tc0_anly} -- there appears to be a pressure dip and significant gas motions (with up to $v \sim 60 \, {\rm km \, s^{-1}}$ at $T \sim 10^{5.3} \, \kai$ and $v \sim 20 \, {\rm km \, s^{-1}}$ close to the floor temperature), there are no sustained, coherent streaming motions as in the other runs. Since hot coronal gas is galaxies has $T\sim 10^6 \, \kai$, this might appear to suggest that there is no streaming in the multi-phase CGM. 
  
\begin{figure*}
    \centering
    \includegraphics[width=\textwidth]{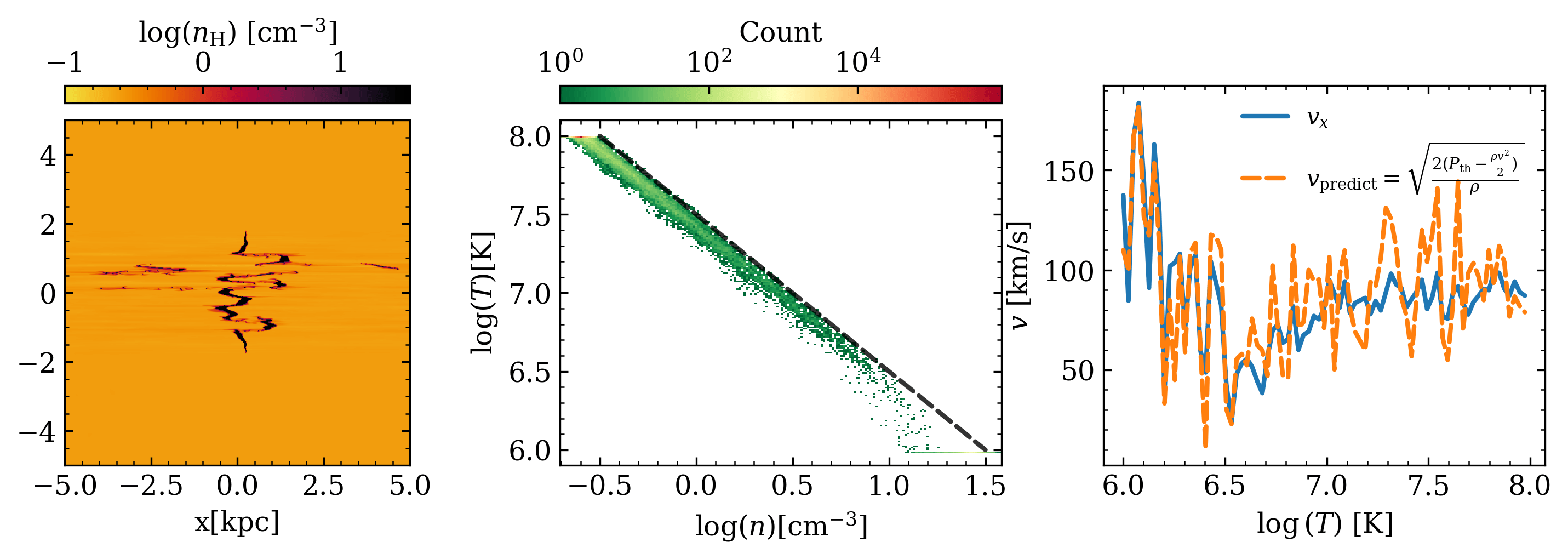}
    \caption{({\it Left}) Number density snapshot, ({\it Middle}) $T-n$ phase diagram, and ({\it Right}) gas velocity in $x$ as a function of temperature for the cooling cloud setup with $T_{\rm{floor}}=10^6$K and outflow boundary conditions. The dashed line in the $T-n$ phase plot shows an isobar down to $T \sim 10^{6.5}$K. Streaming is restored here even though the equivalent thermal instability setup, i.e. \texttt{mhd-tf6}, does not show streaming. 
    Notice that the pressure dip is small but the resulting velocities in the hot gas are still high. }
    \label{fig:mhd-tfloor6-CC}
\end{figure*}

\begin{figure*}
    \centering
    \includegraphics[width=\textwidth]{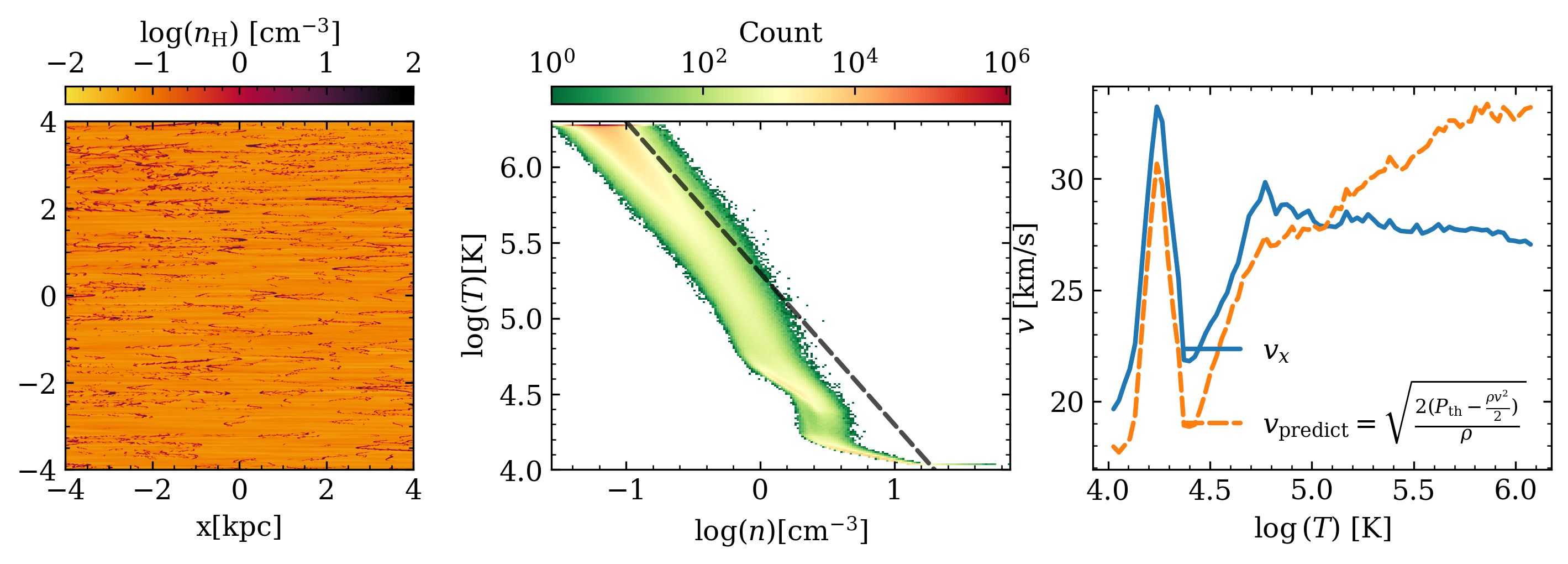}
    \caption{Same as Fig. \ref{fig:mhd-tfloor6-CC} but for the \texttt{mhd-tc6} case where $T_i=5\times10^5$ K. Streaming is restored here as well.}
    \label{fig:mhd-tceil6-ti5e5}
\end{figure*}

  Instead, the lack of streaming here appears to be an artifact of our thermal instability setup. The relative {\it amount} of hot/cold gas appears to be the culprit. The two `no streaming' cases, \mhda\  ($T \sim 10^6-10^8 \, \kai$; blue curves in Fig. ~\ref{fig:tc0_anly}) and \mhde\ ($T \sim 10^4-10^6 \, \kai$; orange curves in Fig.~\ref{fig:tc0_anly}) have the largest ($\sim 98 \, \%$) and smallest ($\sim 2\%$) mass fraction of floor temperature gas respectively. By contrast, in all the runs where streaming occurs, the mass fraction of cold and hot gas is comparable to within a factor of a few. This makes sense: when one phase completely dominates the mass budget, there is little mixing and out of pressure balance intermediate temperature gas to drive motion. When the cold gas fraction is very high (\mhda; blue curves) there is too much inertia and too little free energy in the hot component to drive gas motions. By contrast, when cold gas fractions are very low (\mhde; orange curves), pressure dips and higher velocity gas motions can develop locally around gas clumps, but the net amount of cooling is insufficient to drive a sustained `wind' in a flux tube. In contrast, when we do not simulate linear thermal instability but simply initialize comparable amounts of cold and hot gas, or adopt inflow (rather than periodic) boundary conditions so that the supply of hot gas is unlimited (e.g., the `slab' setup of \S\ref{sec:counter-streaming}), then streaming motions {\it do} develop for both these temperature ranges. This is shown, for instance, in Fig. \ref{fig:mhd-tfloor6-CC}, which is the cooling cloud setup\footnote{Such a situation can certainly develop organically, via non-linear density perturbations, or if pre-existing cold gas is injected into hot gas (e.g., via stripping of a satellite galaxy, or entrainment in a galactic wind).} for $T_{\rm floor}=10^6\ \rm K$, except with $T_i=7\times10^6\rm K$ instead $T_i=2\times10^6\ \rm K$). Here, streaming motions are able to develop in the entire cloud even though the pressure dip is relatively small (the $T-n$ phase plot is mostly isobaric). 
   Similarly, Fig. \ref{fig:mhd-tceil6-ti5e5} shows that the \texttt{mhd-tc6} case can also stream if $T_i=5\times10^5\ \rm K$ instead of $2\times10^6\ K$, i.e. not very close to the ceiling temperature. This demonstrates that as long as $T_i$ is sufficiently far away from both $T_{\rm ceil}$ and $T_{\rm floor}$, streaming can occur in these cases and is independent of the exact choice of $T_{i}$ otherwise.

We conclude that the non-streaming runs, where either the cold or hot gas fraction is small $f_{\rm c} \ll 1, f_{\rm h} \ll 1$, are artifacts of our initial or boundary conditions. As a practical matter, observationally they would no longer be considered multi-phase. We also note that there is no well-established theory for the relative abundant of hot and cold gas in the non-linear saturated state of thermal instability; this is beyond the scope of this paper. It is certainly sensitive to other physics besides the cooling curve. For instance, if we incorporate thermal conduction (see \S\ref{sec:conduction}), the energy transfer between phases means that the relative amount of cold and hot gas become more comparable, and the abundance of intermediate temperature gas is also increased. Streaming is indeed restored for both of our non-streaming runs. 


\begin{figure*}
    \includegraphics[width=\linewidth]{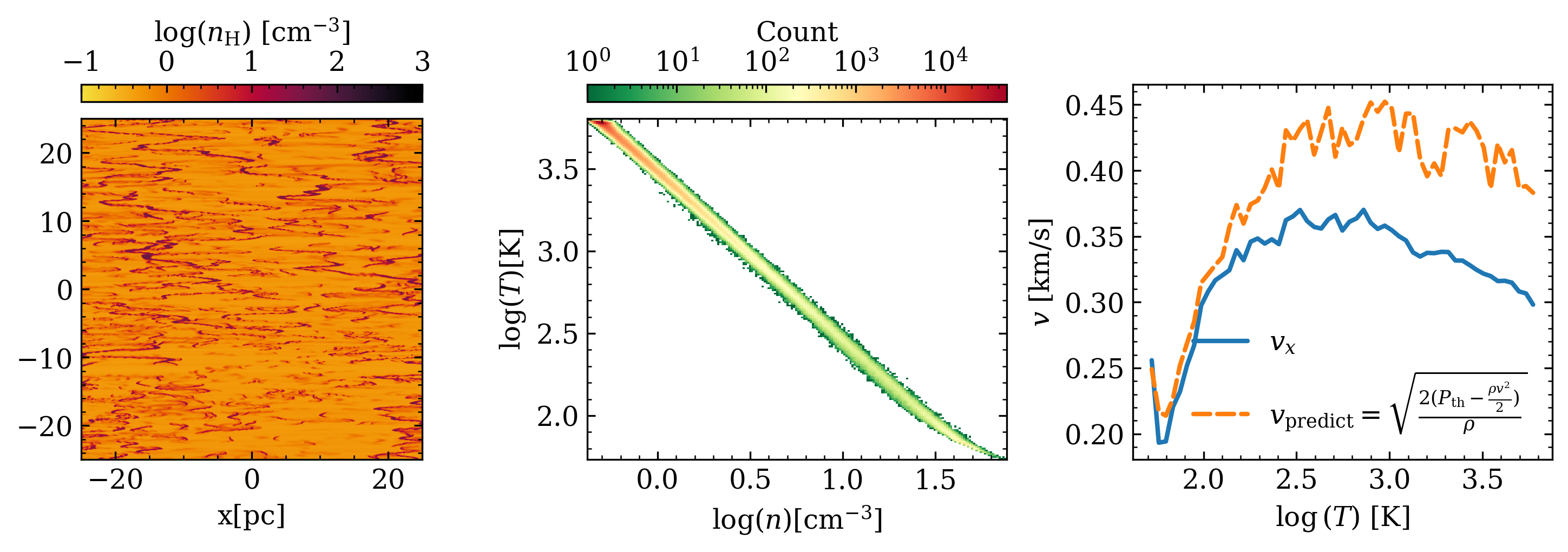}
    \caption{Same as Fig. \ref{fig:mhd-tfloor6-CC} but in the ISM no conduction thermal instability setup. While cold filaments are still formed, the streaming velocity is suppressed by 1-2 orders of magnitude with this different cooling curve. The suppressed velocities are a result of a negligible pressure dip in the cold molecular gas, as seen from the phase plot. Here, cooling is always isobaric.}
    \label{fig:ism_plot}
\end{figure*}

\begin{figure}
     \centering
\includegraphics[width=0.48\textwidth]{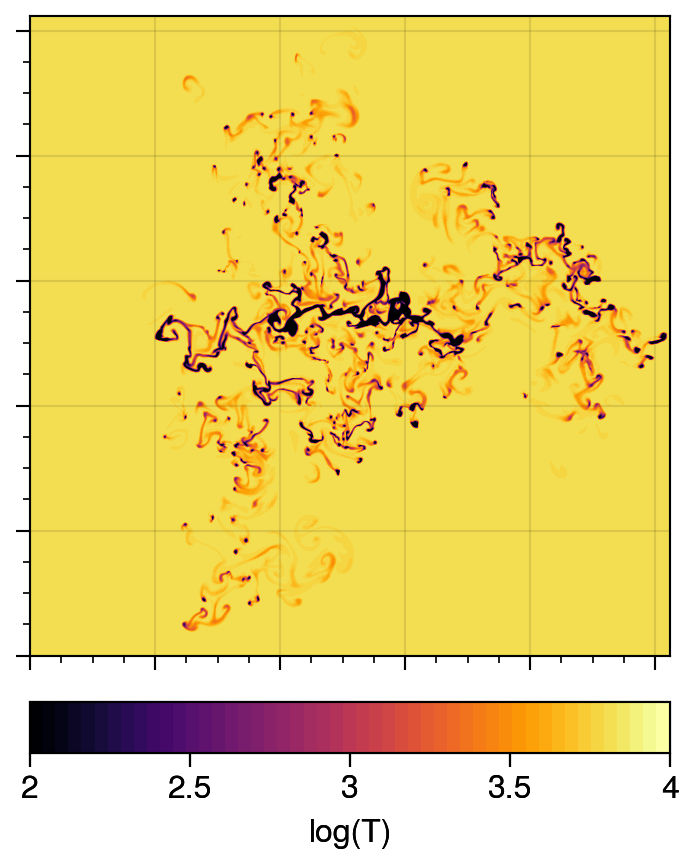}
       \caption[]{Temperature snapshot for the hydrodynamic cooling cloud run with ISM conditions ($T_{\rm floor}=10^2K, T_{\rm ceil}=10^4K$). Similar to the CGM case, the cloud shatters, and then eventually re-establishes pressure balance with its surroundings.
}
\label{fig:go_hd_ism}.
\end{figure}

However, there {\it is} at least one case where streaming does not develop, which we believe is physically meaningful. We simulate the thermal instability setup in the temperature range relevant for the ISM ($T \sim 10 \ \rm K- 10^4 \ \rm K$), using the ISM cooling curve from \cite{KoyamaInutsuka02}; see the cooling time plot in Fig \ref{fig:ism_tcool}. We initialize the number density $n_H = 2 \ \rm{cm^{-3}}$ and initial temperature $T_i=1.5\times 10^2$ K, with plasma $\beta=1$. Fig. \ref{fig:ism_plot} shows the results from this run. Streaming is highly suppressed here (with velocities $v \sim 0.35$ km/s, highly subsonic), compared to the \texttt{mhd-fid} case. Furthermore, there is no isochoric dip, as is the case in the ICM simulation cases. Rather, cooling takes place isobarically throughout the entire temperature/density range. Consistent with our findings, streaming motions were not been reported in similar setups which examine ISM conditions \citep{choi12,jennings21}. In our ISM simulation, the amount of cold and warm gas is comparable, in contrast to previous non-streaming cases, where the cold gas fraction was either very small or very large. Thus when outflow conditions are implemented along with a pre-existing cold mass clump in the simulation box, we still find that the ISM case shows no streaming. We have run hydrodynamic simulations of a cloud cooling down from to molecular temperatures surrounded by $T\sim 10^4$K (analogous to the shattering setup in \S\ref{sec:mhd-shattering}), and found that the cloud still shatters into small fragments, as shown in Fig. \ref{fig:go_hd_ism}. Unlike the case in \S\ref{sec:mhd-shattering}, however, the equivalent MHD setup does not show sustained streaming motions. Note that the only difference here is the different cooling curve for the $10 \, {\rm K} -10^4$ K temperature range, typical of the ISM. While we will explore this case further in future work, this shows the sensitivity of streaming to the cooling curve. 

Overall, we conclude that as long as there are comparable amounts of hot and cold gas, and $T_{\rm floor}=10^4$K, gas will develop streaming motions for $T_{\rm ceil} \sim 10^6 \, \kai -10^8 \, \kai$, i.e., in CGM, group and galaxy cluster environments. However, under ISM conditions ($T\sim 10 \, \kai-10^4 \, \kai$), molecular gas does not stream.

\subsection{Thermal Conduction}
\label{sec:conduction}

\begin{figure*}
     \centering    \includegraphics[width=\textwidth]{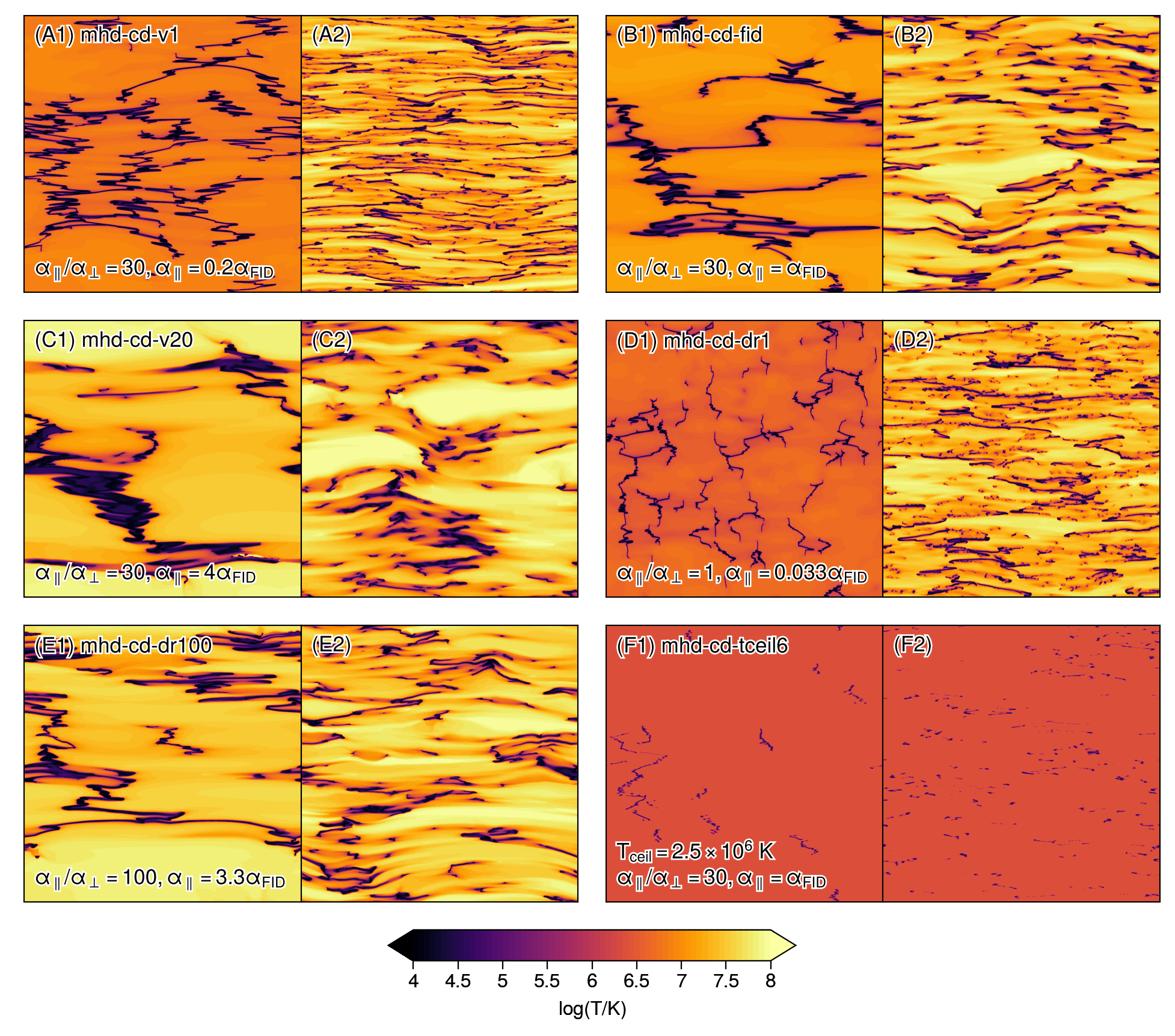}
       \caption[]{Conduction runs with different combinations of heat diffusivity ($\alpha_{\parallel}, \alpha_{\perp}$). We use the parallel heat diffusion coefficient $\alpha_{\parallel}$ of \mhdcdb\ to define $\alpha_{\rm FID}=1.5\times10^{28}~{\rm cm^2\cdot s^{-1}}$. Two snapshots of gas temperature are shown for each run: the left ones, i.e., (A1), (B1), ..., (F1) show early snapshots that mark the end of the initial cold gas contraction; and the right ones, (A2), (B2), ..., (F2), correspond to the stable stages at later times when the cold clumps stream consistently.}
     \label{fig:tsl_tc}
\end{figure*}

\begin{figure*}
    \centering  
    \includegraphics[width=\textwidth]{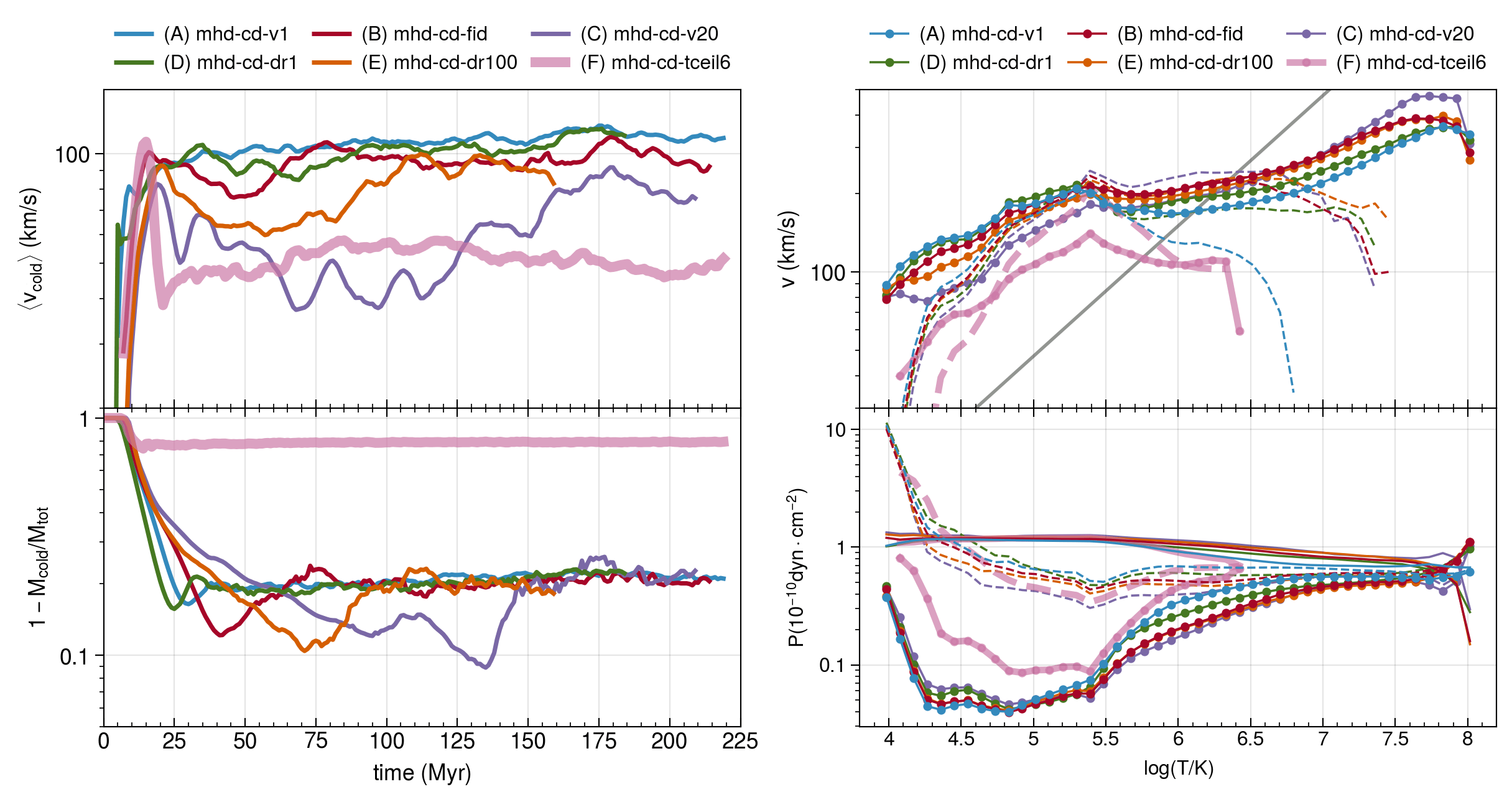}
    \caption{Same as Fig.~\ref{fig:tc0_anly} but for runs with conduction, i.e., (A)~--~(F). The profiles in the right panels are stack-averaged over snapshots $t=100\sim~150{\rm Myr}$ every $1~{\rm Myr}$.}
    \label{fig:tc_anly}
\end{figure*}

\begin{figure*}
    \centering \includegraphics[width=\textwidth]{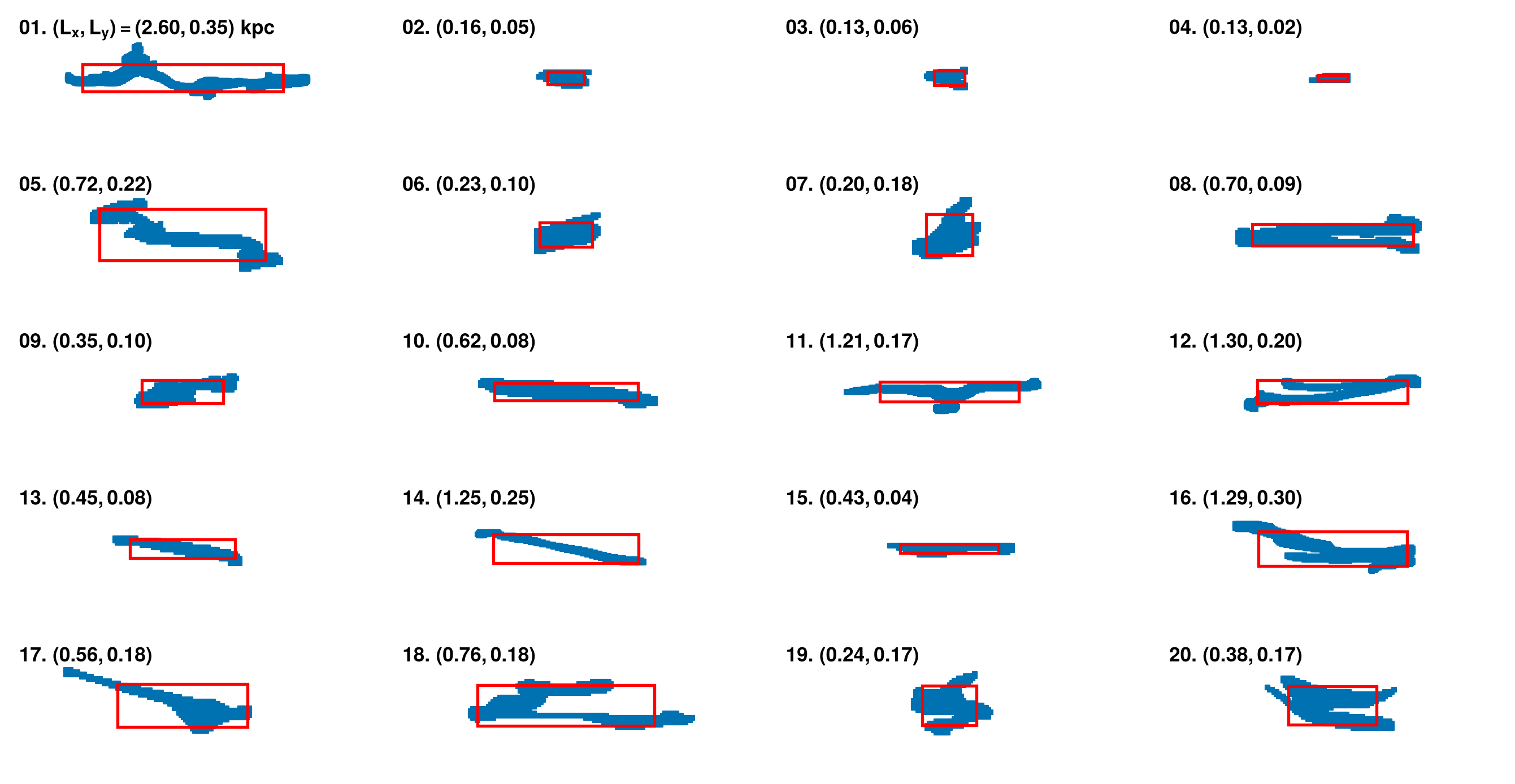}
    \caption{The gallery of identified cold clumps. From top to the bottom, each row shows four clumps that are randomly chosen from simulation snapshots at $t=150, 160, ..., 200~{\rm Myr}$, respectively. Clumps with $L_{x}<0.1~{\rm kpc}$ are not included. 
    The characteristic scales $L_{x}$ and $L_{y}$ are denoted at the upper left corner of each panel; and the red rectangles with width of $L_{x}$ and height of $L_{y}$ are drawn on top of the clump silhouettes.
    The $L_{x}$ ($L_{y}$) are estimated by three times the standard deviation of all $x$ ($y$) coordinates within each clump.}
    \label{fig:clump_gallery}
\end{figure*}

\begin{figure}
    \centering
    \includegraphics[width=\columnwidth]{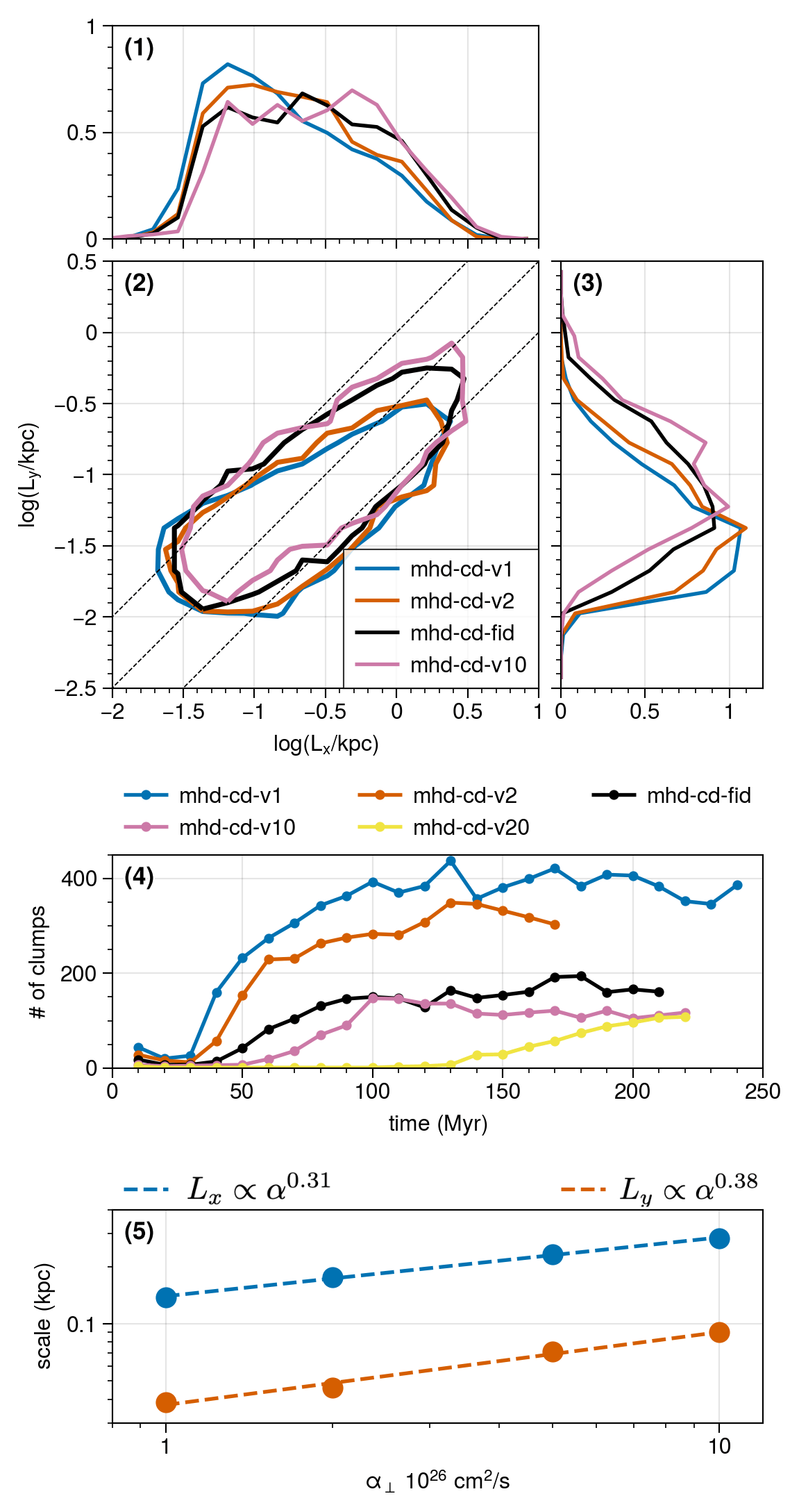}
    \caption{Statistics of $L_x$ and $L_y$ of runs with different ($\alpha_{\parallel}, \alpha_{\perp}$) with fixed $\alpha_{\parallel}/\alpha_{\perp}=30$. Panel~{\bf (1)}~--~{\bf (3)}: Corner plots showing the distributions of clump characteristic scales, $L_x$ and $L_y$. Panel~{\bf (4)}: time evolution of the number of clumps. Panel (5) median of $L_x$ and $L_y$ v.s. $\alpha_{\perp}$. Note that since we simulate at fixed $\alpha_{\parallel}/\alpha_{\perp}$, here $\alpha_{\perp}$ is a proxy for the overall normalization of $\alpha$. 
    }  
    \label{fig:clump_stats}
\end{figure}


In this section we consider the effects of thermal conduction.  As explained in \S\ref{sec:methods}, we adopt constant heat diffusivities in order to get numerically resolved cold clouds. To model anisotropic thermal conduction in MHD, we set an isotropic diffusivity for conduction perpendicular to the field lines, $\alpha_{\rm \perp}=\alpha_{\rm iso}$, and a stronger diffusivity along the field lines, $\alpha_{\rm \parallel}$. We study how the strength of the diffusivities affects the streaming motion. 
Panel (a1), (a2), (b1), (b2),..., (e1), (e2) of Fig.~\ref{fig:tsl_tc} show the gas temperature maps of the conduction runs with different combinations of $\alpha_{\parallel}$, $\alpha_{\perp}$. The runs \mhdcda, \mhdcdb, \mhdcdc\ (panels a, b, and c) have progressively larger $\alpha_{\parallel}$ with fixed $\alpha_{\parallel}/\alpha_{\perp}=30$; the runs \mhdcdd, \mhdcdb, and \mhdcde\  
(panels d, b, and e) have the same $\alpha_{\perp}$ but vary  $\alpha_{\parallel}/\alpha_{\perp}$ from 1, 30 to 100. All these runs adopt the same cooling curve and parameters as \mhd: $T_{\rm floor}=10^{4}~\kai$, $T_{\rm ceil}=10^{8}~\kai$, and $T_{\rm init}=2\times10^{6}~\kai$. The only difference is that we now include conduction. The run \mhdcdf\ (panel f) is the same as \mhdcdb\, except with $T_{\rm ceil}=2.5\times10^{6}~\kai$.

The overall evolution of the conduction runs are similar to the non-conduction runs. 
The seeded fluctuations quickly cool to $T_{\rm floor}$ and are elongated along the field lines. Subsequent mass flows, and attendant increases in density, arises predominantly along the magnetic field lines. Thus, similar to the non-conduction runs, cold clumps acquire a ``folded'' morphology due to the thin shell instability when contraction reaches the maximum, as shown by the first panels (a1, b1, ..., f1) of Fig.~\ref{fig:tsl_tc}. 
The second panels (a2, b2, ..., f2) illustrate the stable stages following the initial contraction, where the folded cold clouds elongate and break up to smaller clumps that stream consistently along the magnetic fields (e.g., the left column of Fig.~\ref{fig:rtp_all}). This stable streaming stage is qualitatively similar to that in the non-conduction runs. The main difference is that the cold clumps are now larger and certainly better resolved once conduction is included; they do not break down to grid scale. The results are also qualitatively similar to the cooling cloud run \mhdcc, where we start with a large, well-resolved cool cloud. 

As shown by the left panels of Fig.~\ref{fig:tc_anly}, the velocity and mass of the cold gas in the different conduction runs are broadly similar to that of the non-conduction run \mhd\ (excluding \mhdcdf\, which has a different temperature ceiling), at the factor of $\sim 2$ level. While the run with the least thermal conduction has a somewhat higher cold gas velocity than the non-conduction run ($\sim 100 \, {\rm km \, s^{-1}}$ for \mhdcda\ compared to $\sim 70 \, {\rm km \, s^{-1}}$ for \mhd), increasing thermal conduction appears to lower cold gas velocities ($\sim 100,80,50 \, {\rm km \, s^{-1}}$ for \mhdcda, \mhdcdb, \mhdcdc\ respectively). This is likely due to the larger sizes of cold filaments as we increase conduction, which increases the inertia of cold gas in a given flux tube. At least for this setup, the cold gas fraction of $\sim 80\%$ is robust to changes in conduction strength and anisotropy, and similar to the corresponding value ($\sim 90\%$) in the non-conduction run \mhd.   
The right column of Fig.~\ref{fig:tc_anly} indicate that including thermal conduction does not significantly change the velocity and pressure phase structure of the multiphase media. The hot gas velocity is higher in conduction runs, but still subsonic; and Eq.~\ref{eq:bernoulli} still holds for the cooling gas, as shown by the dashed lines. 

However, as is already apparent from comparing Figs. \ref{fig:tsl_tc0_cc} and \ref{fig:tsl_tc}, thermal conduction does have a strong impact on the size of cold clouds, which are now larger and numerically resolved. We use a clump finding algorithm to determine the size distributions of the resolved cold clouds in conduction runs. Each isolated $T=10^{5}~\kai$ isothermal contour is identified as a cold clump. In this way we avoid from over-fitting the temperature structures of the cold gas. The sizes of the cold clumps are quantified by the characteristic scales $L_{x}=3\sigma_{\rm lx}$, $L_y=3\sigma_{\rm ly}$, where $\sigma_{\rm lx}$ ($\sigma_{\rm ly}$) is the standard deviations of the $x$($y$) coordinates of all grid cells within a clump. 
Fig.~\ref{fig:clump_gallery} presents a randomly-picked subset of identified clumps that have $L_{x}>0.1~{\rm kpc}$. The red rectangles with width equals to $L_{x}$ and height equals to $L_y$ are overplotted on the clump silhouettes. The clumps are horizontally elongated; and $L_{x}$ and $L_{y}$ describe the length and thickness of the clumps fairly well. A few of the clumps are misaligned with the grid axis (e.g., clump 14, 17) or have the folded ``V-''shape (clump 12, 18); and $L_{y}$ overestimate the thickness for these clumps.

Panel (1)~--~(3) of Fig.~\ref{fig:clump_stats} show the probability distribution functions (pdfs) of $L_{x}$ and $L_y$ in the form of corner plots of the runs \mhdcda, \mhdcdg, \mhdcdb, and \mhdcdh, which have progressively larger $\alpha_\perp$ with fixed $\alpha_\parallel/\alpha_\perp$. The 2D contours in panel (2) enclose regions in the ($L_x$, $L_y$) space containing $\sim90\%$ of all clumps. The clump size data is obtained by stacking snapshots from $t=100~{\rm Myr}$ to the end of the simulations, every $10~{\rm Myr}$. As shown by panel (4), the number of clumps are stable after $t=100~{\rm Myr}$ except for \mhdcdc, which is excluded from the corner plot analysis. 
As shown by the corner plots, the pdfs of clump length generally flatten from $\sim0.05~{\rm kpc}$ to $\sim0.5~{\rm kpc}$; and skew towards small scales for weaker diffusivity runs, resulting in larger number of clumps (panel 4). 
At fixed $\alpha_{\parallel}/\alpha_{\perp}$, we find $L_{\rm x}, L_{\rm y} \propto \alpha^{1/3}$, which differs from the expected $L \propto \alpha^{1/2}$ scaling.
We have also run a suite of simulations where we vary $\alpha_{\parallel}/\alpha_{\perp}$ (specifically, we vary $\alpha_{\parallel}$ at fixed $\alpha_{\perp}$). While we find $L_x \propto \alpha_{\parallel}^{1/3}$ as before, we also find a weak scaling $L_y \propto (\alpha_{\parallel}/\alpha_{\perp})^{0.2}$, which we interpret as due to the distortion and bending of field lines, so that the magnetic field is misaligned with the x-axis, allowing $\alpha_{\parallel}$ to influence $L_y$. 

We caution that in order to numerically resolve the parallel Field length, we have resorted to a temperature independent heat diffusion coefficient $\alpha$, which is quite different from the temperature dependent Spitzer conduction coefficient $\kappa \propto \alpha \rho \propto T^{5/2}$ (which corresponds to $\alpha \propto T^{3/2}$ for isobaric cooling). Our conduction coefficient is much larger than the Spitzer value at low temperatures. A particular danger is that this artificially boosted conduction coefficient results in $\lambda_{\rm F} > c_{\rm s} t_{\rm cool}$ at low temperatures, whereas in reality $\lambda_{\rm F} < c_{\rm s} t_{\rm cool}$ should always hold\footnote{This can be easily verified directly, and also understood analytically, from the fact that the Field length is the geometric mean of the electron elastic ($\lambda_{\rm ee}$) and inelastic ($\lambda_{\rm cool}$) mean free paths: $\lambda_{\rm F} \sim (\lambda_{\rm ee} \lambda_{\rm cool})^{1/2} \sim v_e (t_{\rm ee} t_{\rm cool})^{1/2}$, where $v_e$ is the electron thermal velocity and $t_{\rm ee}$ is the electron Coulomb equilibration time.}. A situation where $\lambda_{\rm F} > c_{\rm s} t_{\rm cool}$ leads to a situation where the temperature scale height is larger than the lengthscale over which pressure balance can be maintained, and very large pressure dips and turbulent gas motions can arise. Although our artificial heat diffusion coefficient obviously affects clump sizes, we have checked that it does not affect our broad conclusions about the effect of conduction on MHD streaming dynamics in the parameter regime we have tested, in particular the size of pressure dips, which are due to magnetic pressure support. This is also clear from the fact that pressure dips and streaming velocities are similar in the non-conduction and conduction runs. However, this artificial assumption can affect high $\beta$ runs where magnetic fields are weak, as we soon see in \S\ref{sec:beta}.  

In summary, the effect of conduction is as follows. Although they differ in detail, qualitatively the velocity and pressure phase structure of conduction and non-conduction runs are similar. The main effect of conduction is to increase the size of filaments as one might expect. This is good news insofar as it enables numerically converged gas morphology, instead of clump size scaling with resolution. However, in detail clump sizes are difficult to understand quantitatively. They are {\it not} of order the Field length (equation \ref{eq:field}), as in thermal instability simulations where the clumps are largely static, but considerably larger. The scaling with diffusion coefficient, $L_{\rm x} \propto \alpha_{\parallel}^{1/3}$, $L_{\rm y} \propto \alpha_{\perp}^{1/3}$, also differs from the expected $L \propto \alpha^{1/2}$ scaling. However, given this scaling, the relative lengths $L_x/L_y \sim (\alpha_{\parallel}/\alpha_{\perp})^{1/3} \sim (30)^{1/3} \sim 3$ is roughly as expected.




\begin{figure}
    \centering
    \includegraphics[width=\linewidth]{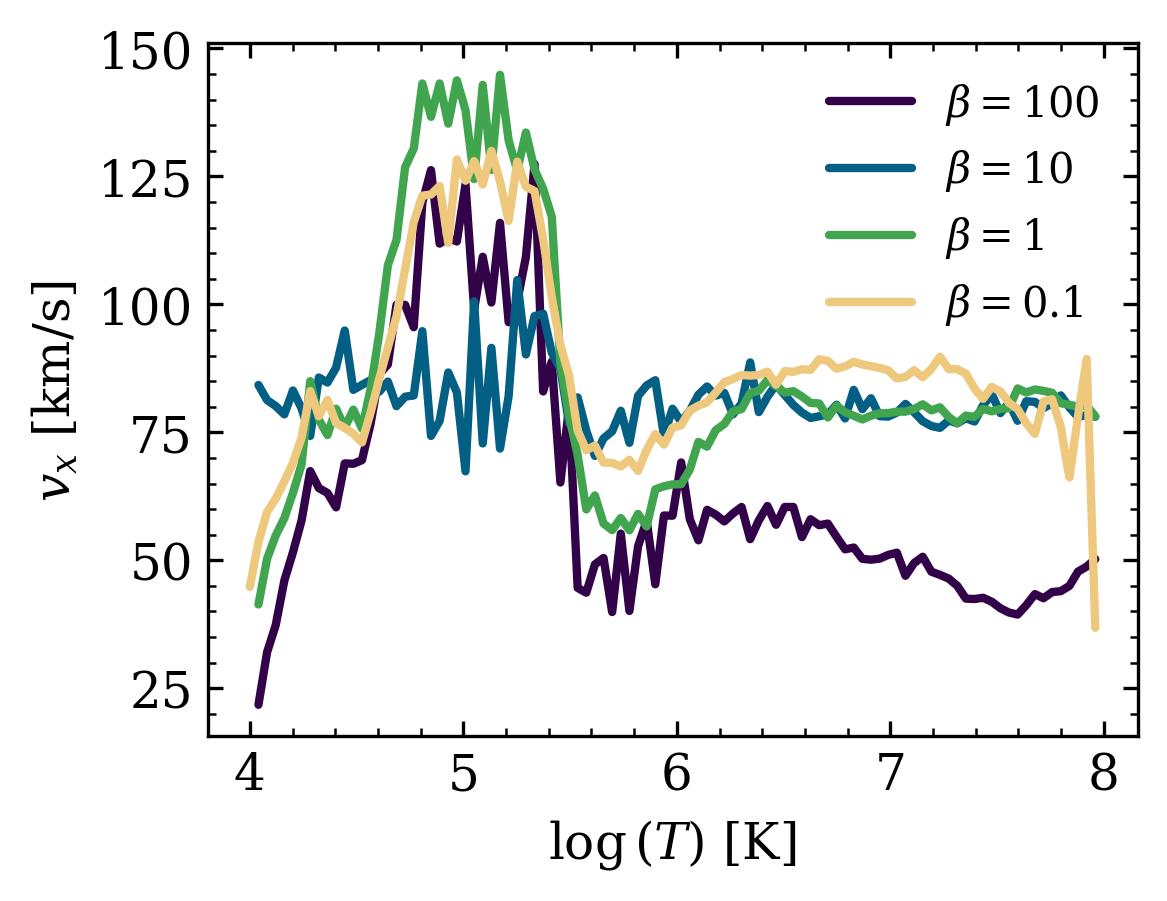}
    \caption{Gas velocity in the $x$ direction as a function of temperature for the fiducial no-conduction runs with varying plasma $\beta$. Streaming velocities are largely consistent across different $\beta$ suggesting an independence of streaming criterion on the magnetic field strength for a large range of reasonable $\beta$. The highest $\beta=100$ run, however, shows lower velocities at high temperature.}
    \label{fig:beta_dependence}
\end{figure}
\begin{figure*}
    \centering
    \includegraphics[width=\linewidth]{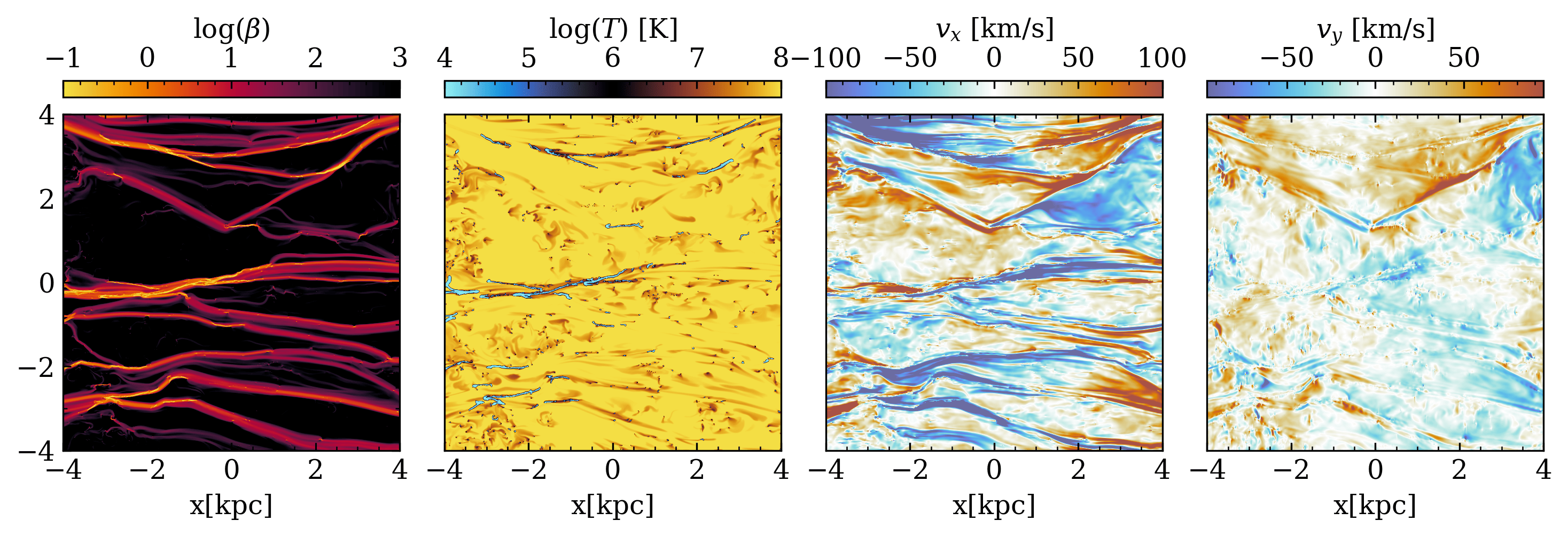}
    \caption{Plasma $\beta$, temperature, and velocity ($x$ and $y$) snapshots of the \texttt{mhd-b100} simulation. Here, more extended cold filaments are formed, along with significant bending of field lines. The cold gas is at much lower $\beta(\sim0.1)$, as compared to the background gas, ($\beta\sim100-1000$) thus making it magnetically supported. The streaming velocities are oriented in both directions ($x$ and $y$) in contrast to the low $\beta$ simulations, where cold gas is directed along the field lines only in $x$.}
    \label{fig:beta100}
\end{figure*}
\subsection{Plasma $\beta$}
\label{sec:beta}

Is the presence of streaming dependent on a strong initial magnetic field? Fig. \ref{fig:beta_dependence} shows that across a broad range of initial plasma $\beta_i=0.1,1,10, 100$, streaming still occurs, and the streaming velocities remain roughly independent of $\beta_i$. The reason for this robustness to $\beta$ is that even at high $\beta_i$, compression and flux-freezing results in the the cold gas having low $\beta$, with values similar to the other lower initial $\beta$ simulation cases. The only significant change with $\beta_i$ is that in highest $\beta$ run ($\beta_i=100$), we find somewhat lower gas velocities at high temperatures $T>10^6$ K, because this high temperature gas is not yet magnetically supported and has somewhat higher density, compared to runs with lower $\beta_i$.  The morphology, dynamics and magnetic field orientation of the \texttt{mhd-b100} case can be seen in Fig. \ref{fig:beta100}. The condensed cold filaments reside in separated ``horizontal strips'' where fields are highly amplified, so the cold clumps are still dominated by magnetic fields ($\beta_{\rm cold}\ll1$). Due to strong compression, these strips occupy a small volume fraction. In these regions, the amplified magnetic fields are aligned so the cold gas still streams in a similar manner as the $\beta_i=1$ case. 
Notably, in the \texttt{mhd-b100} case these strips are unstable  where they merge into a single narrow strip which contain most of the cold gas. 

Additionally, we also find that high $\beta_i$ runs have higher cold gas mass fractions. This is because stronger magnetic fields result in lower density magnetically supported gas, which reduces cooling at the interface of cold and hot gas. The high $\beta$ backgrounds naturally result in more cold gas condensing out, since cooling is the most effective in these cases.  
\begin{figure*}
    \centering
    \includegraphics[width=\linewidth]{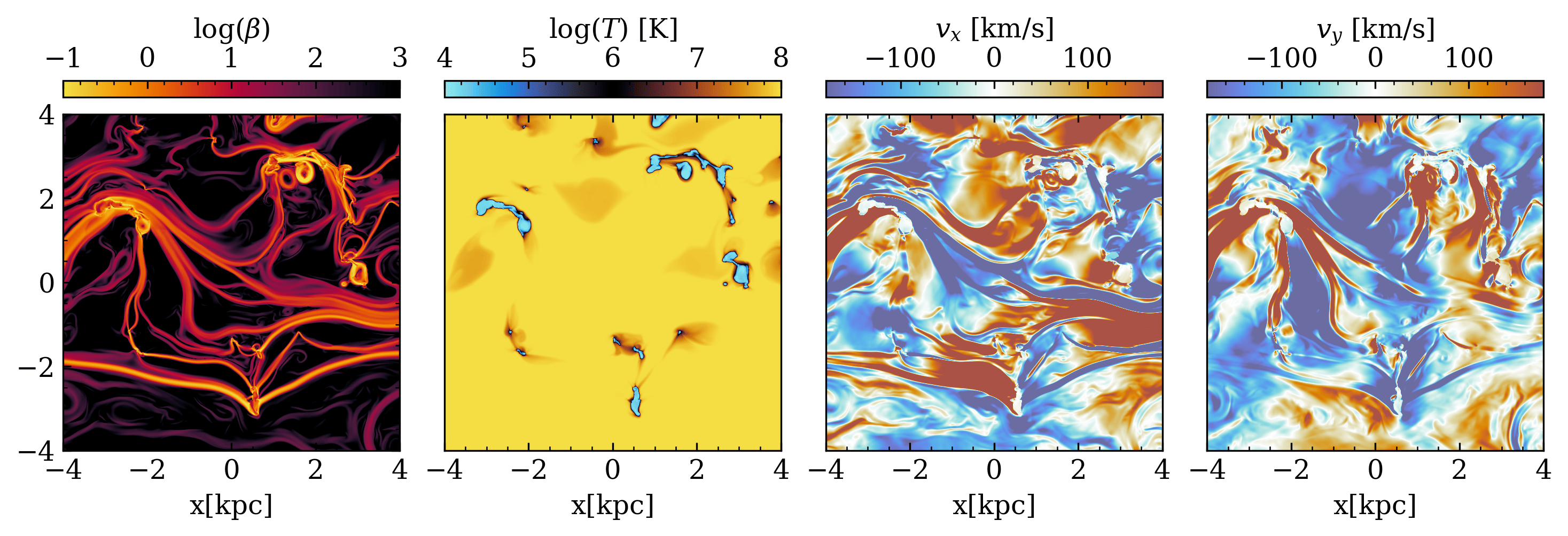}
    \caption{Same as Fig. \ref{fig:beta100} but with anisotropic conduction included. The addition of conduction with high initial $\beta$ causes more disordered motion along with significant bending of magnetic field lines.}
    \label{fig:high-beta-cd}
\end{figure*}
In conduction runs with our fiducial heat diffusion coefficients, while streaming is present at lower $\beta_i$, at the highest $\beta_i \sim 100$, magnetic fields are strongly distorted during the compression, and the ordered streaming motion is absent (see Fig \ref{fig:high-beta-cd}). 
As discussed in \S\ref{sec:counter-streaming}, in hydrodynamic simulations, evaporation and condensation due to thermal conduction in a multi-phase medium can drive cloud disruption, gas motions and turbulence, for similar reasons as the Darrieus–Landau instability in premixed combustion \citep{nagashima05,inoue06,kim13,iwasaki14,jennings21}: convex (concave)
cold gas surfaces have converging (diverging) heat flux, and drive
evaporation (condensation) respectively. When magnetic fields are weak, as in our high $\beta$ simulations, these gas motions stochastically re-orient the background field and prevent coherent streaming. As noted in \S\ref{sec:conduction}, our conduction coefficient at the floor temperature $T \sim 10^4$K is artificially high, so it is not clear if this behavior will still hold with a more realistic Spitzer conduction coefficient.

\subsection{Numerical Resolution; Convergence}
\label{sec:converge}

As we discuss in \S\ref{sec:methods}, when conduction is implemented, our grid size is comparable to the minimum Field length (equation \ref{eq:field}), which is therefore marginally resolved. However, another important lengthscale is the cooling length, $c_s t_{\rm cool}$, the lengthscale on which cooling gas is able to maintain sonic contact and thermal pressure balance with its surroundings (\citealt{mccourt18}; see discussion in \S\ref{sec:mhd-shattering}). Due to the large dynamic range, our simulations are not always able to resolve the cooling length, $l_{\rm cool}=c_s t_{\rm cool}$ of the multiphase gas, which falls drastically as gas cools. Its minimum value $\sim 10^{-3}-10^{-2}~{\rm pc}$ at $T \sim 10^4$K (from the $n \sim 3-30 \, {\rm cm^{-3}}$ typical range of gas densities at $T \sim 10^{4}$K in our fudicial simulation \mhd; see Fig. \ref{fig:rtp_all}) is far below the numerical resolution of our simulations. 
In hydrodynamic simulations of turbulent mixing layers, increasing resolution reduces pressure decrements caused by gas cooling. The latter almost vanish if ${\rm min}(l_{\rm cool})\sim \Delta x$ \citep{fielding20}. However, these pressure decrements due to poor resolution are much smaller ($\Delta P \ll P$) than the ones we observe, which we argue are physical. 
Nonetheless, it is still important to check whether the pressure decrements and the streaming motions we see are affected by numerical resolution.

\begin{figure}
     \centering
\includegraphics[width=\columnwidth]{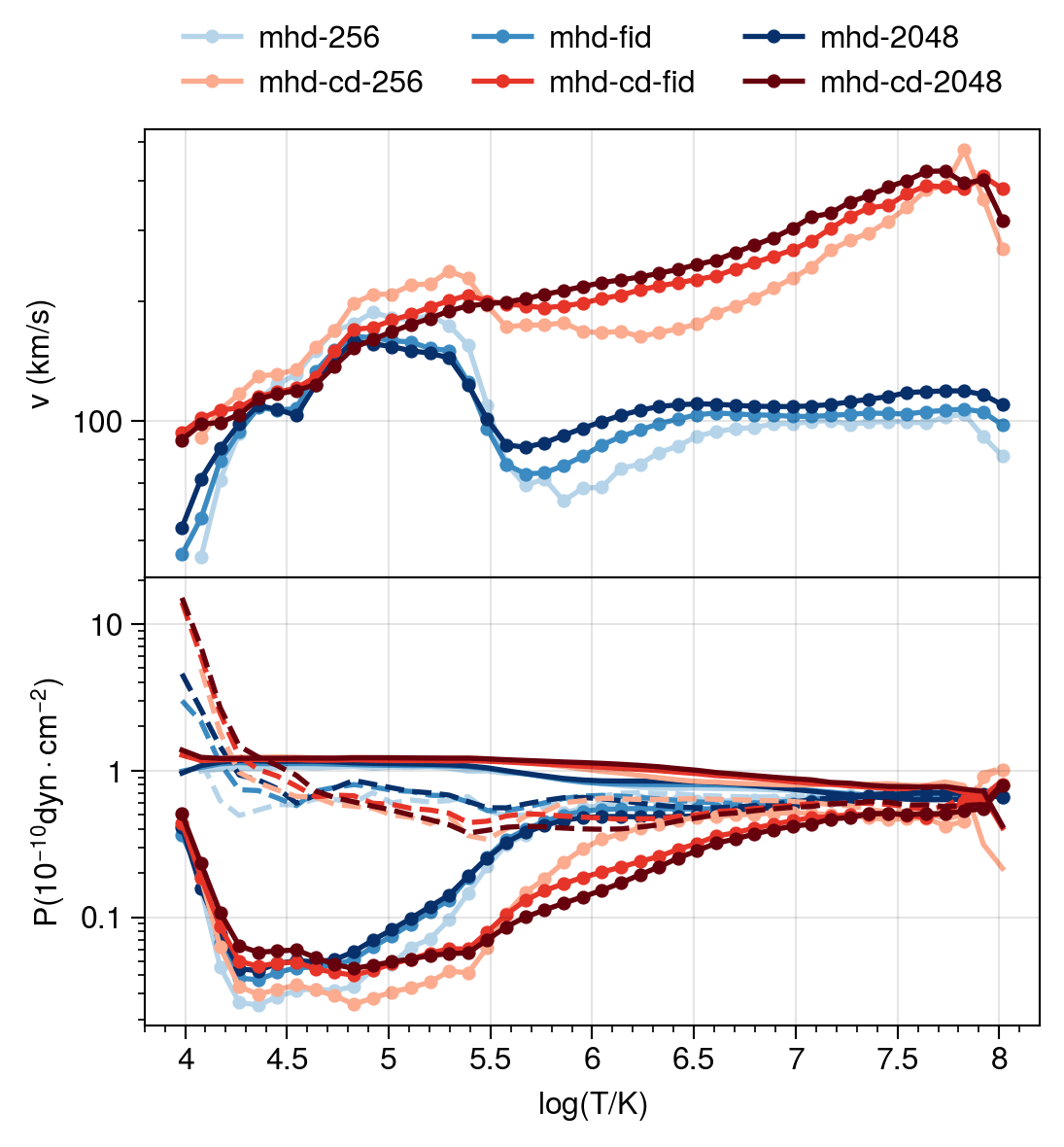}
       \caption[]{ Convergence performance of gas velocity (top) and pressure components (bottom), as a function of temperature. Red (blue) lines show results of runs with (without) thermal conduction, with deeper color representing higher resolution, which are $256^2, 1024^2$, and $2048^2$ respectively. In the bottom panel, $P_{\rm mag}$, $P_{\rm th}$, and $P_{\rm th}+\frac{1}{2}\rho \sigma^2$ are shown as the solid, dot dashed, and dashed lines, respectively.
       All profiles are averaged within each temperature bin and over $t=50\sim70~{\rm Myr}$, during which all runs have reached the stable states.}
\label{fig:conv_test}.
\end{figure}
\begin{figure*}
     \centering
\includegraphics[width=0.49\textwidth]{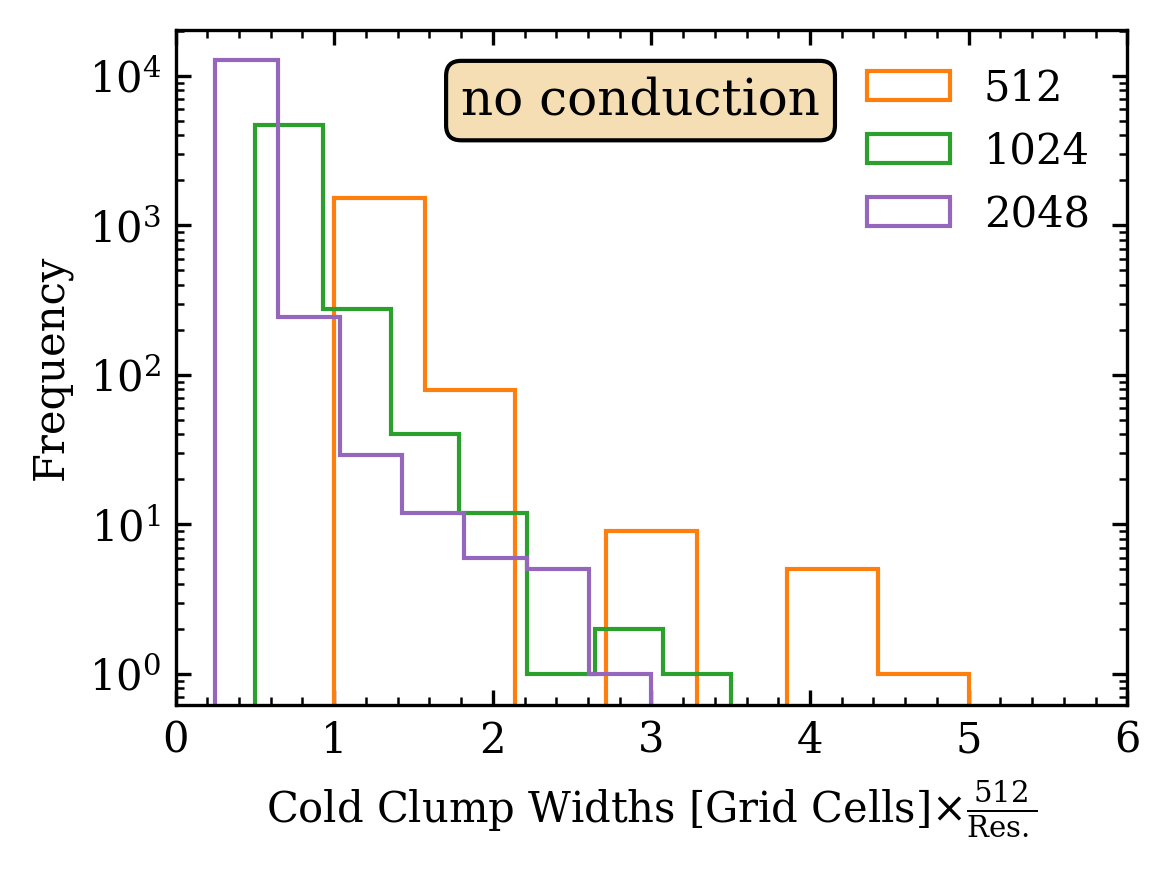}
\includegraphics[width=0.49\textwidth]{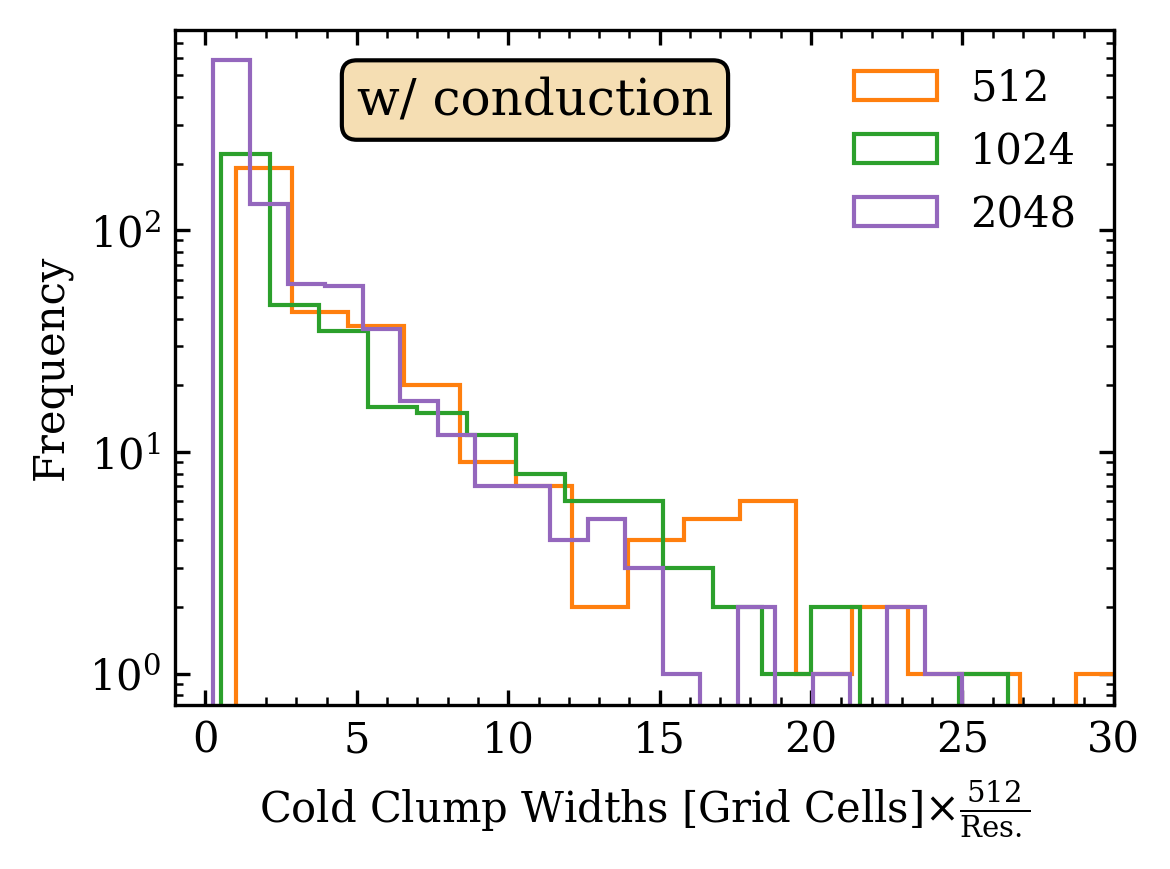}

       \caption{The PDFs of the cold clump widths for the no conduction (Left) and with conduction (Right) cases in the thermal instability setup for varying resolution from $512^2$ to $2048^2$. The widths are scaled by the physical length of the clumps corresponding to the $512^2$ simulation (for example, the 2048 case is scaled by a factor of $512/2048=1/4$). The width PDFs are well converged when conduction is included, but are unconverged without conduction -- observe that the histograms in the left panel are offset by a factor $\sim 2$, i.e. clump sizes scale directly with grid scale. Indeed, without conduction a high fraction of cold clumps contract to a single grid cell in the transverse direction, regardless of resolution.}
\label{fig:temp_slc_conv}.
\end{figure*}

First, we conduct standard convergence tests. Fig.~\ref{fig:conv_test} shows the gas thermal pressure as functions of temperature of runs with $N=256^2, 1024^2,$ and $2048^2$. The pressure decrements and velocities are fairly well converged in the higher-resolution cases for both conduction runs (\texttt{mhd-cd-fid}, \texttt{mhd-cd-2048}), and non-conduction runs (\texttt{mhd-cd-2048},\texttt{mhd-2048}). Even the lowest resolution runs (\texttt{mhd-cd-256} and \texttt{mhd-256} respectively) are not too far off. 
Thus, gas dynamics seems to be fairly well-converged. By contrast, whether gas morphology is well-converged depends on whether conduction operates (Fig. \ref{fig:temp_slc_conv}). The cold clump width PDFs are well converged when conduction is included, but are unconverged without conduction. Note that the histograms in the left panel of Fig. \ref{fig:temp_slc_conv} are offset by a factor $\sim 2$ at successive resolutions, i.e. clump sizes scale directly with grid size. Indeed, without conduction a high fraction of cold clumps contract to a single grid cell in the transverse direction, regardless of resolution. The fact that dynamics can be converged even when morphology is not also holds in other multi-phase contexts. For instance, in turbulent radiative mixing layers, the surface brightness and mass inflow rates are converged, even when the fractal front morphology--where the unresolved interfaces without conduction generally occupy one grid cell-- is not \citep{tan21}. There, convergence arises because the eddy turnover time sets the mixing rate, and thus only the largest eddies (which set this mixing time) need to be resolved.  Here, we speculate that convergence arises because magnetic pressure support arrests further compression (hence transition to isochoric cooling as seen in the phase diagram of Fig. \ref{fig:rtp_all}), allowing the anisotropic pressure tensor, which is behind the crucial thermal pressure dips, to be resolved. However, such statements need to be checked in future detailed simulation studies.  

\begin{figure*}
    \centering
    \includegraphics[width=\linewidth]{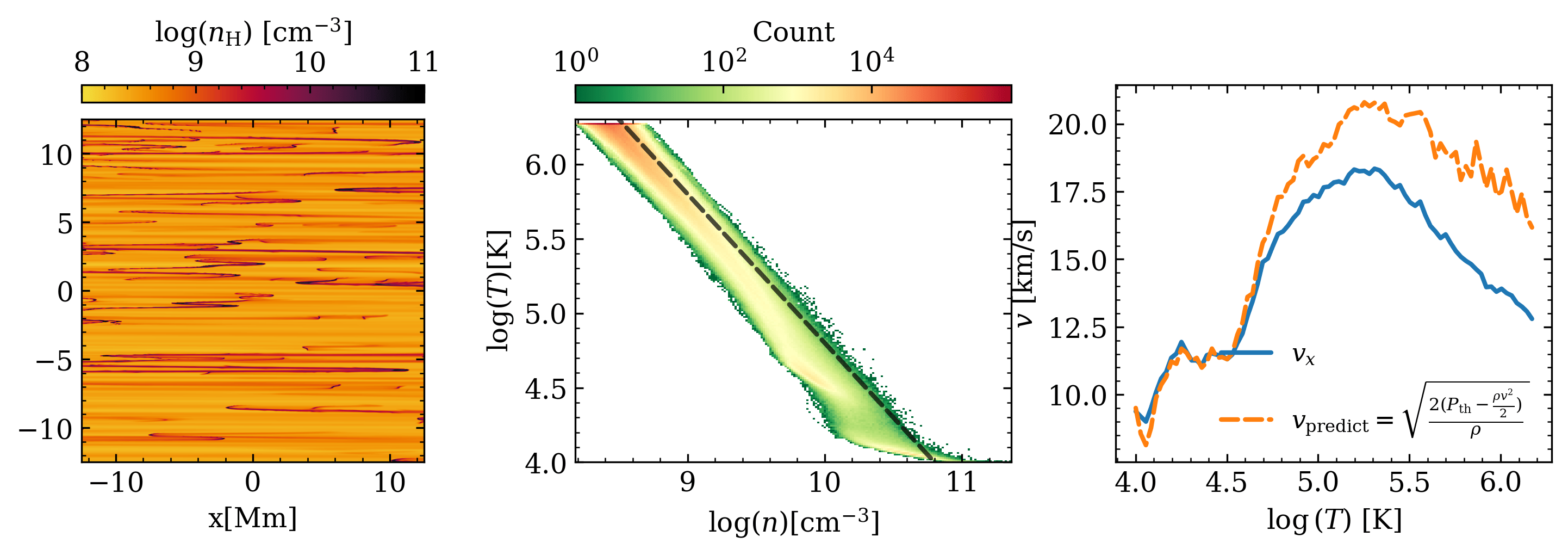}
    \caption{Thermal instability setup with solar corona conditions, where $c_{\rm s} t_{\rm cool}$ of the cold gas is resolved. As before, the dashed line in the $T-n$ phase plot is an isobar. Streaming motions are still observed, caused by the pressure dip indicating that the existence of streaming is independent of resolution, even though the sizes of cold filaments are unconverged.}
    \label{fig:solar_case}
\end{figure*}

Another worry might be that the thermal pressure dips we observe are due to $c_s t_{\rm cool}$ being unresolved, rather than anisotropic magnetic pressure support as we have claimed. We have run smaller box simulations where $L/c_s t_{\rm cool}$ is much smaller, so that $c_s t_{\rm cool}$ is explicitly resolved, and we still find pressure decrements and streaming. This is seen in Fig. \ref{fig:solar_case}, which is the thermal instability setup for solar coronal conditions without conduction. Here, the entire box size is $L=10^5\ \rm{km}$ with resolution $N_x=N_y=4096$ cells, with $T_i=10^6\ \rm{K}, n_i=10^{9} \, {\rm cm^{-3}}, T_{\rm ceil}=10^6\ \rm K, T_{\rm floor} = 10^4\ \rm K$. The background environment has the same cooling curve as the CGM / ICM \citep{claes20} for the same temperature range ($10^4-10^6\ \rm K$), so the background gas has $t_{\rm cool}(T=10^6K)\sim3\ \rm{hrs}$, and the cold gas (coronal rain, as discussed in \S\ref{sec:corona}) has $c_{\rm s}t_{\rm cool}(T=10^4 \rm K)\sim 50\ \rm km$, compared to the resolution of $\sim 24\ \rm km$. Notice that streaming motions are still present, along with the isochoric pressure dip, even when $c_{\rm s}t_{\rm cool}$ is resolved at all temperatures. Although a detailed study of coronal rain in the solar atmosphere is outside the scope of this paper and will be presented in future work, this serves as an existence proof that pressure dips and streaming still occur if $c_s t_{\rm cool}$ is resolved. 



In summary, the pressure decrements and the streaming motions of the cold gas appear well-converged in our current simulations, but gas morphology is not, unless conduction is explicitly included (and the relevant clump sizes are resolved). Furthermore, pressure decrements and cold gas streaming still occurs if $c_s t_{\rm cool}$ is explicitly resolved. However, we note that convergence still has to be carefully explored and characterized in a wider region of parameter space, varying $\beta$, and quantities such as $L/c_s t_{\rm cool}$, $L/v_A t_{\rm cool}$ (where L is the box size). We plan to do so in future work. 

\section{Discussion}
\label{sec:discussion}

In \S\ref{sec:previous}, we place our work in context by discussing some previous work. Next, we discuss possible applications of our simulations. As we have highlighted, streaming motions do not appear for thermal instability in the ISM temperature range $T \sim 100 \ \rm K -10^4$K. Instead, it only appears in a bistable medium where the cooler gas is $T \sim 10^4$K. We therefore discuss potential applications in the solar corona (\S\ref{sec:corona}) and CGM (\S\ref{sec:CGM}), where the temperature range is $T \sim 10^4 \,{\rm K}- 10^6$K. Finally, in \S\ref{sec:future}, we discuss future directions.  

\subsection{Why has multi-phase MHD streaming not been seen more widely?}
\label{sec:previous}

To our knowledge, with the exception of some simulations of solar coronal rain (see \S\ref{sec:corona}), self-sustained MHD streaming motions in a multi-phase medium have not been reported in the literature, either as an analytic prediction or in MHD simulations. Here, we explore differences in setup which can explain this. 

The biggest reason for the absence of MHD streaming in the thermal instability literature appears to be the fact that most simulations focus on the ISM. For instance, \citet{choi12}, which simulates thermal instability in this regime, finds field-aligned cold filaments, with weak motions ($v\sim 0.2 \, {\rm km \, s^{-1}}$) attributed to evaporative flows driven by thermal conduction. These velocities are similar to those in our non-conduction runs (Fig \ref{fig:ism_plot}). \citet{jennings21} find similar results. We have confirmed in our own MHD simulations with the ISM cooling curve that streaming does not occur; we will study the underlying reason in future work. 

What about MHD simulations of thermal instability in conditions relevant to the CGM or ICM? The closest analog to our simulations is \citet{sharma10}, who run 2D MHD simulations to study thermal instability in ICM considering weak magnetic fields (initial plasma $\beta_i \sim 250$), thermal conduction and adiabatic cosmic rays. They implement field-aligned Spitzer conduction $\kappa \propto T^{5/2}$ and a small isotropic heat diffusivity, and emphasize the need to resolve the Field length both along and across B-fields to achieve numerically converged cold gas morphology. Their choice of a realistic conduction coefficient implies a huge dynamic range -- the Field length at a few times $10^4$K is $\sim 10^8$ times smaller than that at $T \sim 10^7$K. For this reason, they set a temperature floor at $T \sim 2 \times 10^6$K. Their results are similar to our Fig. \ref{fig:high-beta-cd}, a similarly high $\beta$ run with anistropic thermal conduction: while the cold medium forms field-aligned filaments, and there are gas motions, there are no sustained streaming motions. As in Fig. \ref{fig:high-beta-cd}, the combination of high $\beta$ and artificially high thermal conduction at the floor temperature result in bent B-fields and disordered motion which preclude streaming. 
While streaming under ICM conditions requires more study, it is quite possible that the very weak conduction at a floor temperature $T\sim 10^4$K with Spitzer conduction will behave similarly to the high $\beta$,  no conduction case (Fig. \ref{fig:beta100}), where coherent streaming is present. 


In summary, our work is in agreement with existing literature when we adopt similar assumptions. The streaming motions we find appear in a different portion of parameter space. Another very relevant corpus of previous work is simulations of solar coronal rain, which we now discuss. 

\subsection{Application: Coronal Rain}
\label{sec:corona}

In his original paper, \citet{parker53} envisioned the solar corona and chromosphere as a prime candidate for his new theory of thermal instability. It was also a key application in Field's seminal paper \citep{field65}. Indeed, perhaps the most direct and dramatic application of our simulations is coronal rain (see \citealt{antolin20,antolin22} for recent reviews). These are blobs of cool (${\rm T} \sim 10^4$K), dense ($n \sim 10^{10}-10^{13} \, {\rm cm^{-3}}$) gas which can spontaneously appear out of the hot ($T \sim 10^6$K) corona on a time scale of minutes, with its motion guided by the B-field lines of coronal lines. Interestingly, coronal rain is seen not only as downflow as the dense ``raindrops'' fall due to gravity, but also as upward motion towards the loop apex. The infall velocities are $\sim 100 \, {\rm km \, s^{-1}}$, with downward acceleration significantly lower than the effective gravity ($a \approx 1/3 g_{\rm eff}$). Closely related (with similar temperatures and densities) but much less common are solar prominences, also $\sim 100$ times cooler and denser than coronal gas, which are typically much larger magnetically supported structures which continuously drain and refill on timescales of days to weeks \citep{vial15}. 

While the solar corona is a multi-phase medium with similar temperature contrasts ($T \sim 10^{4}-10^6$K) as the CGM, it enjoys observations with remarkable spatial and temporal resolution which would be impossible in the CGM. The formation and dynamics of coronal rain is tracked in {\it real time} in both emission and absorption in a broad range of spectral lines on timescales of seconds onwards. The same structures can host many rain events over the course of several days. The observed lines include Ly$\alpha$, H$\alpha$, as well as lines associated with transition temperatures such as HeII 304 \AA, OV and OVI, SiIV and SiVII. The characteristic widths are $100-600$ km, depending on wavelength and resolution, while the lengthscales range from $\sim 1000-20 \, 000$ km \citep{antolin15}. Unlike the CGM, direct B-field measurements are also available (e.g., \citealt{schad16,kriginsky21}). Magnetic fields are significantly stronger ($\beta \sim 0.01-1$) compared to expectations for the CGM. 

These remarkable observations of coronal rain formation and evolution makes it the most direct application of our simulation. In upcoming work, we will present simulations directly tuned to solar parameters. However, our present calculations (which were more scaled to CGM parameters), already show many features of the solar observations, since the same physics is at play. In particular, from ultra-high resolution direct imaging \citep{alexander13}, coronal rain shows clear evidence for counter-streaming flows (known as `siphon flows'), with velocity differences of up to $\sim 80-100 \, {\rm km \, s^{-1}}$. Bi-directional counter-streaming flows are also seen in H$\alpha$ \citep{zirker98,lin03-solar,ahn10}. MHD simulations of thermal instability in the solar context do already show the development of fast counter-streaming flows in neighboring flux tubes \citep{fang13-solar,fang15,xia17,zhou20}, with fragmentation of the initial pancake-like cool gas into field-aligned filaments attributed to the Rayleigh-Taylor \citep{xia17} and thin-shell instabilities \citep{claes20}, in simulations with and without gravity respectively. Counter-streaming flows in close accord with observational constraints have been demonstrated in 2D MHD simulations of a curved flux sheet anchored in the solar chromosphere, extending up to the solar corona \citep{zhou20}. The counter-streaming flows were attributed to spatially and temporally stochastic turbulent heating, localized at the solar surface. However, this cannot be the primary explanation, given that we see similar counter-streaming flows in our simulations, which only have uniform heating. \citet{zhou20} suggest that while the hot coronal gas has alternating unidirectional flows, counter-streaming in cold gas is due to longitudinal oscillations of the filament threads. While we see alternating unidirectional flows in both phases, note that we employ different boundary conditions (periodic, rather than a finite loop geometry). Other work attributes the siphon flows to thermal pressure gradient induced inflows onto condensing gas \citep{xia17,claes20}, in agreement with our results. However, the details of fragmentation, as well as the development and maintainence of counter-streaming flows, were not analyzed or explained. 

Our calculations adopt a background unstratified plasma in thermal equilibrium. The solar corona obviously is subject to a strong gravitational field. Indeed, coronal loops are also thought to exist in a state of thermal non-equilibrium (TNE), with heating and cooling cycles where the loops fills with evaporating gas from the loop footprints, which either accumulates to form a prominence or evacuates under the action of gravity and coronal rain formation \citep{antolin20,antolin22}. One manifestation of TNE is the observed  long-period intensity pulsations in lines corresponding to hot coronal gas $T \gsim 10^6$K \citep{auchere14,froment15}. However, there is sufficient separation of scales that local, unstratified thermal instability can be a good approximation \citep{claes19,claes20}. Coronal rain appears in a matter of minutes, compare to the few to tens of hours timescale for long period intensity pulsations. Also, at the top of the magnetically supported loop, the effective gravity is initially zero. There are many interesting related questions (e.g., the physics behind the observed parallel and transverse scales of observed rain; the low acceleration of infall, with a terminal velocity set only by the clump density contrast) that we will pursue in future work. However, Fig. \ref{fig:solar_case} serves as an existence proof that self-sustained counter-streaming motions can occur at the temperatures and densities of the solar corona, in simulations where $c_s t_{\rm cool}$ is explicitly resolved. 



\subsection{Application: Cold Clouds in a Wind, ICM filaments} 
\label{sec:CGM}

In principle, the CGM (or the ICM) should be a prime target for the magnetized thermal instability and `shattering' that we have simulated. In practice, direct application of our results is complicated by the action of gravity and compensating buoyancy forces. Since magnetic fields are expected to be much weaker ($\beta \sim 10-100$) than in the solar corona, it is unclear whether magnetic tension can halt infall, as for the top of coronal loop arcs. Linear thermal instability is damped by buoyancy: in hydrodynamic simulations, $t_{\rm cool}/t_{\rm ff} \lsim 10$ is required for cooling to overcome the damping effect of buoyant oscillations \citep{mccourt12,sharma12}. However, MHD simulations of stratified thermal instability find this constraint to be considerably weakened, independent of field orientation, as magnetic tension suppresses buoyant oscillations \citep{ji18}. Once cold gas forms, since it is $\sim 100$ times overdense, hydrodynamic simulations show that they fall quasi-ballistically, until they reach terminal velocities $v_{\rm t} \approx g t_{\rm grow}$ (where $t_{\rm grow} = m/\dot{m}$ is the mass growth time) set by drag from the accretion of low momentum hot gas \citep{tan23-gravity}. MHD simulations of cloud infall finds similar results, with magnetic drag further reducing infall velocities when B-fields are transverse to gravity \citep{kaul2025}. 

However, there is at least one situation in the multi-phase CGM where there is little relative motion between cold and hot gas, at least in the background state: cold clouds entrained in a hot wind. Such clouds are seen in quasar-line absorption observations outflowing at high velocity. Theoretically, in recent years there has been an emerging new consensus paradigm for cloud entrainment, which focuses on the role of condensation (\citealt{gronke18,gronke20-cloud,li20,schneider20,kanjilal21,abruzzo22}; see \citealt{FGOH23} for a recent review). Hydrodynamic instabilities cause the cold gas and hot gas to mix. If the mixed gas cools faster than it is produced, the cloud will survive and grow. The cooled gas retains its initial momentum, and in time cool gas both grows in mass and comoves with the hot gas. This form of mixing-induced thermal instability enforces kinematic coupling between phases, as gas is converted from hot to cold.

How do B-fields affect condensation? 
As the cloud sweeps up field lines, magnetic draping quickly amplifies B-fields to equipartition with ram pressure, resulting in a high magnetized `skin' \citep{lyutikov06,dursi07}. This has two effects: (i) magnetic drag increases momentum transfer between hot and cold phases, speeding up acceleration \citep{dursi08,mccourt15}; (ii) B-fields suppress the KH instability via magnetic tension \citep{jones97}. In plane-parallel shearing layer KH simulations, the latter strongly suppresses condensation \citep{ji19}. B-fields should therefore drastically change cloud-wind interactions. Mass growth {\it is} suppressed indeed in radiative MHD simulations of cold streams, similar to KH simulations \citep{kaul2025}. However, for clouds, although morphology changes dramatically and is much more filamentary, hydro and MHD cloud survival and mass growth are similar \citep{gronke20-cloud,li20,sparre20,hildalgo23}. The physical origin of this disconnect is not understood. 
 
 One difference between clouds and streams is the mode of gas mixing, due to geometry. Mixing in cold streams is via the Kelvin-Helmholtz instability -- if shear drops to zero, so does mass growth. By contrast, mass growth in clouds peaks when they are entrained and shear plummets: cooling driven pressure fluctuations induce pulsations, which drive mixing and condensation \citep{gronke20-cloud, gronke22-coag}. Thus, suppressing shear or KH mixing has little effect. We suspect that thermal pressure gradients also drive mixing and cooling in magnetized entrained clouds. 
In MHD cloud-crushing simulations after the cloud is entrained, we see field-aligned `streaming' due to cooling-induced pressure gradients in the tail of the cloud, very similar to what is seen in our static `slab' simulations (\S\ref{sec:counter-streaming}). The common origin of thermal pressure induced inflow and mixing in both the hydrodynamic case (where it manifests as acoustic cloud pulsations), and the MHD case (where it manifests as counter-streaming motions) is an important clue for the physical origin of their similar mass growth rates. However, the latter does not follow trivially, as MHD clouds fragment into low density magnetically supported filaments with larger surface area, but weaker cooling. This will be the subject of a future paper.  

Streaming due to thermal pressure gradients does not affect the bulk motion of the cloud, since streaming is equal and opposite; there is no net force or change in the center of mass. However, it does result in small-scale motions which would produce line shifts and (if unresolved) line broadening. In general this cannot be distinguished from the effects of turbulence in the CGM. However, in spatially resolved observations, it might be possible (just as counter-streaming is seen in the solar corona). For instance, it would be interesting to consider if counter-streaming flows might be observable in HVCs or IVCs in the Milky Way. 

Another potential application are ICM filaments, which can encompass very narrow and long thread-like structures, consisten with significant magnetic support \citep{fabian08}. The large temperature contrasts ($T \sim 10^4-10^8$K) are conducive to high velocity streaming, and the high $\beta\sim 100$ of ICM plasma need not be an impediment since flux-freezing strongly amplifies B-fields in the cold gas (\S\ref{sec:beta}). The simulations in this paper ignore gravity. There are MHD simulations of thermal instability in stratified halos which do incorporate gravity \citep{ji18,wibking25}. However, while those works do show that magnetic support can ameliorate the effects of gravity, they largely focused on how magnetic fields changed the criteria for linear thermal instability, and also used artificial (power-law) cooling curves. The potential for streaming should be studied in simulations which take gravity, ambient turbulence, and more realistic assumptions about conduction (\S\ref{sec:conduction}) into account.


\subsection{Future Directions}
\label{sec:future}

This is a first study of `magnetized shattering', which leads to rapid streaming of multi-phase gas along B-field lines. There is much more work to be done. Possible extensions and refinement of this work include: 
\begin{itemize}

\item{{\it 3D simulations.} In this paper, we have conducted mostly 2D simulations, although we also reported the results from a 3D simulation, and confirmed that rapid multi-phase streaming still operates. Indeed, the thermal pressure dip and streaming velocities are remarkably similar between 2D and 3D  (Fig \ref{fig:3d_pressure}), although there some morphological differences between 3D and 2D runs, with more small-scale cold clumps in 3D runs. Nonetheless, given the different properties of magnetic fields in 2D and 3D, it would be important to revisit and reconfirm many of the key results of this paper in 3D.}

\item{{\it Tangled magnetic fields; turbulence.} In our idealized study, we have assumed initially uniform fields. The anisotropic magnetic pressure in this configuration is ultimately responsible for producing field aligned cold filaments and streaming. Tangled B-fields could change this -- for instance, in ISM thermal instability calculations, long cold filaments aligned with B-fields are no longer produced \citep{choi12}, and it seems likely that streaming would also be stifled. It would be worth doing a careful series of calculations where the relative strength of the mean guide field and random field, as well as plasma $\beta$, are varied. A potential outcome might be that streaming only occurs in local patches where the coherence length of the magnetic field is large compared to other lengthscales, and the mean field is dominant. Observationally, long thin cold gas filaments are commonly seen in ISM and ICM environments, and are seen to be aligned with the local magnetic field (e.g., \citealt{mcclure06}). 

A related issue is that of driven turbulence, which drives gas motions, tangles magnetic fields, mixes gas phases, and provides turbulent pressure support. It would be interesting to understand when streaming occurs in MHD simulations of multi-phase driven turbulence (e.g., \citealt{das23}). We suspect it should operate in sub-Alfvenic turbulence in the presence of a strong mean guide field, but this of course requires further study and is beyond the scope of this paper. Also note that in our high $\beta$ simulations, radiative cooling itself drives turbulence (similar to hydrodynamic simulations; \citealt{iwasaki14}) and tangles magnetic fields. However, in the non-linear end state, much of the cold gas resides in regions where the magnetic field has been amplified by compression and twisting, with a strong mean field where streaming takes place. }

\item{{\it Cosmic rays.} The streaming motion here depends on a source of non-thermal pressure support, magnetic fields, which is also anisotropic. Cosmic rays (CRs) -- which often reach energy equipartition with thermal gas and magnetic fields -- are another potential source of non-thermal pressure support. However, in contrast to magnetic fields, in the strong scattering limit necessary for tight coupling, cosmic ray pressure is isotropic. Similar to the thermal gas, it operates in the direction of its pressure gradient, and the combined CR and gas pressure force is just $- \nabla (P_{\rm g} + P_{\rm CR})$. If CRs are tightly coupled to the gas and behave quasi-adiabatically, so that their contribution to pressure support rises in dense regions (just as for B-fields), the combined (CR+thermal) pressure dip along field lines will be smaller, reducing streaming velocities. However, in preliminary work including CRs, we have found that for reasonable assumptions for CR transport (CR streaming and/or diffusion), CRs rapidly leave cold filaments and have no net dynamical effect, with CR pressure fairly uniform over the simulation box. Thus, results are similar to MHD only simulations. However, the detailed dependence of streaming on CR transport should be carefully quantified in separate work. }

\item{{\it Temperature-dependent conduction.} In this paper, we assumed temperature independent heat diffusion coefficient $\alpha$, which is related to the customary conduction coefficient $\kappa$ (where the heat flux ${\mathbf F} = - \kappa \nabla T$) via $\kappa \propto \alpha \rho$, so that for isobaric cooling, $\kappa \propto T^{-1}$. By contrast, Spitzer conduction has a temperature dependence $\kappa \propto T^{5/2}$, i.e. it falls rapidly at lower temperatures. We did this so that at least the parallel Field length $\lambda_{\rm F} \sim (\alpha_{\parallel} t_{\rm cool})^{1/2}$ is resolved. Otherwise, for a more realistic conduction coefficient, the Field length drops rapidly with temperature. The fact that $\lambda_{\rm F} > c_s t_{\rm cool}$ with our assumptions (whereas the reverse is true in reality) can lead to artifacts in high $\beta$ environments (see discussion at end of \S\ref{sec:conduction}). We will investigate these issues, and the origin of the $L \propto \alpha^{1/3}$ scaling we observe, in future work. } 

\item{{\it Simulations in more realistic settings.} The simulations in this paper are highly idealized. Now that we understand the basic physics, some next steps would be to understand when and how it occurs in simulations with more astrophysically relevant and realistic setups -- for instance, those discussed in \S\ref{sec:corona} and \S\ref{sec:CGM}: coronal rain, in clouds entrained in galactic winds, the multi-phase CGM and ICM, and so forth, where we can include geometry, dynamics, and other physics (gravity, winds, turbulence) specific to those physical settings.}

\end{itemize}

\section{Conclusions}
\label{sec:conclusions}

In this work, we study the dynamics of radiatively cooling, magnetized multi-phase gas, which form organically via thermal instability, starting from either linear or non-linear initial density perturbations. We find a novel pattern of gas motion: both ambient hot gas and field-aligned cold clumps exhibit long-lived, self-sustained streaming motions parallel to the magnetic field, at velocities up to the sound speed of the hot phase. This can be up to $\sim 50-100 \, {\rm km \, s^{-1}}$ for CGM and solar corona conditions ($T_{\rm hot} \sim 10^6$K), and up to an order of magnitude higher for ICM conditions ($T_{\rm hot} \sim 10^8$K). Strikingly, neighboring flux tubes counter-stream in opposite directions. We delve in the physical origin of these counter-streaming motions. Our conclusions are as follows:  

\begin{itemize}
    \item {\it Streaming motions arise when cooling gas falls out of thermal pressure balance. Strong thermal pressure gradients are sustained by cooling and anisotropic magnetic pressure support.} 
    Rapidly cooling gas falls out of thermal pressure balance with its surroundings, and can even cool isochorically. In hydrodynamics, this drives `shattering', where cold gas fragments to small pieces until pressure balance is re-established \citep{mccourt18,gronke20-mist,yao25}. However, in MHD, magnetic fields are amplified via flux freezing in compressed gas. Cooling gas receives magnetic pressure support perpendicular to field lines, but not parallel to field lines, where thermal pressure gradients are unopposed. The latter accelerate gas along B-field lines until ram pressure makes up for the thermal pressure deficit, $\rho v^2 \sim \Delta P$. This results in streaming velocities $v \sim \sqrt{\Delta P/\rho}$, which can be as large as the hot gas sound speed when pressure deficits are large $\Delta P \sim P$, and gas cools quasi-isochorically.

    \item{{\it Counter-streaming motions arise from a cooling-induced, MHD version of the thin-shell instability.} Neighboring flux tubes stream in opposite directions (and there is no overall net streaming, as required by momentum conservation). Around cool filaments, magnetic tension forces cause the mostly horizontal flows to diverge away from convex heads and converge towards the concave tails. In particular, magnetic fields draped around a convex head deflects gas to the concave tail of a neighbor, which is thus pushed from behind. This amplifies initial corrugations, and sets up long-lived counter-streaming flows.}

    \item{\it Streaming is relatively robust, both physically and numerically (with some notable exceptions).} Streaming appears to be a robust phenomenon. It occurs both with and without thermal conduction (though the {\it size} of filaments is sensitive to conduction; see below); streaming velocities only change weakly when conduction is included. It still occurs with weak initial B-fields ($\beta_{i} \sim 100$), where one might expect magnetic pressure support to be negligible; streaming velocities change only weakly with $\beta_i$ (Fig. \ref{fig:beta_dependence}). This is because $\beta$ is still consistently low (and magnetic pressure support important) in compressed cool gas. It occurs with scale-free power law cooling curves where there is no isochoric thermal instability, and in both CGM and ICM temperature ranges. In detail, the size of pressure dips and velocity of streaming motions is sensitive to the cooling curve and temperature range, among other factors. 
    Streaming appears robust to numerical resolution (Fig. \ref{fig:conv_test}), and occurs both at very low resolution and in high resolution simulations where $c_s t_{\rm cool}$ is explicitly resolved (Fig. \ref{fig:solar_case}). While we mostly run 2D simulations, it also appears (with similar pressure dips and velocities) in a 3D simulation (Fig. \ref{fig:3d_pressure}). 

 However, it is not completely universal. Most importantly, it appears strongly suppressed with ISM cooling curves ($T\sim 10-10^4$K; Fig \ref{fig:ism_plot}), even though cooling blobs `shatter' in hydrodynamic simulations (Fig \ref{fig:go_hd_ism}). We address this case in future work. It also does not appear in some more contrived setups, which we view as less physically important. For instance, it does not appear in high $\beta$ sets with strong conduction at $T \sim 10^4$K (Fig. \ref{fig:high-beta-cd}), though we expect realistically conduction should be negligible at $T \sim 10^4$K; if so, streaming does occur (Fig. \ref{fig:beta100}). It is also suppressed in setups where the cold gas mass fraction is very high or very low (\texttt{mhd-tf6,mhd-tc6}), though such extremes are not observed in realistic settings and sensitive to our assumed initial or boundary conditions. 
 

    \item {\it Thermal conduction sets the physical size of cold filaments.} 
Although streaming dynamics does not appear sensitive to resolution, cool gas morphology is --  without conduction, clump sizes scale with resolution. However, once conduction is included, both the width and length of the filaments are numerically converged (Fig. \ref{fig:temp_slc_conv}). However, both the normalization (clump sizes much larger than the Field length) and scaling ($L \propto \alpha^{1/3}$, rather than $L \propto \alpha^{1/2}$, where $\alpha$ is the heat diffusivity) differ from the case when clumps are static, and will be the subject of future work. 

\item{\it Streaming may already have been seen in the solar corona; it is also relevant to the CGM and ICM.} Counter-streaming motions of cool gas filaments with velocities of the similar amplitude as our simulations have already been seen in coronal rain. We will study this important case in more detail in upcoming work. Streaming may also place an important role in the dynamics and growth of cool magnetized filaments in galactic winds and in the ICM. 
    
\end{itemize}

\section*{Acknowledgements}

We thank Brent Tan, Rony Keppens, Ethan Vishniac, participants of the KITP workshop `Turbulence in Astrophysical Environments' for helpful discussions. We particularly thank Patrick Antolin for helpful discussions and comments regarding coronal rain. We acknowledge NASA grant 19-ATP19-0205 and NSF grant AST240752 for support. This research was also supported in part by grant NSF PHY-2309135 to the Kavli Institute for Theoretical Physics (KITP). We acknowledge use of the Stampede2 supercomputer through allocation PHY240194. 

\section*{Data Availability}
The data underlying this article will be shared on reasonable request to the corresponding author.



\bibliographystyle{mnras}
\bibliography{master_references,References} 




\appendix
\section{B-field Orientation: Numerical Effects}
\label{app:bfields}

Our calculations initially utilized a setup where B-fields were aligned diagonally, at 45 degrees relative to the grid, following \citet{sharma10}. However, we found that our results differed compared to simulations where the B-field is grid aligned. We have already shown that the grid aligned setup is robust and---for important quantities like pressure dips and streaming velocities--numerically converged. We attribute this difference to excessive numerical diffusion in the diagonal B-field case. Note that since almost all streaming motion is field-aligned, it is also diagonal to the grid. We document this here as a caution to other researchers, so they can be aware of this potential numerical pitfall. Sometimes, one avoids setups which are aligned with the numerical grid to avoid the carbuncle instability (e.g. see \citet{gronke20-mist}).

\begin{figure}
     \centering
\includegraphics[width=\columnwidth]{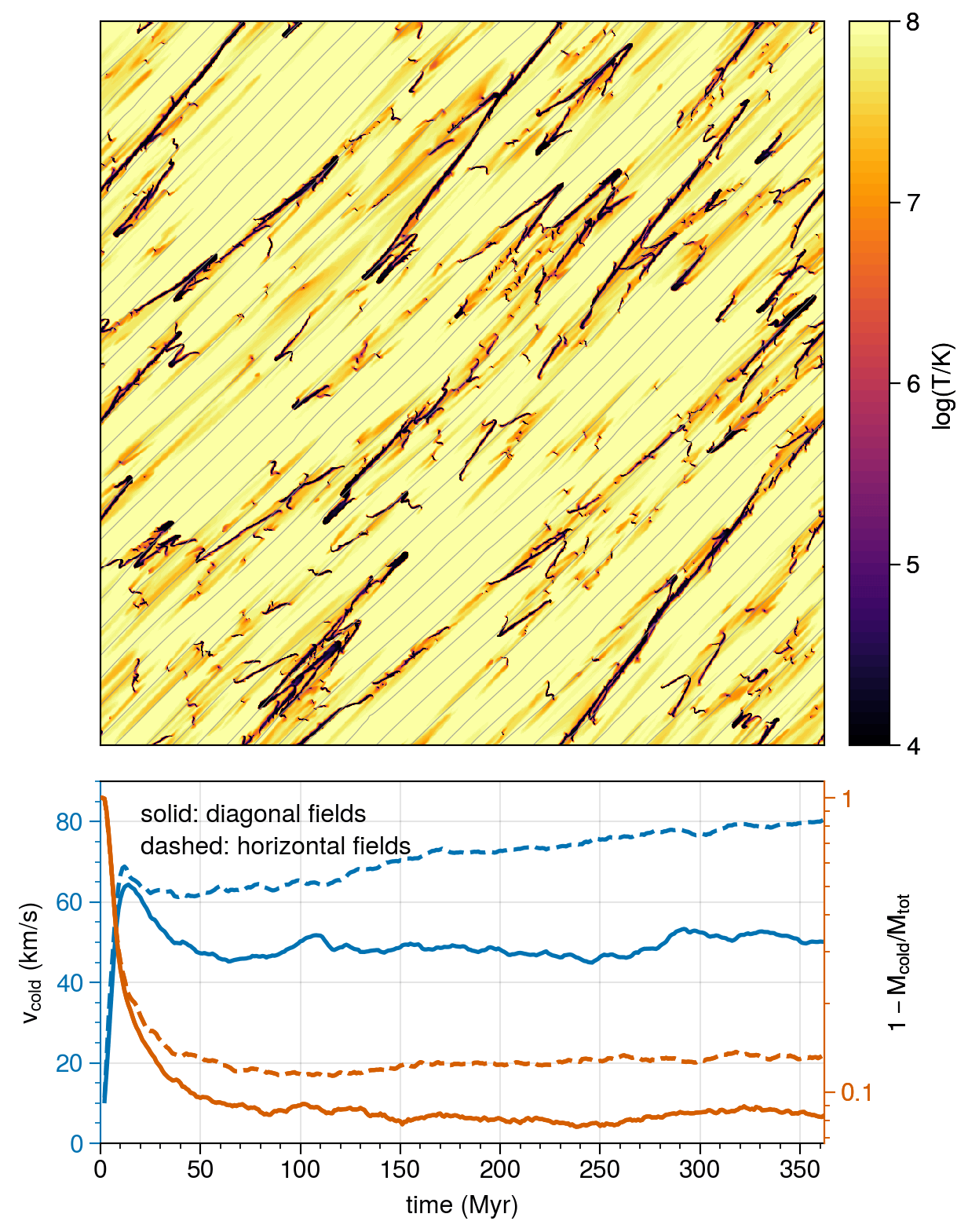}
       \caption[]{Top panel: temperature map of the run \texttt{mhd-bxy} at $t=0.36~{\rm Gyr}$. The gray streamlines denote the magnetic fields. Bottom: time evolution of cold gas velocity (blue) and mass fraction (orange). Results of \texttt{mhd-bxy} are shown as solid lines; and those of \mhdc\ are shown as dashed lines.}
\label{fig:bxy}.
\end{figure}

Figure \ref{app:bfields} shows an example. The run \texttt{mhd-bxy} has the same setup as \mhdc\, except that it adopts diagonal magnetic fields. Instead 
of collapsing to grid-scale cold clumps, as in \mhdc (Fig.~\ref{fig:tsl_tc0_cc} panel c) -- as one expects in the absence of thermal conduction --  \texttt{mhd-bxy} forms long filaments along the field lines  (top panel of Fig.~\ref{fig:bxy}). Since the field orientation is the only thing that changed, the difference must be purely numerical. 

\bsp	
\label{lastpage}
\end{document}